\PassOptionsToPackage{table}{xcolor}
\documentclass[format=acmsmall, review=false, screen=true]{acmart}

\usepackage{soul}
\usepackage{amsfonts,amsmath}
\usepackage{mathrsfs}
\usepackage{bbm}
\usepackage{mdframed}
\usepackage{arydshln}
\usepackage{listings}
\usepackage{lstautogobble}
\usepackage[ruled, linesnumbered]{algorithm2e}
\usepackage[noend]{algpseudocode}
\usepackage{amsthm}
\usepackage{booktabs} 
\usepackage{adjustbox}
\usepackage{rotating}
\usepackage[flushleft]{threeparttable}
\usepackage{caption}
\usepackage{multirow}
\usepackage{subcaption}
\usepackage{comment}
\usepackage{graphicx}
\usepackage{enumerate}
\usepackage{nicefrac}
\usepackage{xfrac}
\usepackage{tikz}
\usepackage{tkz-euclide}
\usetikzlibrary{fit,shapes,arrows,patterns}
\usepackage{pdflscape}
\usepackage{afterpage}
\usepackage{pifont}
\usepackage{units}
\usepackage{hyperref}

\PassOptionsToPackage{usenames,dvipsnames,svgnames,table}{xcolor}

\DeclareMathOperator{\sign}{sign}

\DeclareMathOperator*{\argmin}{arg\,min}
\DeclareMathSymbol{\mh}{\mathord}{operators}{`\-}

\DeclareMathOperator{\round}{round}

\makeatletter
\newcommand{\removelatexerror}{\let\@latex@error\@gobble}

\newtheorem{definition}{Definition}
\newtheorem{insight}{Insight}

\newtheorem{assumption}{Assumption}

\newcommand{\ignore}[1]{}

\definecolor{dkgreen}{rgb}{0,0.6,0}
\definecolor{mauve}{rgb}{0.58,0,0.82}

\newcommand{\specialcell}[2][c]{\begin{tabular}[#1]{@{}c@{}}#2\end{tabular}}

\newenvironment{rotatepage}%
{\clearpage\pagebreak[4]\global\pdfpageattr\expandafter{\the\pdfpageattr/Rotate 90}}%
{\clearpage\pagebreak[4]\global\pdfpageattr\expandafter{\the\pdfpageattr/Rotate 0}}%

\newcommand{\vthead}[1]{\makebox[1em][l]{\rotatebox{90}{#1}}}

\newcommand{\pie}[1]{
	\hspace{-.2ex}\hskip-\tabcolsep
	\begin{tikzpicture}[scale=0.8, baseline=-.5ex]
	\draw (0,0) circle (1ex);
	\fill[white] (0,0) circle (1ex);
	\ifthenelse{\equal{#1}{0}}{
	}{
		\fill (0,1ex) arc (90:int(#1*90+90):1ex) -- (0,0) -- cycle;
	}
	\end{tikzpicture}
	\hspace{-1.6ex}\hskip-\tabcolsep
}

\newcommand{\pieg}{
	\hspace{-.2ex}\hskip-\tabcolsep
	\begin{tikzpicture}[scale=0.8, baseline=-.5ex]
	\draw (0,0) circle (1ex);
	\fill[black, draw] (0,0) circle (0.5ex);
	\end{tikzpicture}
	\hspace{-1.6ex}\hskip-\tabcolsep
}

\newcommand{\emptycirc}{%
	\begin{tikzpicture}[scale=0.8, baseline=-.5ex]
	\draw (0,0) circle (1ex);
	\end{tikzpicture}
} 
\newcommand{\dotcirc}{%
	\begin{tikzpicture}[scale=0.8, baseline=-.5ex] 
	\draw (0,0) circle (1ex);
	\fill[black, draw] (0,0) circle (0.5ex);
	\end{tikzpicture}
}
\newcommand{\fullcirc}{
	\begin{tikzpicture}[scale=0.8, baseline=-.5ex] 
	\draw (0,0) circle (1ex);
	\fill (0,0) circle (1ex);
	\end{tikzpicture}
}

\AtBeginDocument{%
  \providecommand\BibTeX{{%
    \normalfont B\kern-0.5em{\scshape i\kern-0.25em b}\kern-0.8em\TeX}}}

\setcopyright{acmcopyright}
\copyrightyear{2021}
\acmYear{2021}

\acmJournal{CSUR}

\begin{document}

\title{Arms Race in Adversarial Malware Detection: A Survey}

\author{Deqiang Li}
\affiliation{
	\institution{Nanjing University of Science and Technology}
	\country{China}
}

\author{Qianmu Li}
\affiliation{
	\institution{Nanjing University of Science and Technology}
	\country{China}
}

\author{Yanfang (Fanny) Ye}
\affiliation{%
	\institution{Case Western Reserve University}
	\streetaddress{10900 Euclid Ave.}
	\city{Cleveland}
	\state{Ohio}
	\postcode{44106}
	\country{USA}
}

\author{Shouhuai Xu}
\affiliation{%
	\institution{University of Colorado Colorado Springs}
	\streetaddress{1420 Austin Bluffs Pkwy}
	\city{Colorado Springs}
	\state{Colorado}
	\postcode{80918}
	\country{USA}
}
\email{sxu@uccs.edu}
\thanks{Work was partly done when Shouhuai Xu was affiliated with University of Texas at San Antonio, One UTSA Circle, San Antonio, Texas 78249, USA} 

\renewcommand{\shortauthors}{D. Li, Q. Li, F. Ye, and S. Xu}

\begin{abstract}
Malicious software (malware) is a major cyber threat that has to be tackled with Machine Learning (ML) techniques because millions of new malware examples are injected into cyberspace on a daily basis. However, ML is vulnerable to attacks known as adversarial examples. In this paper, we survey and systematize the field of Adversarial Malware Detection (AMD) through the lens of a unified conceptual framework of assumptions, attacks, defenses, and security properties. This not only leads us to map attacks and defenses to partial order structures, but also allows us to clearly describe the attack-defense arms race in the AMD context. We draw a number of insights, including: knowing the defender's feature set is critical to the success of transfer attacks; the effectiveness of practical evasion attacks largely depends on the attacker's freedom in conducting manipulations in the problem space; knowing the attacker's manipulation set is critical to the defender's success; the effectiveness of adversarial training depends on the defender's capability in identifying the most powerful attack. We also discuss a number of future research directions. 
\end{abstract}

\ccsdesc[500]{Security and privacy~Malware and its mitigation}
\ccsdesc[500]{Theory of computation~Adversarial learning}

\keywords{Malware Detection, Adversarial Machine Learning, Evasion Attacks, Poisoning Attacks}

\maketitle

\section{Introduction} \label{sec:intro}

Malware (malicious software) is a big cyber threat and has received a due amount of attention. For instance, Kaspersky reports that 21,643,946 unique malicious files were detected in the year 2018, 24,610,126 in 2019, and 33,412,568 in 2020 \cite{GREAT:Kaspersky,kasparsky:Online}.
A popular defense against malware is to use  {\em signature-based} detectors \cite{egele2012survey}, where a signature is often extracted by malware analysts from known malware examples. This approach has two drawbacks: signatures are tedious to extract and can be evaded \cite{cisco:Online} by a range of techniques (e.g., encryption, repacking, polymorphism \cite{moser2007limits,shaid2015memory,5633410,song2008bitblaze,aprville2014hide,bilge2012before,jung:avpass-bh}). This incompetence has motivated the use of Machine Learning (ML) based malware detectors, which can be automated to some degree and can possibly detect \emph{new} malware examples (via model generalization or knowledge adaptation  \cite{DBLP:journals/csur/YeLAI17,tamersoy2014guilt, hou2017hindroid,Zhu:2016:FAE:2976749.2978304,bartos2016optimized,DBLP:conf/ijcai/HouYSA18,DBLP:conf/kdd/FanHZYA18,DBLP:journals/tifs/DuSCCX18,Daniel:NDSS,mariconti2016mamadroid,ye2009sbmds,kendall2007practical, KIM201883}). More recently, Deep Learning (DL) has been used for malware detection (see, e.g., \cite{raff2017malware,vinayakumar2018deepmalnet,saxe2015deep}). 

While promising, ML-based malware detectors are vulnerable to attacks known as {\em adversarial examples} \cite{Biggio:Evasion, rndic_laskov,huang2011adversarial}. 
There are two kinds of attacks. One is {\em evasion attack}, where the attacker perturbs test examples to {\em adversarial examples} to evade malware detectors \cite{szegedyZSBEGF13,grosse2017adversarial,Biggio:Evasion,al2018adversarial,DBLP:conf/ijcai/HouYSA18,papernotMGJCS16, demontis2017yes, wang_2017}. The other is {\em poisoning attack}, where the attacker manipulates the training dataset for learning malware detectors \cite{chen2018automated,suciu2018does,demontis2018intriguing}. These attacks usher in the new field of Adversarial Malware Detection (AMD) \cite{316904628,dang2017evading,demontis2017yes,Biggio:Evasion,rndic_laskov,xu2014evasion,Chen:2017:SES,wang_2017,grosse2017adversarial,suciu2018does,demontis2018intriguing}. 

The state-of-the-art in AMD is that there are some specific results scattered in the literature but there is no systematic understanding.
This is true despite that there have been attempts at systematizing the related field of Adversarial Machine Learning (AML) \cite{barreno2010security,huang2011adversarial,DBLP:journals/corr/abs-1810-00069,yuan2019adversarial}, which however cannot be automatically translated to AMD. This is so because malware detection 
has three {\em unique} characteristics which are not exhibited by the other application domains (e.g., image or audio processing). (i) There are no common, standard feature definitions because both attackers and defenders can define their own features to represent computer files. As a consequence, attackers can leverage this ``freedom'' in feature definition to craft adversarial examples. (ii) Malware features are often discrete rather than continuous and program files are often highly structured with multiple modalities. This means that arbitrarily perturbing malware files or their feature representations might make the perturbed files no more executable. 
This also means that the discrete domain makes perturbation a {\em non-differentiable} and {\em non-convex} task. 
(iii) Any meaningful perturbation to a malware example or its feature representation must preserve its malicious functionality. For example, the Android Package Kit (APK) requires that the used permissions are publicized in the AndroidManifest.xml, meaning that removing permissions in this manifest file would incur a runtime error. The preceding (ii) and (iii) make both the attacker's and defender's tasks more challenging than their 
counterparts where small perturbations are not noticeable (e.g., images).


\noindent{\bf Our Contributions}. We propose a conceptual framework for systematizing the AMD field through the lens of {\em assumptions}, {\em attacks}, {\em defenses}, and {\em security properties}. In specifying these, we seek rigorous definitions whenever possible, while noting that these definitions have been scattered in the literature. Rigorous definitions are important because they can serve as a common reference for future studies. The framework allows us to map the known attacks and defenses into some partial order structures and systematize the AMD attack-defense arms race. 

We make a number of observations, including: (i) the indiscriminate attack that treats malicious examples as equally important has been extensively investigated, but targeted and availability attacks are much less investigated; (ii) the evasion attack is much more extensively studied than the poisoning attack; (iii) there is no silver-bullet defense against evasion and poisoning attacks; (iv) sanitizing examples is effective against black-box and grey-box attacks, but not white-box attacks; (v) AMD security properties have been evaluated empirically rather than rigorously; (vi) there is no theoretical evidence to support that the effectiveness of defense techniques on the training set can generalize to other adversarial examples.

We draw a number of insights, including: (i) knowing defender's feature set is critical to the success of transfer attacks, highlighting the importance of keeping defender's feature set secret (e.g., by randomizing defender's feature set);
(ii) the effectiveness of practical evasion attacks largely depends on the attacker's degree of freedom in conducting manipulations in the problem space (i.e., a small degree of freedom means harder to succeed); (iii) effective defenses often require the defender to know the attacker's manipulation set, explaining from one perspective why it is hard to design effective defenses;
(iv) effectiveness of adversarial training depends on the defender's capability in identifying the most powerful attack.

Finally, we discuss a number of future research directions, which hopefully will inspire and encourage many researchers to explore them.

\smallskip

\noindent{\bf Related Work}.
The closely related prior work is Maiorca et al. \cite{Maiorca:2019:TAM:3359984.3332184}, which surveys previous studies in adversarial malicious PDF document detection. In contrast, we consider the broader context of AMD and propose novel partial orders to accommodate AMD assumptions, attacks, defenses, and properties.
There are loosely-related prior studies, which survey prior AML studies (but not focusing on AMD), including \cite{yuan2019adversarial,barreno2010security,barreno2006can,liu2018survey, biggio2014security, papernot2016towardsF,DBLP:journals/corr/abs-1810-00069,serban2018adversarial}.
For example, Yuan et al. \cite{yuan2019adversarial} survey attack methods for generating adversarial examples, while briefly discussing evasion attacks in the AMD context; 
Barreno et al. \cite{barreno2010security,barreno2006can} propose a taxonomy of AML attacks
(causative vs. exploratory attacks, integrity vs. availability attacks, and targeted vs. indiscriminate attacks); Biggio et al. \cite{biggio2014security} propose a defense framework for protecting 
Support Vector Machines (SVMs) from evasion attacks, poisoning attacks and privacy violations; Papernot et al. \cite{papernot2016towardsF} systematize AML security and privacy with emphasis on demonstrating the trade-off between detection accuracy and robustness.

\smallskip

\noindent{\bf Paper Outline}.
Section \ref{sec:fw} describes our survey and systematization methodology and framework.
Section \ref{sec:systematizstion-AMD-studies} applies  our framework to systematize the literature AMD studies.
Section \ref{sec:cd} discusses future research directions. 
Section \ref{sec:conc} concludes the paper. 
\section{Survey and Systematization Methodology} \label{sec:fw}

\begin{table}[htbp!]
	\centering\caption{Main notations used in the paper \label{table:notations}}
	\begin{adjustbox}{width=0.99\textwidth}
	\begin{tabular}{l|p{.8\textwidth}}
		\hline
		Notation & Meaning\\\hline
		$\mathbb R$ ($\mathbb R_{+}$) & the set of (positive) real numbers \\
		$\mathcal{A},\mathcal{I}$ & attacker and defender (treated as algorithms)\\
		$\mathbb{P}$ & the probability function \\
		$z, z'\in\mathcal{Z}$ & $\mathcal{Z}$ is example space; $z'$ is obtained by perturbing $z$\\
		$S$ & defender $\mathcal{I}$'s feature set for representing files \\
		${\mathbf x}, {\mathbf x}'\in \mathcal X$ & $\mathcal X=\mathbb{R}^d$ is $d$-dimensional feature space; ${\mathbf x},{\mathbf x}' \in \mathcal{X}$ are respectively feature representations of $z,z'\in \mathcal{Z}$ \\
		$\mathcal{Y}$, $y$ & $\mathcal{Y}$ is the label space of binary classification, $\mathcal{Y}=\{+/1,-/0\}$; $y\in \mathcal{Y}$ \\
		$\mathcal{D}=\mathcal Z \times \mathcal Y$ & the file-label (i.e., example-label) space \\
		$D_{train}\subset \mathcal D,n$ & the training set in file-label space; $n=|D_{train}|$  \\
		$D_{test}$ & the test set in file-label space \\
		$D_{poison},D'_{poison}$& $D'_{poison}$ is set of adversarial file-label pairs obtained by perturbing non-adversarial files in  $D_{poison}\subset\mathcal{D}$ \\
		$\mathcal{O}(z,z')$ & $\mathcal{O}(z,z'):\mathcal{Z}\times\mathcal{Z}\to\{{\tt true}, {\tt false}\}$ is an oracle telling if two files have the same functionality or not \\
		$\delta$ & a manipulation for perturbing files {\em with} preserving their functionalities \\
		$\mathcal{M}$, $\mathcal{Z}_\mathcal{M}\subseteq \mathcal{Z}$ & $\mathcal{M}$ is manipulation set in the problem space;
		$\mathcal{Z}_\mathcal{M}$ is set of adversarial files generated using $\mathcal{M}$\\
		$\mathbf{M}$, $\mathcal{X}_\mathbf{M}\subseteq \mathcal{X}$ & $\mathbf{M}$ is feature manipulation set;
		$\mathcal{X}_\mathbf{M} $ is set of adversarial feature vectors generated using $\mathbf{M}$ \\
		$\Gamma(z,z')$ & $\Gamma(z,z'):\mathcal Z \times \mathcal{Z} \to \mathbb R_+$ measures the degree of manipulation for perturbing $z \in \mathcal{Z}$ into $z'\in \mathcal{Z}$ \\
		$C(\mathbf{x},\mathbf{x}')$& $C(\mathbf{x},\mathbf{x}'):\mathcal X \times \mathcal X \to \mathbb R_+$ is the function measuring the cost incurred by changing feature vector $\mathbf{x}$ to $\mathbf{x}'$ \\
		$\delta_z \in \mathcal{M}$ &  $\delta_z$ is a set of manipulations of $z$ w.r.t. $z'$ \\
		$\delta_{\mathbf{x}} \in \mathbf{M}$ &  $\delta_{\mathbf{x}}={\mathbf x}' - {\mathbf x}$ is a perturbation vector of $\mathbf{x}$ w.r.t. $\mathbf{x}'$\\
		$\phi:\mathcal Z \to \mathcal{X} $ & feature extraction function;  $\mathbf{x}\leftarrow\phi(z)$, $\mathbf{x}'\leftarrow\phi(z')$ \\
		$\varphi,f$  & $\varphi:\mathcal{X} \to \mathbb{R}$ is classification function; $f:\mathcal Z \to \mathbb{R}$ is classifier $f=\varphi(\phi(\cdot))$; by abusing notation a little bit, we also use ``$+\leftarrow f(z)$'' to mean that $f$ predicts $z$ as malicious when
		$f(z)\geq \tau$ for a threshold $\tau$\\
		$F_{\theta}:\mathcal X \to \mathbb R$& machine learning algorithm  
		with parameters $\theta$ \\
		$L:\mathbb{R}
		\times\mathcal{Y}\to\mathbb{R}$ & loss function measuring prediction error of $F_\theta$ \\
		${\sf EL},{\sf WR},{\sf AT}$ & defense techniques: Ensemble Learning, Weight Regularization, Adversarial Training \\
		${\sf VL},{\sf RF},{\sf IT},{\sf CD},{\sf SE}$ & defense techniques: Verifiable Learning, Robust Feature, Input Transformation, Classifier ranDomization, Sanitizing Examples \\
		${\sf BE},{\sf OE},{\sf BP},{\sf OP}$ & attack tactics: basic and optimal evasion; basic and optimal poisoning\\
		{\sf GO},{\sf SF},{\sf MI} & attack techniques: Gradient-based Optimization, Sensitive Features, MImicry\\
		{\sf TR},{\sf HS},{\sf GM},{\sf MS} & attack techniques: TRansferability, Heuristic Search, Generative Model, Mixture Strategy \\
		$A_1,\ldots,A_5$ & the 5 attributes under $\mathcal{I}$'s control; they are known to $\mathcal{A}$ at respective degrees
		$a_1,\ldots,a_5$ \\
		$A_6,\ldots,A_9$ & the 4 attributes under $\mathcal{A}$'s control; they are known to $\mathcal{I}$ at respective degree $a_6,\ldots,a_9$ \\
		${\sf RR},{\sf CR},{\sf DR},{\sf TR}$ & security properties: Representation Robustness, Classification Robustness, Detection Robustness, Training Robustness\\
		\hline
	\end{tabular}
	\end{adjustbox}
\end{table}

\noindent{\bf Terminology, Scope and Notations}. 
In the AMD context, a defender $\mathcal{I}$ aims to use ML to detect or classify computer files as benign or malicious; i.e., we focus on {\em binary classification}. An attacker $\mathcal{A}$ attempts to make malicious files evade $\mathcal{I}$'s detection by leveraging {\em adversarial files} (interchangeably, {\em adversarial examples}). Adversarial malware examples are often generated by perturbing or manipulating malware examples, explaining why we will use the two terms, perturbation and manipulation, interchangeably. Adversarial attacks can be waged in the training phase of a ML model (a.k.a., poisoning attack) or in the test phase (a.k.a., evasion attack). It is worth mentioning that the {\em privacy violation} attack \cite{huang2011adversarial} is waged in addition to the preceding two attacks because $\mathcal{A}$ can always probe $\mathcal{I}$'s detectors.
A file, benign and malicious alike, is adversarial if it is intentionally crafted to (help malicious files) evade $\mathcal{I}$'s detection, and non-adversarial otherwise.
We focus on $\mathcal{I}$ using supervised learning to detect malicious files, which may be adversarial or non-adversarial because they co-exist in the real world with no self-identification.  This means that we do not consider the large body of malware detection literature that does not cope with AMD, which has been addressed elsewhere (e.g., \cite{DBLP:conf/sp/RossowDGKPPBS12}). 
Table \ref{table:notations} summarizes the main notations used in the paper. 

\subsection{Brief Review on ML-based Malware Detection} \label{sec:ml}

Let $\mathcal Z$ be the {\em example space} of benign/malicious adversarial/non-adversarial files. Let $\mathcal{Y}=\{+,-\}$ or $\mathcal{Y}=\{1,0\}$ be the {\em label space} of binary classification, where $+$/$1$ ($-$/$0$) means a file is malicious (benign). Let $\mathcal D =\mathcal Z \times \mathcal Y$ be the file-label (example-label) space. For training and evaluating a classifier in the absence of adversarial files, $\mathcal I$ is given a set $D\subset \mathcal{D}$ of non-adversarial benign/malicious files as well as their {\em ground-truth} labels. $\mathcal I$ splits $D$ into three disjoint sets: a training set $D_{train}=\{(z_i,y_i)\}_{i=1}^n$, a validation set for model selection, and a test set for evaluation. 
Each file $z_i\in \mathcal{Z}$ is characterized by a set $S$ of features and represented by a numerical vector $\mathbf{x}_i=(x_{i,1},\ldots,x_{i,d})$ in the $d$-dimensional {\em feature space} $\mathcal X=\mathbb R^d$, which accommodates both continuous and discrete feature representations \cite{abou2004n,Laskov13detectionof,tamersoy2014guilt,kolosnjaji2017empowering,hardy2016dl4md,hou2017hindroid}.
The process for obtaining feature representation $\mathbf{x}_i$ of $z_i\in \mathcal{Z}$ is called {\it feature extraction}, denoted by a function $\phi:\mathcal{Z}\to \mathcal{X}$ with ${\bf x}_i\leftarrow \phi(S,z_i)$. Because $\phi$ can be hand-crafted (denoted by $\phi_c$), automatically learned (denoted by $\phi_a$), or a hybrid of both \cite{bengio2013representation}, we unify them into $\phi$ such that $\phi(S,z)=\phi_a(\phi_c(S,z))$; when only manual (automatic) feature extraction is involved, we can set $\phi_a$ ($\phi_c$) as the identity map.
There are two kinds of features: {\em static features} are extracted via static analysis (e.g., strings, API calls
\cite{6891250,anderson2012improving}); {\em dynamic features} are extracted via dynamic analysis (e.g., instructions, registry activities
\cite{kendall2007practical,dornhackl2014malicious}).

\begin{figure}[!htbp]
	\centering
	\scalebox{0.32}{
	\includegraphics{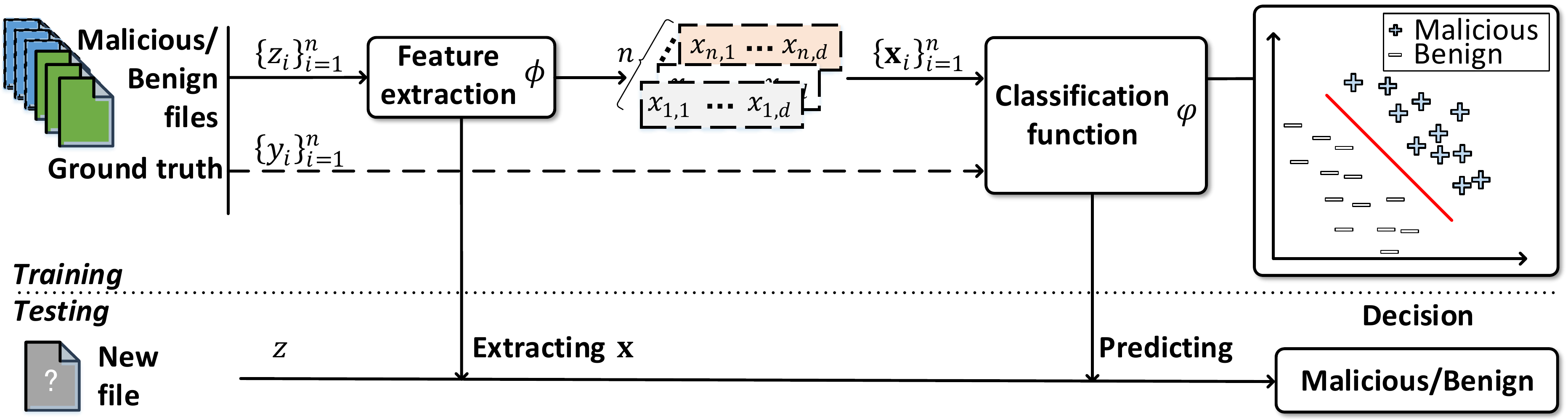}
	}
\caption{Illustration of ML-based malware detector.
}
	\label{fig:ml}
\end{figure}

As highlighted in Figure \ref{fig:ml},
$\mathcal{I}$ uses $\{(z_i,y_i)\}_{i=1}^n$ to learn a malware detector or classifier $f:\mathcal{Z}\to [0,1]$, where $f(z)=\varphi(\phi(S,z))$ is composed of feature extraction function $\phi: \mathcal{Z}\to \mathcal{X}$ and classification function $\varphi:\mathcal{X} \to [0,1]$.
Note that $f(z)\in [0,1]$, namely $\varphi({\bf x})\in [0,1]$ with ${\bf x}\leftarrow \phi(S,z)$, can be interpreted as the probability that $z$ is malicious (while noting that calibration may be needed \cite{DBLP:conf/icml/Niculescu-MizilC05}). For a given threshold $\tau\in[0,1]$, we further say (by slightly abusing notations) $z$ is labeled by $f$ as $+$, or $+\leftarrow f(z)$, if $f(z)\geq \tau$, and labeled as $-$ or $-\leftarrow f(z)$ otherwise. In practice, $f$ is often specified by a learning algorithm $F$ with learnable parameter $\theta$ (e.g., weights) and a hand-crafted feature extraction $\phi_c$; then, $\theta$ is tuned to minimize the empirical risk associated with a loss function $L:[0,1] \times \mathcal Y \to \mathbb{R}$ measuring the prediction error of $F_\theta$ \cite{vapnik1992principles} (e.g., cross-entropy \cite{goodfellow2016deep}), namely

\begin{equation}
\min \limits_{\theta}~\mathcal{L}(\theta, D_{train}) = \min \limits_{\theta}~\frac{1}{n}\sum_{(z_i, y_i)\in D_{train}}\left(L(F_\theta(\phi_c(S,z_i)), y_i)\right). \label{eq:rsk_emp}
\end{equation}

\noindent{\bf Example 1: The Drebin malware detector}. Drebin is an Android malware detector trained from static features \cite{Daniel:NDSS}. Table \ref{table:drebin} summarizes Drebin's feature set, which includes 4 subsets 
\begin{table}[htbp!]
	\centering\caption{Drebin features \label{table:drebin}}
	\begin{tabular}{c|p{.32\textwidth}}
		\hline
		\multicolumn{2}{c}{Feature set} \\\hline
		\multirow{4}{*}{Manifest} 
		& $S_1$~~Hardware components \\
		& $S_2$~~Requested permissions\\
		& $S_3$~~App components \\
		& $S_4$~~Filtered intents \\\hline
		\multirow{4}{*}{Dexcode}
		& $S_5$~~Restricted API calls\\
		& $S_6$~~Used permissions\\
		& $S_7$~~Suspicious API calls\\
		& $S_8$~~Network addresses\\
		\hline
	\end{tabular}
\end{table}
of features $S_1, S_2, S_3, S_4$ extracted from the AndroidManifest.xml and another 4 subsets of features $S_5, S_6, S_7, S_8$ extracted from the disassembled DEX code files (recalling that DEX code is compiled from a program written in some language and can be understood by the Android Runtime). Specifically, $S_1$ contains features related to the access of an Android package (APK) to smartphone hardware (e.g., camera, touchscreen, or GPS module); $S_2$ contains features related to APK's requested permissions listed in the manifest prior to installation; $S_3$ contains features related to application components (e.g., {\em activities}, {\em service}, {\em receivers}, and {\em providers}.); $S_4$ contains features related to APK's communications with the other APKs; $S_5$ contains features related to critical system API calls, which cannot run without appropriate permissions or the {\em root} privilege; $S_6$ contains features corresponding to the used permissions; $S_7$ contains features related to API calls that can access sensitive data or resources in a smartphone; and $S_8$ contains features related to IP addresses, hostnames and URLs found in the disassembled codes.
The feature representation is binary, meaning $\phi=\phi_c:\mathcal{Z} \mapsto \{0,1\}^d$ with $|S|=d$ and ${\bf x}=(x_1,\ldots,x_d)$, where $x_i=1$ if the corresponding feature is present in the APK  $z$ and $x_i=0$ otherwise. A file $z$ in the feature space looks like the following: 
\[
\mathbf x = \phi(z) \to
\begin{pmatrix}
\cdots \\ 0\\ 1\\
\cdots \\ 1\\ 0\\
\cdots\\
\end{pmatrix}
\begin{array}{ll}
\cdots & \multirow{4}{*}{\hspace{-1mm}\bigg \} $S_2$ }\\
\texttt{\small permission::SEND\_SMS} \\
\texttt{\small permission::READ\_CONTACTS}\\
\cdots & \multirow{4}{*}{\hspace{-1mm}\bigg \} $S_5$ }\\
\texttt{\small api\_call::getDeviceID}\\
\texttt{\small api\_call::setWifiEnabled}\\
\cdots & \\
\end{array}
\]
Drebin uses a linear Support Vector Machine (SVM) to learn classifiers.

\begin{figure}[!htbp]
	\centering
	\includegraphics[width=.7\textwidth]{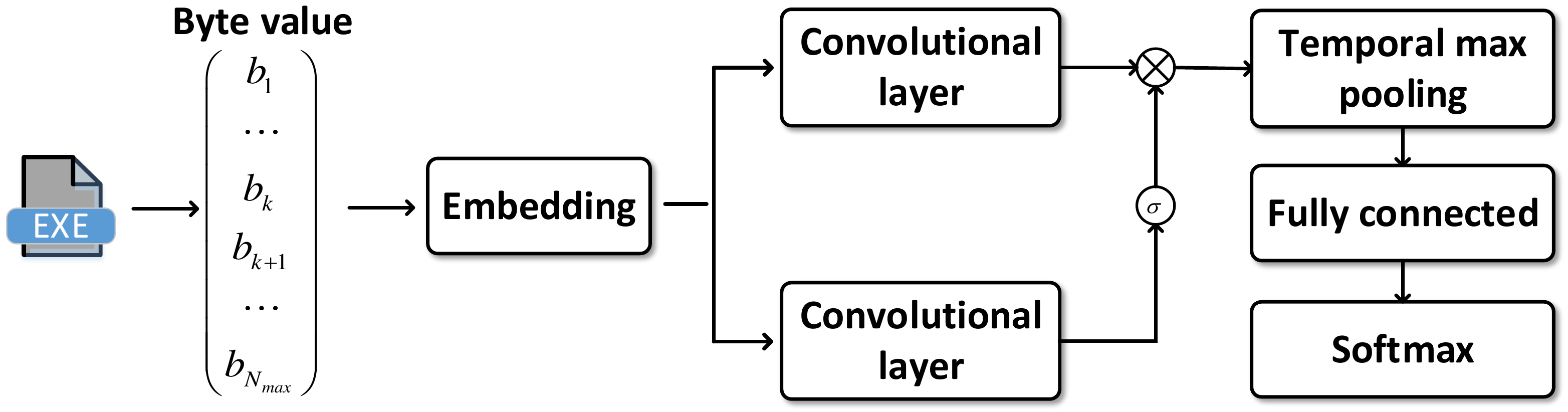}
	\caption{MalConv architecture  \cite{raff2017malware}.}
	\label{fig:malconv}
\end{figure}

\smallskip
\noindent{\bf Example 2: The MalConv malware detector}.
MalConv \cite{raff2017malware} is Convolutional Neural Network (CNN)-based Windows Portable Executable (PE) malware detector learned from raw binary programs (i.e., end-to-end detection)
\cite{kim2014convolutional}. 
Figure \ref{fig:malconv} depicts its architecture.
The sequence of binary code is transformed into byte values (between 0 to 255) with the maximum length bounded by $N_{max}$ (e.g., $N_{max}=2^{21}$ bytes or 2MB). Each byte is further mapped into a real-valued vector using the embedding \cite{collobert2011natural}. The CNN layer and pooling layer learn abstract representations. The embedding, CNN and pooling layers belong to feature extraction $\phi_a$, and the fully-connected and softmax layers belong to the classification operation $\varphi$.

\subsection{Framework}

We systematize AMD studies through the lens of four aspects: (i) the assumptions that are made; (ii) the attack or threat model in terms of attacker ${\mathcal A}$'s {\em objective} and ${\mathcal A}$'s {\em input}, with the latter including ${\mathcal A}$'s information about the defender $\mathcal{I}$ and ${\mathcal A}$'s own; (iii) the defense in terms of $\mathcal{I}$'s {\em objective} and $\mathcal{I}$'s {\em input}, with the latter including $\mathcal{I}$'s information about $\mathcal{A}$ and $\mathcal{I}$'s own; (iv) the security properties that are at stake. These four aspects are respectively elaborated below.

\subsubsection{Systematizing Assumptions} \label{sec:assumption}

Five assumptions have been made in the AMD literature. Assumption \ref{assumption:iid} below says that the data samples in $D$ are Independent and Identically Distributed (IID), which is a strong assumption and researchers have started to weaken it \cite{smutz2016tree,DBLP:journals/corr/GrosseMP0M17}.

\begin{assumption}[{\sf IID} assumption; see, e.g., \cite{shalev2014understanding}]
\label{assumption:iid}
Computer files in training data and testing data are independently drawn from the same distribution.
\end{assumption}

Assumption \ref{assumption:oracle} below is adapted from AML context, where humans can serve as an oracle $\mathcal{O}$ for determining whether two images are the same \cite{wang2016theoretical}. In the AMD context, $\mathcal{O}$ can be instantiated as (or approximated by) malware analysts \cite{rndic_laskov,Chen:2017:SES,al2018adversarial,khasawneh2017rhmd} or automated tools (e.g., Sandbox \cite{dang2017evading,316904628}), with the latter often using heuristic rules produced by malware analysts (e.g., YARA \cite{yara:Online}).

\begin{assumption}[{\sf Oracle} assumption; adapted from \cite{wang2016theoretical}]
\label{assumption:oracle}
There is an oracle $\mathcal{O}:\mathcal{Z}\times\mathcal{Z}\to\{{\tt true},{\tt false}\}$ that tells if two files $z,z'\in\mathcal{Z}$ have the same functionality or not; ${\tt true}\leftarrow {\mathcal{O}}(z,z')$ if and only if $z$ and $z'$ have the same functionality. 
\end{assumption}

Assumption \ref{assumption:measurability} below says that there is a way to measure the degree of manipulations by which one file 
is transformed to another.

\begin{assumption}[{\sf Measurability} assumption \cite{kolosnjaji2018adversarial, dang2017evading}]
\label{assumption:measurability}
There is a function $\Gamma(z,z'):\mathcal Z \times \mathcal Z \to \mathbb R_+$ that measures the degree of manipulations according to which a file $z'\in \mathcal{Z}$ can be derived from the file $z \in \mathcal{Z}$.
\end{assumption}

Since Assumption \ref{assumption:measurability} is often difficult to validate, $\Gamma(z,z')$ may be replaced by a function that quantifies the degree of manipulation that can turn feature representation $\mathbf{x}$ into $\mathbf{x}'$, where
$\mathbf{x}=\phi(S,z)$ and $\mathbf{x}'=\phi(S,z')$. This leads to:

\begin{assumption}[{\sf Smoothness} assumption \cite{bengio2013representation}]
\label{assumption:smoothness}
\ignore{
This assumption says that for any $z,z'\in \mathcal{Z}$ and $\delta \in \Delta$, it holds that $C(\phi(z),\phi(z'))\approx 0$ when $(z'\leftarrow \mathcal{A}(z,\delta)) \wedge (\Gamma(z,z')\approx 0) \wedge ({\tt true}\leftarrow {\mathcal{O}(z,z')})$.
}
There is a function $C(\mathbf{x},\mathbf{x}'):\mathcal X \times \mathcal X \to \mathbb R_+$ such that $C(\phi(S,z),\phi(S,z'))\approx 0$ when $ (\Gamma(z,z')\approx 0) \wedge ({\tt true}\leftarrow {\mathcal{O}(z,z')})$.
\end{assumption}

Assumption \ref{assumption:inverse} below says that the inverse of feature extraction, $\phi^{-1}$, is solvable so that a perturbed representation $\mathbf{x}'$ can be mapped back to a legitimate file. 
\begin{assumption}[{\sf Invertibility} assumption \cite{DBLP:LiL20}] \label{assumption:inverse}
Feature extraction $\phi$ is invertible, meaning that given $\mathbf{x}'$, the function $\phi^{-1}:\mathcal{X}\to\mathcal{Z}$ produces $z'=\phi^{-1}(\mathbf{x}')$. 
\end{assumption}
Recall that the feature extraction function $\phi$ may be composed of a hand-crafted $\phi_c$ and an automated $\phi_a$, where $\phi_c$ may be neither differentiable nor invertible \cite{Biggio:Evasion,pierazzi2020problemspace}. This means $\mathbf{x}'$ may not be mapped to a legitimate file. Researchers tend to relax the assumption by overlooking the interdependent features \cite{rndic_laskov,DBLP:LiL20}, while suffering from the side-effect $\mathbf{x}'\neq \phi(\phi^{-1}(\mathbf{x}'))$ \cite{rndic_laskov,pierazzi2020problemspace}.

\subsubsection{Systematizing Attacks} \label{sec:sys-attack}

We systematize attacks from two perspectives: {\em attacker's objective} (i.e., what the attacker attempts to accomplish) and {\em attacker's input} (i.e., what leverages the attacker can use). Whenever possible, we seek rigorous definitions to specify the attacker's input, while noting that these definitions have been scattered in the literature. We believe this specification is important because it can serve as a common reference for future studies. To demonstrate this, we discuss how to apply it to formulate a partial-order structure for comparing attacks.

\smallskip

\noindent{\bf Attacker's Objective}.
There are three kinds of objectives: (i)
{\sf Indiscriminate}, meaning $\mathcal{A}$ attempts to cause as many false-negatives as possible \cite{wang_2017,rosenberg2017generic,grosse2017adversarial,anderson2017evading,yang2017malware,YeFOSINT-SI-2018}; 
(ii) {\sf Targeted}, meaning $\mathcal{A}$ attempts to cause specific false-negatives (i.e., making certain malicious files evade the detection \cite{demontis2018intriguing,suciu2018does});
(iii) {\sf Availability}, meaning $\mathcal{A}$ attempts to frustrate defender $\mathcal{I}$ by rendering $\mathcal{I}$'s classifier $f$ unusable (e.g., causing substantially high false-positives \cite{chen2018automated,biggio2012poisoning,demontis2018intriguing,provos2007virtual,munoz2017towards}).

\begin{table}[!htbp]
\caption{Attributes for specifying $\mathcal{A}$'s and $\mathcal{I}$'s input.}
	\begin{minipage}{\columnwidth}
		\centering
		\begin{tabular}{l|c|c}
		\hline
Attributes & Attacker $\mathcal{A}$'s input & Defender $\mathcal{I}$'s input \\\hline\hline
\multicolumn{3}{l}{Attributes under $\mathcal{I}$'s control but may be known to $\mathcal{A}$ to some extent} \\\hline
$A_1$: Training set $D_{train}$ & $a_1\in [0,1]$ & 1 \\\hline
$A_2$: Defense technique & $a_2\in \{0,1\}$ & 1 \\\hline
$A_3$: Feature set $S$ & $a_3\in [0,1]$ & 1\\\hline
$A_4$: Learning algorithm $F_\theta$ & $a_4\in [0,1]$ & 1 \\\hline
$A_5$: Response  & $a_5\in \{0,1\}$ & 1 \\\hline\hline
\multicolumn{3}{l}{Attributes under $\mathcal{A}$'s control but may be known to $\mathcal{I}$ to some extent} \\\hline
$A_6$: Manipulation set $\mathcal{M}$ & 1 & $a_6\in [0,1]$ \\\hline
$A_7$: Attack tactic & 1 & $a_7\in \{0,1\}$ \\\hline
$A_8$: Attack technique & 1& $a_8\in \{0,1\}$ \\\hline
$A_9$: Adversarial examples & 1 & $a_9\in [0,1]$ \\\hline
		\end{tabular}
	\end{minipage}
	\label{tab:attr}
\end{table}

\smallskip

\noindent{\bf Attacker's Input}.
Table \ref{tab:attr} highlights the attributes we define to describe $\mathcal{A}$'s input, including: five attributes $A_1,\ldots,A_5$ that are under $\mathcal{I}$'s control (indicated by 1) but may be known to $\mathcal{A}$ at some extent $a_1,\ldots,a_5$, respectively; and four attributes $A_6,\ldots,A_9$ that are under $\mathcal{A}$'s control (indicated by 1). These attributes are elaborated below.

({\bf i}) $A_1$: it describes $\mathcal{I}$'s training set $D_{train}$ for learning classifier $f$. 
We use $a_1\in [0,1]$ to represent the extent at which $D_{train}$ is known to $\mathcal{A}$.
Let $\hat{D}_{train}$ be the training files that are known to $\mathcal{A}$. Then, $a_1=|\hat{D}_{train}\cap{D}_{train}|/|{D}_{train}|$.

({\bf ii}) $A_2$: it describes $\mathcal{I}$'s techniques, which can be Ensemble Learning (${\sf EL}$), Weight Regularization (${\sf WR}$), Adversarial Training (${\sf AT}$), Verifiable Learning (${\sf VL}$), Robust Feature ({\sf RF}), Input Transformation (${\sf IT}$), Classifier ranDomization (${\sf CD}$), Sanitizing Examples ({\sf SE}). Let $A_2\in\{${\sf EL},{\sf WR},{\sf AT},{\sf VL},{\sf RF},{\sf IT},{\sf CD},{\sf SE}$\}$ and $a_2\in \{0,1\}$ such that $a_2=0$ means $\mathcal{A}$ does not know $\mathcal{I}$'s techniques and $a_2=1$ means $\mathcal{A}$ knows $\mathcal{I}$'s technique. The techniques are defined as follows.
Definition \ref{definition:el} says that $\mathcal{I}$ constructs multiple classifiers and uses them collectively in malware detection.
\begin{definition} [ensemble learning or {\sf EL} \cite{zhou2012ensemble}] \label{definition:el}
Let $\mathcal{H}$ be $\mathcal{I}$'s classifier space.
Given $K$ classifiers $\{f_i\}_{i=1}^{K}$ where $f_i\in\mathcal{H}$ and $f_i:\mathcal{Z}\to[0,1]$, let $f_i$ be assigned with weight $\omega_i$ with $\sum_{i=1}^K{\omega_i}=1$ and $\omega_i\geq 0$. Then, $f=\sum_{i=1}^{K}\omega_i f_i$.
\end{definition}

Definition \ref{definition:wr} says that $\mathcal{I}$ uses regularization (e.g., $\ell_2$ regularization \cite{10.1145/1015330.1015435} or {\em dropout} \cite{srivastava2014dropout}) to decrease model's sensitivity to adversarial examples.

\begin{definition} [weight regularization or {\sf WR} \cite{goodfellow2016deep}] \label{definition:wr}
Given a regularization item $\Omega$ (e.g., constraints imposed on the learnable parameters), the empirical risk is
$\min \limits_{\theta}~\left[\mathcal{L}(\theta, D_{train})+\Omega(\theta)\right]$, where $\mathcal{L}$ is defined in Eq. \eqref{eq:rsk_emp}.
\end{definition}

Definition \ref{definition:at} says that $\mathcal{I}$ proactively makes its classifier $f$ perceive some information about adversarial files. That is, $\mathcal{I}$ augments the training set by incorporating adversarial examples that may be produced by $\mathcal{I}$, $\mathcal{A}$, or both.
\begin{definition} [adversarial training or {\sf AT} \cite{grosse2017adversarial}] \label{definition:at}
Let $D'$ denote a set of adversarial file-label pairs.
Then, $\mathcal{I}$ tunes model parameters by minimizing the empirical risk: $\min \limits_{\theta}~\left[\mathcal{L}(\theta, D_{train}) + \beta \mathcal{L}(\theta, D')\right]$, where $\beta\geq 0$ denotes a balance factor.
\end{definition}

Definition \ref{definiton:vl} says that $\mathcal{I}$ intentionally over-estimates the error incurred by $\mathcal{A}$'s manipulations and then minimizes it.

\begin{definition}[verifiable learning or {\sf VL} \cite{DBLP:conf/icml/WongK18}] \label{definiton:vl}
Given $(z,y)\in D_{train}$ and a manipulation set $\hat{\mathcal{M}}$ known by $\mathcal{I}$, let $z(\hat{\mathcal{M}})$ denote the upper and lower boundaries on $\hat{\mathcal{M}}$. Then, this defense technique minimizes the following loss function derived from Eq.\eqref{eq:rsk_emp}:
$L(F_\theta(\phi_c(S,z)), y)+\beta L(F_\theta(\phi_c(S,z(\hat{\mathcal{M}}))), y).$
\end{definition}

Definition \ref{eq:robust-feature-defense-technique} says that $\mathcal{I}$ uses a set of features $S^*\subseteq S$ that can lead to higher detection capability against adversarial example attacks.
\begin{definition} [robust feature or {\sf RF}; adapted from \cite{zhang2016adversarial}]
\label{eq:robust-feature-defense-technique}
Given a training set $D_{train} \cup D'$ that contains (adversarial) file-label pairs, the set of robust feature set $S^*$ is  $$S^*=\argmin_{\tilde{S}\subset S}\sum_{(z,y)\in {D_{train}\cup D'}}L(\widetilde{F}_\theta(\phi_c(\tilde{S},z)),y),$$ 
where $\widetilde{F}_\theta$ is $F_\theta$ or a simplified learning algorithm that is computationally faster than $F_\theta$ \cite{zhang2016adversarial}.
\end{definition}

Definition \ref{def:it} says that $\mathcal{I}$ aims to use non-learning methods (e.g., de-obfuscation as shown in Proguard \cite{10.1145/2976749.2978422}) to offset $\mathcal{A}$'s manipulations.

\begin{definition} [input transformation or {\sf IT}, adapted from \cite{YeFOSINT-SI-2018}] \label{def:it}
Let ${\tt IT}:\mathcal{Z}\to\mathcal{Z}$ denote an input transformation in the file space. Given file $z$ and transformation ${\tt IT}$, the classifier is $f=\varphi(\phi({\tt IT}(z)))$.
\end{definition}
Definition \ref{definition:cd} says that $\mathcal{I}$ randomly chooses $m$ classifiers and uses their results for prediction. That is, $\mathcal{I}$ aims to randomize the feature representation used by $f$, the learning algorithm, and/or response to $\mathcal{A}$'s queries (to prevent $\mathcal{A}$ from inferring information about $f$). 
\begin{definition} [classifier randomization or {\sf CD}; adapted from \cite{khasawneh2017rhmd}] \label{definition:cd}
Given $\mathcal{I}$'s classifier space $\mathcal{H}$ and an input file $z$, $\mathcal{I}$ randomly selects $m$ classifiers from $\mathcal{H}$ with replacement, say $\{f_i\}_{i=1}^m$. Then,
	$f=\frac{1}{m}\sum_{i=1}^m~f_i(z)$. 
\end{definition}

Instead of enhancing malware detectors, Definition \ref{definition:se} provides an alternative that detects the adversarial examples for further analysis.
\begin{definition}[sanitizing examples or ${\sf SE}$; adapted from \cite{chen2018automated,carlini2017adversarial}] \label{definition:se}
$\mathcal{I}$ aims to detect adversarial files by using function 
${\sf flag}:\mathcal{Z}\to \{{\tt yes},{\tt no}\}$ to flag a file as adversarial ({\tt yes}) or not ({\tt no}).
\end{definition}

({\bf iii}) $A_3$: it describes $\mathcal{I}$'s feature set $S$. We use $a_3\in [0,1]$ to represent the extent at which $\mathcal{A}$ knows about $S$. Let $\hat{S}$
denote the features that are known to $\mathcal{A}$. Then,
$a_3=|\hat{S}\cap S|/|S|$.

({\bf iv}) $A_4$: it describes $\mathcal{I}$'s learning algorithm $F_\theta$, the set of trainable parameters $\theta$, and hyperparameters (which are set manually, e.g., $\beta$ in Definition \ref{definition:at})
\cite{narayanan2016adaptive,demontis2018intriguing}. We use $a_4\in [0,1]$ to represent that $\mathcal{A}$ knows an $a_4$ degree about $A_4$, where $a_4=0$ means $\mathcal{A}$ knows nothing and $a_4=1$ means $\mathcal{A}$ knows everything.

({\bf v}) $A_5$: it describes $\mathcal{I}$'s response to $\mathcal{A}$'s query to $f$ (if applicable), which is relevant because $\mathcal{A}$ can learn useful information about $f$ by observing $f$'s responses \cite{suykens1999least}.
We define $a_5\in \{0,1\}$ such that $a_5=0$ means there is a limit on the response that can be made by $\mathcal{A}$ to $f$
(referred as ${\sf LQ}$) and $a_5=1$ means there is no limit (referred as ${\sf FQ}$).

({\bf vi}) $A_6$: it describes $\mathcal{A}$'s manipulation set in the problem space, which describes perturbations for generating adversarial files (adapted from {\em perturbation set} in the AML literature \cite{tramer2019adversarial}):
\begin{eqnarray*}
\mathcal{M}=\{\delta: (z'\leftarrow \mathcal{A}(z,\delta)) \wedge ({\tt true}\leftarrow \mathcal{O}(z,z')
) \wedge (z \in \mathcal{Z}) \wedge (z'\neq z)\}.
\end{eqnarray*}
$\mathcal{M}$ is application-specific. For instance, an Android Package Kit (APK) permits adding codes or renaming class names \cite{grosse2017adversarial,demontis2017yes,chen2018android,DBLP:LiL20}, a Windows Portable Executable (PE) permits adding codes or changing PE {\em section} names \cite{anderson2018learning,kolosnjaji2018adversarial,DBLP:conf/itasec/DemetrioBLRA19,DBLP:journals/corr/abs-2008-07125}, and a Portable Document Format (PDF) file permits appending dead-code at its end \cite{rndic_laskov} or add new instructions \cite{DBLP:conf/ndss/CarmonyHYBZ16,316904628}. This means that a perturbation $\delta\in\mathcal{M}$ can be a tuple specifying an operator (e.g., addition or removal), an object (e.g., a feature used by $\mathcal{I}$), and other kinds of information (e.g., perturbation location in a file). 

Since it is often infeasible to enumerate the entire manipulation set, $\mathcal{A}$ may leverage an empirical one $\widetilde{\mathcal{M}}$ \cite{grosse2017adversarial,demontis2017yes,chen2018android,rndic_laskov,DBLP:LiL20,pierazzi2020problemspace}, which can be defined in the problem or feature space. Manipulations in the problem space must not violate the relevant constraints (e.g., adding APIs in an APK should not cause the request of unauthorized permissions). Manipulations in the feature space facilitate efficient computing via  gradient-based methods as long as the inverse feature mapping $\phi^{-1}$ is available.
Furthermore, we can use the manipulation set $\mathcal{M}$ to define a {\em feature manipulation set} $\mathbf{M}$:
\begin{eqnarray}
\mathbf{M}=\{\delta_{\mathbf{x}}=\mathbf{x}'-\mathbf{x}: (\mathbf{x}=\phi(z)) \land (\mathbf{x}'=\phi(z')) \land (z'\leftarrow \mathcal{A}(z,\delta)) \land (\delta \in \mathcal{M}) \land (z \in \mathcal{Z})
\}. \label{eq:manipulations}
\end{eqnarray}
In order to compute $\mathbf{M}$ efficiently, one strategy is to estimate a feature-space analog of $\widetilde{\mathcal{M}}$, denoted by $\widetilde{\mathbf{M}}$ \cite{rndic_laskov,smutz_2012}. This however demands resolving the invertibility Assumption \ref{assumption:inverse}.

({\bf vii}) $A_7$: it describes $\mathcal{A}$'s attack tactics. We consider two tactics: classifier {\em evasion} and classifier {\em poisoning}. For the evasion attack, we consider three variants: basic evasion (${\sf BE}$), optimal evasion 1 (${\sf OE1}$) and optimal evasion 2 (${\sf OE2}$).
For the poisoning attack, we consider two variants: basic poisoning (${\sf BP}$) and optimal poisoning (${\sf OP}$). Correspondingly, we have $A_7\in\{{\sf BE},{\sf OE1},{\sf OE 2},{\sf BP},{\sf OP}\}$.
These tactics are elaborated below, while noting that they do not explicitly call oracle $\mathcal{O}$ because definitions of manipulation sets $\mathcal{M}$ already assure that manipulations preserve functionalities of non-adversarial files.

As shown in Definition \ref{defintion:basic_evs}, the basic evasion attack is that $\mathcal{A}$ uses a set of perturbations $\delta_z\subseteq\mathcal{M}$ to manipulate a malicious file $z$, which is classified by $\mathcal{I}$'s classifier $f$ as $+\leftarrow f(z)$, to an adversarial file $z'$ such that $-\leftarrow f(z')$.
\begin{definition}[basic evasion or ${\sf BE}$ \cite{grosse2017adversarial}] \label{defintion:basic_evs}
$\mathcal{A}$ looks for $\delta_z\subseteq\mathcal{M}$ to achieve the following for $z\in \mathcal{Z}$ with $+\leftarrow f(z)$:
\begin{eqnarray*}
-\leftarrow f(z') ~\text{where}~
(z'\leftarrow \mathcal{A}(z,\delta_z))   \wedge
(\delta_z\subseteq\mathcal{M}).
\end{eqnarray*}
\end{definition}
As shown in Definition \ref{def:opt_evasion1}, the attacker attempts to minimize the degree of perturbations. In other words, this attack tactic is the same as {\sf BE}, except that $\mathcal{A}$ attempts to minimize the manipulation when perturbing a non-adversarial file $z\in \mathcal{Z}$ into an adversarial file $z'\in \mathcal{Z}$.
\begin{definition}[optimal evasion 1 or ${\sf OE1}$; adapted from \cite{carliniW16a}] \label{def:opt_evasion1}
$\mathcal{A}$ attempts to achieve the following for $z\in \mathcal{Z}$ with $+\leftarrow f(z)$:
\begin{eqnarray*}
\min \limits_{z'}
\Gamma(z', z) 
~\text{s.t.}~(z'\leftarrow \mathcal{A}(z,\delta_z))\land (\delta_z\subseteq\mathcal{M}) \land (-\leftarrow f(z')). 
\end{eqnarray*}
\end{definition}

As shown in Definition \ref{def:opt_evasion2}, the attacker attempts to maximize $\mathcal{I}$'s loss for waging high-confidence evasion attacks, while noting the small perturbations may be incorporated.

\begin{definition}[optimal evasion 2 or ${\sf OE2}$; adapted from \cite{Biggio:Evasion}] \label{def:opt_evasion2}
$\mathcal{A}$ attempts to achieve the following for $z\in \mathcal{Z}$ with $+\leftarrow f(z)$:
\begin{eqnarray*}
\max\limits_{z'}
L(F_\theta(\phi_c(S,z')),+)
~\text{s.t.}~(z'\leftarrow \mathcal{A}(z,\delta_z))\land (\delta_z\subseteq\mathcal{M}) \land (-\leftarrow f(z')).
\end{eqnarray*}
\end{definition}

Let $D'_{poison}\subset \mathcal{D}$ be a set of adversarial file-label pairs obtained by manipulating non-adversarial files in $D_{poison}$. Let $D'_{train}\leftarrow D_{train} \cup D'_{poison}$ be the contaminated training data for learning a classifier $f'$ with parameters $\theta'$. As shown in Definition \ref{def:basic poisoning attack}, the basic poisoning attack is that the attacker aims to make $f'$ mis-classify the files in a dataset $D_{target}$, while accommodating the attacks that $\mathcal{A}$ manipulates labels of the files in $D_{poison}$ \cite{paudice2018label}.
\begin{definition}[basic poisoning or ${\sf BP}$ \cite{barreno2010security}]
\label{def:basic poisoning attack}
Given a set $D_{target}$ of files where $+\leftarrow f(\dot{z})$ for $\dot{z}\in D_{target}$ and a set $D_{poison}$ of non-adversarial files, $\mathcal{A}$ attempts to perturb files in $D_{poison}$ to adversarial ones
$D'_{poison}=
\{(\mathcal{A}(z,\delta_z), \mathcal{A}(y)): ((z,y)\in D_{poison})\land (\delta_z\subseteq \mathcal{M})\land $ $ (\mathcal{A}(y)\in\{+,-\}) \}$ such that classifier
$f'$ learned from $D'_{train}\leftarrow D_{train} \cup D'_{poison}$
mis-classifies the files in $D_{target}$.
Formally, the attacker intents to achieve the following for $\forall~\dot{z} \in D_{target}$: 
$-\leftarrow f'(\dot{z})$ where $f'$ is learned from $D'_{train}\leftarrow D_{train}\cup D'_{poison}$.
\end{definition}

As shown in Definition \ref{def:optimal poisoning attack}, the optimal poisoning attack is the same as {\sf BF}, except that $\mathcal{A}$ attempts to
maximize the loss when using classifier $f'$ with parameter $\theta'$ to classify
files in $D_{target}$. 
Definition \ref{def:optimal poisoning attack} can have multiple variants by considering bounds on $|D'_{poison}|$ \cite{steinhardt2017certified} or bounds on the degree of perturbations $|\delta_z|$ \cite{suciu2018does}. 

\begin{definition}[optimal poisoning or ${\sf OP}$ \cite{munoz2017towards}]
\label{def:optimal poisoning attack}
Given $D_{poison}$, $\mathcal{A}$ perturbs $D_{poison}$ into $D'_{poison}$ for achieving:
\begin{align*}
\max \limits_{D'_{poison}}~\mathcal{L}(\theta', D_{target})
~\text{where}~{\theta'} \leftarrow \operatorname*{\arg\min}_{\theta} \mathcal{L}(\theta, D_{train} \cup D'_{poison}).
\end{align*}
\end{definition}

{({\bf viii})} $A_8$: it describes $\mathcal{A}$'s attack techniques, such as Gradient-based Optimization ({\sf GO}), Sensitive Features ({\sf SF}), MImicry ({\sf MI}), TRansferability ({\sf TR}), Heuristic Search ({\sf HS}), Generative Model ({\sf GM}), and Mixture Strategy ({\sf MS}). We denote this by $A_8\in\{${\sf GO}, {\sf SF}, {\sf MI}, {\sf TR}, {\sf HS}, {\sf GM}, {\sf MS}$\}$. Let $\mathcal{A}$ have a classifier $\hat{f}$, which consists of a hand-crafted feature extraction $\hat{\phi}_c$ and a parameterized model $\hat{F}_{\hat\theta}$. Let $\mathcal{A}$ also have an objective function $L_\mathcal{A}:[0,1]\times\mathcal{Y}\to\mathcal{R}$, which measures $\hat{f}$'s error or $\mathcal{A}$' failure in evasion \cite{Biggio:Evasion,carliniW16a}. Note that $\hat{f}$ and $L_\mathcal{A}$ can be the same as, or can mimic (by leveraging $\mathcal{A}$'s knowledge about $\mathcal{I}$'s attributes $A_1,\ldots,A_5$), $\mathcal{I}$'s classifier $f$ and loss function $L$, respectively. 

The attack technique specified by Definition \ref{definition:go} is that $\mathcal{A}$ solves the feature-space optimization problems described in Definitions \ref{def:opt_evasion1}, \ref{def:opt_evasion2} and \ref{def:optimal poisoning attack} by using some gradient-based optimization method and then leverages the invertibility Assumption \ref{assumption:inverse} to generate adversarial malware examples.
\begin{definition}[Gradient-based Optimization or {\sf GO}, adapted from \cite{carliniW16a,munoz2017towards}] \label{definition:go}
Let $\mathbf{x}\leftarrow\hat{\phi}_c(\hat{S},z)$ and $\mathbf{x}'\leftarrow\mathbf{x}+\delta_\mathbf{x}$. The feature-space optimization problem in Definition \ref{def:opt_evasion1} can be written as
	\begin{equation}
		\min_{\delta_\mathbf{x}} C(\mathbf{x}, \mathbf{x}+\delta_\mathbf{x})~~~\text{s.t.}~~~ (\delta_\mathbf{x}\in[\underline{\mathbf{u}},\overline{\mathbf{u}}]) \land (\hat{F}_{\hat{\theta}}(\mathbf{x}')<\tau), \label{eq:a8:evs1}
	\end{equation}
where $\underline{\mathbf{u}}$ and $\overline{\mathbf{u}}$ are respectively the lower and upper bounds on $\mathbf{M}$ (e.g., $\delta_\mathbf{x}\in[-\mathbf{x}, 1-\mathbf{x}]$ for binary representation $\mathbf{x}$). 
The feature-space optimization problem in Definition \ref{def:opt_evasion2} can be written as 
    \begin{equation}
    	\max_{\delta_\mathbf{x}} L_\mathcal{A}\left(\hat{F}_{\hat{\theta}}(\mathbf{x}+\delta_\mathbf{x}), +\right)~~\text{s.t.}~~(\delta_\mathbf{x}\in[\underline{\mathbf{u}},\overline{\mathbf{u}}]). \label{eq:a8:evs2}
    \end{equation}
The feature-space optimization problem specified in Definition \ref{def:optimal poisoning attack} can be written as 
    \begin{align}
    	&\max_{\delta_\mathbf{x}\in[\underline{\mathbf{u}},\overline{\mathbf{u}}]}\mathbb{E}_{(\dot{z},\dot{y})\in D_{target}}L_\mathcal{A}(\hat{F}_{\hat{\theta}'}(\hat{\phi}_c(\hat{S},\dot{z}),\dot{y})),~~ \forall (z,y)\in D_{poison} \label{eq:a8:poi}\\
    	~~\text{where}~~&{\hat\theta}'\leftarrow\argmin_{\hat\theta}\mathbb{E}_{(z_t,y_t)\in{\hat{D}_{train}\cup\{(\hat{\phi}_c^{-1}(\hat{\phi}_c(z)+\delta_\mathbf{x}),y')\}}}L_\mathcal{A}\left(\hat{F}_{\hat\theta}(\hat{\phi}_c(\hat{S},z_t),y_t)\right). \nonumber
    \end{align}
\end{definition}
In order to calculate the gradients of loss function $L_\mathcal{A}$ with respect to $\delta_\mathbf{x}$ in Eqs.\eqref{eq:a8:evs1} and \eqref{eq:a8:evs2}, inequality constraints can be handled by appending penalty items to the loss function in question and box-constraints can be coped with by using gradient projection \cite{carliniW16a,DBLP:journals/tnse/LiLYX21}. Since $\mathbf{x}+\delta_\mathbf{x}$ is continuous, the {\sf GO} attack technique needs to map $\delta_\mathbf{x}$ to a discrete perturbation vector in $\mathbf{M}$, for instance by using the nearest neighbor search \cite{DBLP:journals/tnse/LiLYX21}.
The gradients of loss function $L_\mathcal{A}$ with respect to $\delta_\mathbf{x}$ in Eq.  \eqref{eq:a8:poi} are delicate to deal with. One issue is the indirect relation between $L_\mathcal{A}$ and $\delta_\mathbf{x}$, which can be handled by the chain rule \cite{lecun2015deep}. 
Another issue is the difficulty that is encountered when computing the partial derivatives $\partial{\hat\theta'}/\partial\delta_{\mathbf{x}}$ \cite{munoz2017towards}. 
For dealing with this, researchers often relax the underlying constraints (e.g., by supposing that $\hat{F}_{\hat\theta}$ is a linear model). 

The attack technique specified by Definition \ref{definition:sf} is that $\mathcal{A}$ perturbs malware examples by injecting or removing a small number of features to decrease the classification error measured by the loss function ${L}_\mathcal{A}$ as much as possible.
\begin{definition}[Sensitive Features or {\sf SF}, adapted from \cite{DBLP:conf/mlsys/LeiWCDDW19}] \label{definition:sf}
For evasion attacks, $\mathcal{A}$ aims to maximize the following with respect to a given malware example-label pair $(z,+)$:  $$\max_{z'} L_\mathcal{A}\left(\hat{F}_{\hat\theta}(\hat{\phi}(\hat{S},z')),+\right)~~\text{s.t.}~~(z'\leftarrow \mathcal{A}(z,\delta_z)) \land (\delta_z\subseteq \mathcal{M}) \land (|\delta_z|\leq m),$$ 
where $m$ is the maximum degree of manipulations.

For poisoning attacks, $\mathcal{A}$ aims to maximize the following with respect to the given $D_{poison}$ and $D_{target}$, 
\begin{align}
    \max_{D'_{poison}}\mathbb{E}_{(\dot{z},\dot{y})\in D_{target}}L_\mathcal{A}(\hat{F}_{\hat{\theta}'}(\hat{\phi}_c(\hat{S},\dot{z}),\dot{y})),
\end{align}
where $\hat{\theta}'$ is learned from $\hat{D}_{train}\cup D'_{poison}$ such that $\forall z'\in D'_{poison}$ is obtained via the perturbation $\delta_z$ with obeying $(z'\leftarrow \mathcal{A}(z,\delta_z)) \land (\delta_z\subseteq \mathcal{M}) \land (z\in D_{poison}) \land (|\delta_z|\leq m)$.
\end{definition} 

The attack technique specified by Definition \ref{definition:mi} is that $\mathcal{A}$ perturbs malware example $z$ by mimicking a benign example, while noting that this attack technique can be algorithm-agnostic.
\begin{definition}[MImicry or {\sf MI}, adapted from \cite{rndic_laskov}] \label{definition:mi}
Given a set of benign examples $D_{ben}$ and a malware example $z$, $\mathcal{A}$ aims to achieve the following minimization:
	\begin{eqnarray}
		\min_{\delta_z\in\mathcal{M}} \Gamma(\mathcal{A}(z,\delta_z),z_{ben})~~\text{s.t.}~~ (\exists~z_{ben}\in D_{ben}). \label{eq:mimicry}
	\end{eqnarray}
\end{definition}

The attack technique specified by Definition \ref{definition:mi} can be extended to accommodate the similarity between representations in the feature space \cite{Biggio:Evasion,rndic_laskov}.
The attack technique specified by Definition \ref{definition:tr} is that $\mathcal{A}$ generates adversarial examples against a surrogate model  $\hat{f}$.

\begin{definition}[TRansferability or {\sf TR}, adapted from \cite{papernotMGJCS16}] \label{definition:tr}
$\mathcal{A}$ learns a surrogate model $\hat{f}$ of $f$ from $\hat{D}_{train}$, $\hat{S}$, $\hat{\phi_c}$ and $\hat{F}$. For evasion attacks,
$\mathcal{A}$ achieves $-\leftarrow\hat{f}(z')$ by perturbing malware example $z$ to $z'$ and then attacks $f$ with $z'$. For poisoning attacks, $\mathcal{A}$ contaminates $\hat{f}$ to $\hat{f}'$ such that $-\leftarrow\hat{f}'(\dot{z})~\text{for}~\forall(\dot{z},\dot{y}) \in D_{target}$, by poisoning the training set $\hat{D}_{train}$ with $D'_{poison}$ and the attacks $f$ with $D'_{poison}$.
\end{definition}

The attack technique specified by
Definition \ref{definition:hs} is that $\mathcal{A}$ searches perturbations in $\mathcal{M}$ via some heuristics, while leveraging oracle $\mathcal{O}$'s responses to $\mathcal{A}$'s queries and $f$'s responses to $\mathcal{A}$'s queries. Since $\mathcal{M}$ is defined with respect to the problem space, this attack technique does not need the invertibility Assumption \ref{assumption:inverse}.

\begin{definition}[Heuristic Search or {\sf HS}] \label{definition:hs}
Let $h$ be a function taking $\mathcal{O}$'s response and $f$'s response as input. Given a malware example $z$, $\mathcal{A}$ looks for an $m$-length manipulation path 
\begin{align}
\langle z_{(0)}, z_{(1)},\ldots,z_{(m)}\rangle 
~~\text{s.t.}~~z_{(i+1)}=\mathcal{A}(z_{(i)},\delta_{z,(i)}) \land (\delta_{z,(i)}& \in \mathcal{M}) \land (h(\mathcal{O}, f,z,z_{(i)})\leq h(\mathcal{O}, f,z,z_{(i+1)})) \nonumber
\end{align}
where $z_{(0)}=z$.
\end{definition}

The attack technique specified by Definition \ref{def:generative-model} is that $\mathcal{A}$ uses a generative model $G$ with parameters $\theta_g$ to perturb malware representation vectors and then leverages the invertibility Assumption \ref{assumption:inverse} to turn the perturbed vector into an adversarial malware example.
\begin{definition}[Generative Model or {\sf GM}]\label{def:generative-model}
Given a malware representation vector $\mathbf{x}=\hat{\phi}_c(\hat{S},z)$, $\mathcal{A}$ achieves
$$\max_{\theta_g} L_\mathcal{A}\left(\hat{F}_{\hat{\theta}}(G_{\theta_g}(\mathbf{x})),+\right)~~\text{s.t.}~~ G_{\theta_g}(\mathbf{x})\in [\underline{\mathbf{u}}-\mathbf{x}, \overline{\mathbf{u}}-\mathbf{x}]$$
and leverages the invertibility Assumption \ref{assumption:inverse} to obtain an adversarial example $z'=\hat\phi_c^{-1}(G_{\theta_g}(\mathbf{x}))$.
\end{definition}

The attack technique specified by
Definition \ref{definition:ms} is that $\mathcal{A}$ combines multiple perturbation methods to perturb an example. 
\begin{definition}[Mixture Strategy or {\sf MS} \cite{DBLP:LiL20}] \label{definition:ms}
Let $\mathcal{H}_A$ denote the space of generative methods and $\mathcal{W}_a=\{\mathbf{w}_a:\mathbf{w}_a=(w_{a,1},\ldots,w_{a,K}),w_{a,i}\geq 0\}$ with $i=1,\ldots,K$ denote the weights space. Given a malware example $z$, $\mathcal{A}$ aims to achieve $$\max_{\mathbf{w}_a} L_\mathcal{A}(\hat{F}_{\hat{\theta}}(\hat\phi(z')),+) ~~\text{s.t.}~~(z'=\sum_{i=1}^K{w_{a,i}}g_i(z))\land (\mathcal{O}(z,z')={\sf true})) \land (g_i\in \mathcal{H}_A) \land (\mathbf{w}_a\in \mathcal{W}_a).$$
\end{definition}

({\bf ix}) $A_9$: it corresponds to $\mathcal{A}$'s adversarial files. Given file manipulation set $\mathcal{M}$, the corresponding set of adversarial files is defined as $\mathcal{Z}_{\mathcal M}=\{\mathcal{A}(z, \delta_z):(z \in \mathcal{Z}) \land (\delta_z\subseteq\mathcal{M})\}$. Given feature manipulation set $\mathbf{M}$, the set of adversarial feature vectors is: $\mathcal{X}_{\mathbf{M}}=\{\mathbf{x}':(\mathbf{x}'=\mathbf{x}+\delta_{\mathbf{x}}) \land (\delta_\mathbf{x}\in\mathbf{M})\}.$

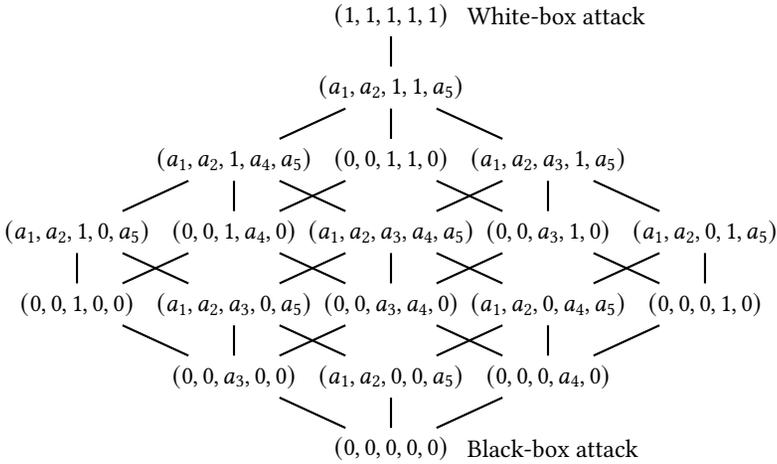
\begin{figure}[htbp!]
	\centering
	\begin{tikzpicture}[thick,scale=0.95, every node/.style={transform shape}]
	\node(top) [label=right:{White-box} attack, label=left:{}] at (0,6) {$(1,1,1,1,1)$};
	
	\node(l5n0)[label=below:{},align=center] at (0,5) {$(a_1,a_2,1,1,a_5)$};
	
	\node(l4n0)[label=below:{},align=center] at (2.2,4) {$(a_1,a_2,a_3,1,a_5)$};
	\node(l4n1)[label=below:{},align=center] at (0,4) {$(0,0,1,1,0)$};
	\node(l4n2)[label=below:{},align=center] at (-2.2,4) {$(a_1,a_2,1,a_4,a_5)$};
	
	\node(l3n0)[label=below:{},align=center] at (4.4,3) {$(a_1,a_2,0,1,a_5)$};
	\node(l3n1)[label=below:{},align=center] at (2.2,3) {$(0,0,a_3,1,0)$};
	\node(l3n2)[label=below:{},align=center] at (0,3) {$(a_1,a_2,a_3,a_4,a_5)$};
	\node(l3n3)[label=below:{},align=center] at (-2.2,3) {$(0,0,1,a_4,0)$};
	\node(l3n4)[label=below:{},align=center] at (-4.4,3) {$(a_1,a_2,1,0,a_5)$};
	
	\node(l2n0)[label=below:{},align=center] at (4.4,2) {$(0,0,0,1,0 )$};
	\node(l2n1)[label=below:{},align=center] at (2.2,2) {$(a_1,a_2,0,a_4,a_5)$};
	\node(l2n2)[label=below:{},align=center] at (0,2) {$(0,0,a_3,a_4,0)$};
	\node(l2n3)[label=below:{},align=center] at (-2.2,2) {$(a_1,a_2,a_3,0,a_5)$};
	\node(l2n4)[label=below:{},align=center] at (-4.4,2) {$(0,0,1,0,0)$};
	
	\node(l1n0)[label=below:{},align=center] at (2.2,1) {$(0,0,0,a_4,0 )$};
	\node(l1n1)[label=below:{},align=center] at (0,1) {$(a_1,a_2,0,0,a_5)$};
	\node(l1n2)[label=below:{},align=center] at (-2.2,1) {$(0,0,a_3,0,0 )$};
	
	\node(bot)[label=right:{Black-box attack}, label=left:{}] at (0,0) {$ (0,0,0,0,0)$};

	\draw(l5n0) -- (top);
	
	\draw(l4n0) -- (l5n0);
	\draw(l4n1) -- (l5n0);
	\draw(l4n2) -- (l5n0);
	
	\draw(l3n0) -- (l4n0);
	\draw(l3n1) -- (l4n0);
	\draw(l3n1) -- (l4n1);
	\draw(l3n2) -- (l4n0);
	\draw(l3n2) -- (l4n2);
	\draw(l3n3) -- (l4n1);
	\draw(l3n3) -- (l4n2);
	\draw(l3n4) -- (l4n2);

	\draw(l2n0) -- (l3n0);
	\draw(l2n0) -- (l3n1);
	\draw(l2n1) -- (l3n0);
	\draw(l2n1) -- (l3n2);
	\draw(l2n2) -- (l3n1);
	\draw(l2n2) -- (l3n2);
	\draw(l2n2) -- (l3n3);
	\draw(l2n3) -- (l3n2);
	\draw(l2n3) -- (l3n4);
	\draw(l2n4) -- (l3n3);
	\draw(l2n4) -- (l3n4);

	\draw(l1n0) -- (l2n0);
	\draw(l1n0) -- (l2n1);
	\draw(l1n0) -- (l2n2);
	\draw(l1n1) -- (l2n1);
	\draw(l1n1) -- (l2n3);
	\draw(l1n2) -- (l2n2);
	\draw(l1n2) -- (l2n3);
	\draw(l1n2) -- (l2n4);
	
	\draw(bot) -- (l1n2);
	\draw(bot) -- (l1n1);
	\draw(bot) -- (l1n0);
	
	\end{tikzpicture}
	
	\caption{A portion of the partial order defined over $(a_1,\ldots,a_5)$.
	} 
	\label{fig:attack_input}
\end{figure}

\smallskip

\noindent{\bf On the Usefulness of the Preceding Specification}. The preceding specification can be applied to formulate a partial order in the attribute space, which allows to compare attacks unambiguously. Figure \ref{fig:attack_input} depicts how vector $(a_1, \cdots, a_5)$ formulates a partial order between the widely-used informal notions of {\em black-box} attack, namely $(a_1,a_2,a_3,a_4,a_5)=(0,0,0,0,0)$, 
and {\em white-box} attack, namely $(a_1,a_2,a_3,a_4,a_5)=(1,1,1,1,1)$;
there are many kinds of grey-box attacks in between.

\subsubsection{Systematizing Defenses}
Similarly, we systematize defenses from two perspectives: {\em defender's objective} (i.e., what the defender aims to achieve) and {\em defender's input} (i.e., what leverages the defender can use). 
We also discuss how to apply the specification to formulate a partial-order structure for comparing defenses.

\smallskip

\noindent{\bf Defender's Objectives}.
$\mathcal{I}$ aims to detect ideally all of the malicious files, adversarial and non-adversarial alike, while suffering from small side-effects (e.g., increasing false-positives).

\smallskip

\noindent{\bf Defender's Input}.
As highlighted in Table \ref{tab:attr}, $\mathcal{I}$'s input includes attributes $A_1,\ldots,A_5$, which are under $\mathcal{I}$'s control, and the extent $a_6,\ldots,a_9$ at which $\mathcal{I}$ respectively knows about attributes $A_6,\ldots,A_9$, which are under $\mathcal{A}$'s control. Note that $A_1,\ldots,A_9$ have been defined above.
\begin{itemize}
\item We define $a_6 \in [0,1]$ to represent the extent at which $\mathcal{I}$ knows $\mathcal{A}$'s manipulation set $\mathcal{M}$. Let $\hat{\mathcal{M}}\subseteq \mathcal{M}$ denote the subset of $\mathcal{A}$'s manipulation set known to $\mathcal{I}$. 
Then, we set $a_6=|\hat{\mathcal{M}}|/|\mathcal{M}|$.

\item We define $a_7\in\{0,1\}$ such that $a_7=0$ means $\mathcal{I}$ does not know $\mathcal{A}$'s attack tactic $A_7\in \{{\sf BE},{\sf OE1},{\sf OE2},{\sf BP},$ ${\sf OP}\}$ and $a_7=1$ means $\mathcal{I}$ knows $\mathcal{A}$'s tactic. 

\item  We define $a_8\in\{0,1\}$ such that $a_8=1~(a_8=0)$ means the defender does (not) know $\mathcal{I}$'s attack technique $A_8\in\{{\sf GO}, {\sf SF}, {\sf TR}, {\sf MI}, {\sf HS}, {\sf GM}, {\sf MS}\}$.

\item We use $a_9=|\hat{\mathcal{Z}}_{\mathcal{M}}| / |\mathcal{Z}_{\mathcal{M}}|$ to represent the extent at which $\mathcal{I}$ knows about $\mathcal{A}$'s adversarial files, where $a_9\in [0,1]$ and $\mathcal{Z}_{\mathcal{M}}$ is $\mathcal{A}$'s adversarial files and $\hat{\mathcal{Z}}_{\mathcal{M}}\subseteq\mathcal{Z}_{\mathcal{M}}$ is known to $\mathcal{I}$.
\end{itemize}

\smallskip

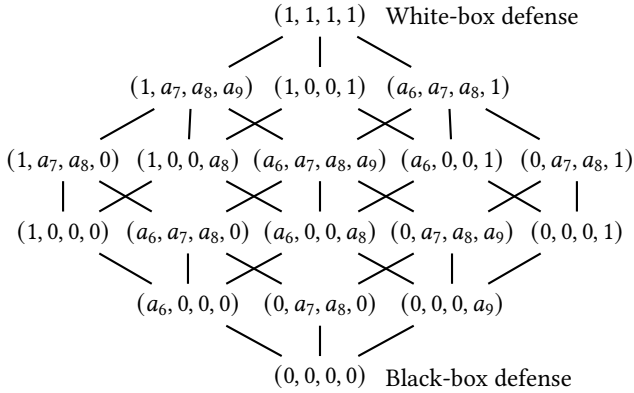
\begin{figure}[!htbp]
	\centering
	\begin{tikzpicture}[thick,scale=0.95, every node/.style={transform shape}]
		\node(top)[label=right:{White-box defense}, label=left:{}] at (0,6) {$( 1,1,1,1)$};
		
		\node(l3n0)[label=below:{},align=center] at (1.8,5) {$( a_6,a_7,a_8,1)$};
		\node(l3n1)[label=below:{},align=center] at (0,5) {$( 1,0,0,1)$};
		\node(l3n2)[label=below:{},align=center] at (-1.8,5) {$( 1,a_7,a_8,a_9)$};
		
		\node(l2n0)[label=below:{},align=center] at (3.6,4) {$( 0,a_7,a_8, 1)$};
		\node(l2n1)[label=below:{},align=center] at (1.85,4) {$( a_6,0,0,1)$};
		\node(l2n2)[label=below:{},align=center] at (0,4) {$( a_6,a_7,a_8,a_9)$};
		\node(l2n3)[label=below:{},align=center] at (-1.85,4) {$( 1,0,0,a_8)$};
		\node(l2n4)[label=below:{},align=center] at (-3.6,4) {$( 1,a_7,a_8,0)$};
		
		\node(l1n0)[label=below:{},align=center] at (3.6,3) {$( 0,0,0,1)$};
		\node(l1n1)[label=below:{},align=center] at (1.85,3) {$( 0,a_7,a_8,a_9)$};
		\node(l1n2)[label=below:{},align=center] at (0,3) {$( a_6,0,0,a_8)$};
		\node(l1n3)[label=below:{},align=center] at (-1.85,3) {$( a_6,a_7,a_8,0)$};
		\node(l1n4)[label=below:{},align=center] at (-3.6,3) {$( 1,0,0,0)$};
		
		\node(l0n0)[label=below:{},align=center] at (1.85,2) {$( 0,0,0,a_9)$};
		\node(l0n1)[label=below:{},align=center] at (0,2) {$( 0,a_7,a_8,0)$};
		\node(l0n2)[label=below:{},align=center] at (-1.85,2) {$( a_6,0,0,0)$};
		
		\node(bot)[label=right:{Black-box defense}, label=left:{}] at (0,1) {$( 0,0,0,0)$};
		
		\draw(bot) --  (l0n0);
		\draw(bot) --  (l0n1);
		\draw(bot) -- (l0n2);
		
		\draw(l0n0) -- (l1n0);
		\draw(l0n0) -- (l1n1);
		\draw(l0n0) -- (l1n2);
		\draw(l0n1) -- (l1n1);
		\draw(l0n1) -- (l1n3);
		\draw(l0n2) -- (l1n2);
		\draw(l0n2) -- (l1n3);
		\draw(l0n2) -- (l1n4);
		
		\draw(l1n0) -- (l2n0);
		\draw(l1n0) -- (l2n1);
		\draw(l1n1) -- (l2n0);
		\draw(l1n1) -- (l2n2);
		\draw(l1n2) -- (l2n1);
		\draw(l1n2) -- (l2n2);
		\draw(l1n2) -- (l2n3);
		\draw(l1n3) -- (l2n2);
		\draw(l1n3) -- (l2n4);
		\draw(l1n4) -- (l2n3);
		\draw(l1n4) -- (l2n4);
		
		\draw(l2n0) -- (l3n0);
		\draw(l2n1) -- (l3n0);
		\draw(l2n1) -- (l3n1);
		\draw(l2n2) -- (l3n0);
		\draw(l2n2) -- (l3n2);
		\draw(l2n3) -- (l3n1);
		\draw(l2n3) -- (l3n2);
		\draw(l2n4) -- (l3n2);
		
		\draw(l3n0) -- (top);
		\draw(l3n1) -- (top);
		\draw(l3n2) -- (top);
	\end{tikzpicture}
	\caption{A portion of the partial order defined over $(a_6,\ldots,a_9)$.
	}
	\label{fig:defense_input}
\end{figure}

\noindent{\bf On the Usefulness of the Preceding Specification}. 
Similarly, the defense specification can be used to formulate a partial order in the attribute space, paving the way for comparing defenses unambiguously. Figure \ref{fig:defense_input} depicts how vector $(a_6,\ldots,a_9)$ formulates a partial order between the widely-used informal notions of black-box defense $(a_6,a_7,a_8,a_9)=(0,0,0,0)$ and white-box defense $(a_6,a_7,a_8,a_9)=(1,1,1,1,1)$; there are many kinds of grey-box defenses in between.

\subsection{Systematizing Security Properties} \label{sec:ml_model_property}

Since $f=\varphi(\phi(\cdot))$, we decompose $f$'s security properties into $\varphi$'s and $\phi$'s. We consider: Representation Robustness ({\sf RR}), meaning that two similar files have similar feature representations; Classification Robustness ({\sf CR}), meaning that two similar feature representations lead to the same label; Detection Robustness ({\sf DR}), meaning that feature extraction function $\phi$ returns similar representations for two files with the same functionality; Training Robustness ({\sf TR}), meaning that small changes to the training set does not cause any significant change to the learned classifier. 
With respect to small perturbations, Definitions \ref{definition:representation-robustness} and
\ref{definition:classification-robustness} below collectively say that when two files $z$ and $z'$ are similar, they would be classified as the same label with a high probability. Since the classification function $\varphi$ is linear, we can obtain a $\epsilon$-robust $\varphi$ analytically, where $\epsilon$ is a small scalar that bounds the perturbations applied to feature vectors \cite{madry2017towards}. This means that the main challenge is to achieve robust feature extraction.

\begin{definition}[{\sf RR} or $(\epsilon,\eta)$-robust feature extraction; adapted from \cite{wang2016theoretical}] 
\label{definition:representation-robustness}
Given constants $\epsilon,\eta\in [0,1]$, and files $z,z' \in \mathcal{Z}$ such that $(\Gamma(z,z')\approx 0) \land ({\tt true}\leftarrow \mathcal{O}(z,z'))$,
we say feature extraction function $\phi$ is $(\epsilon,\eta)$-robust if
$$\mathbb{P}(C(\mathbf{x},\mathbf{x}')\leq \epsilon)=\mathbb{P}(C(\phi(z),\phi(z'))\leq \epsilon)>1 - \eta.$$
\end{definition}

\begin{definition}[{\sf CR} or {$\epsilon$-robust classification} \cite{bastani2016measuring}]
\label{definition:classification-robustness} 
Given constant $\epsilon\in [0,1]$ as in Definition 
\ref{definition:representation-robustness} and
any feature vectors $\mathbf{x},{\bf x}'\in \mathcal{X}$, we say classification function $\varphi$ is $\epsilon$-robust if $$(C(\mathbf{x},\mathbf{x}') \leq \epsilon) \rightarrow ((\varphi({\bf x})>\tau)\land (\varphi({\bf x}')>\tau)).$$
\end{definition}

Definition \ref{definition:representation-robustness2} 
specifies detection robustness, which says that feature extraction function $\phi$ returns similar representations for two different files as long as they have the same functionality.
Note that Definitions \ref{definition:classification-robustness} and
\ref{definition:representation-robustness2} collectively produce a malware detector with detection robustness.

\begin{definition}[{\sf DR} or $(\mathcal{O},\eta)$-robust feature extraction; adapted from \cite{al2018adversarial}] 
\label{definition:representation-robustness2}
Given constant $\eta\in [0,1]$ and two files $z,z' \in \mathcal{Z}$ such that $(\Gamma(z,z')>>0)\land({\tt true}\leftarrow \mathcal{O}(z,z'))$, we say feature extraction $\phi$ is $(\mathcal{O}, \eta)$-robust if $\mathbb{P}(C(\phi(z),\phi(z'))\leq \epsilon)>1 - \eta.$
\end{definition}

Suppose we impose a restriction on the adversarial files set $D'_{poison}$ such that $|D'_{poison}|\leq \gamma|D_{train}|$ for some constant $\gamma\in[0,1]$. Let classifier $f'$ be learned from $D_{train} \cup D'_{poison}$. 
Definition \ref{def:robust-training} says that a classifier $f'$ learned from poisoned training set can predict as accurately as $f$ learned from $D_{train}$ with a high probability. 

\begin{definition}[{\sf TR} or $(\gamma,\zeta)$-robust training; adapted from \cite{steinhardt2017certified}]\label{def:robust-training}
Given classifiers $f$ learned from $D_{train}$ and $f'$ learned from $D_{train} \cup D'_{poison}$ where
$|D_{poison}'| \leq \gamma |D_{train}|$, and small constants $\zeta\in [0,1]$, we say
$f'$ is $(\gamma,\zeta)$-robust if $\forall z\in\mathcal{Z}$
$$\left((f(z)>\tau) \land (|D_{poison}'| \leq \gamma |D_{train}|)\right) \rightarrow \left(\mathbb{P}(f'(z) > \tau)>1-\zeta\right).$$
\end{definition}

\section{Systematizing AMD Arms Race}
\label{sec:systematizstion-AMD-studies}

We systematize attacks according to $\mathcal{A}$'s objective, input, assumptions, the security properties that are broken, and the types of victim malware detectors (e.g., Windows vs. Android). Similarly, we systematize defenses according to $\mathcal{I}$'s objective, input, assumptions, the security properties that are achieved, and the types of enhanced malware detectors (e.g., Windows vs. Android).
We group attacks (defenses) according to the attacker's (defender's) techniques and then summarize them in a table according to the publication date in chronological order. 
For convenience, we will use wildcard $*$ to indicate any value in a domain (e.g., $[0,1]$); we will use $\lor$ to describe $\mathcal{A}$'s and $\mathcal{I}$'s ``broader'' input (if applicable). For example, $(0,1,0,1,0|A_6,\ldots,A_9)\lor (1,0,1,1,1|A_6,\ldots,A_9)$ means that $\mathcal{A}$ has either $(a_1,a_2,a_3,a_4,a_5)=(0,1,0,1,0)$ or
$(a_1,a_2,a_3,a_4,a_5)=(1,0,1,1,1)$. Finally, we will present the attack-defense escalation.

\subsection{Systematizing Attack Literature} \label{sec:attack}

\begin{table}[htbp]
\caption{Summary of AMD attacks (\checkmark means applicable, \protect\fullcirc means 0, \protect\emptycirc means 1, \protect\dotcirc means a value in $[0,1]$).}
	\centering
	\resizebox{0.9\columnwidth}{!}{
	\setlength{\tabcolsep}{0.4em}
	\begin{tabular}{l|ccc|ccccc|cccc|ccccc|cccc|ccc}
	\hline
	\multicolumn{1}{c|}{\specialcell{Attack \\ (in chronological order)}}& \multicolumn{3}{c|}{\specialcell{Attack \\ Objective}} & \multicolumn{9}{c|}{\specialcell{Attack Input}} & 
	\multicolumn{5}{c|}{Assumptions} &
	\multicolumn{4}{c|}{\specialcell{Broken \\ Properties}} & \multicolumn{3}{c}{\specialcell{Malware \\ detector}} \\
	
	& \vthead{\sf Indiscriminate} 
	& \vthead{\sf Targeted}
	& \vthead{\sf Availability}
	
	& \vthead{$A_1$: Training set $D_{train}$}
	& \vthead{$A_2$: Defense technique}
	& \vthead{$A_3$: Feature set}
	& \vthead{$A_4$: Learning algorithm}
	& \vthead{$A_5$: Response}
	& \vthead{$A_6$: Manipulation set}
	& \vthead{$A_7$: Attack tactic}
	& \vthead{$A_8$: Attack technique}
	& \vthead{$A_9$: Adversarial example set}
	
	& \vthead{{\sf IID} assumption}
	& \vthead{{\sf Oracle} assumption}
	& \vthead{{\sf Measurability} assumption}
	& \vthead{{\sf Smoothness} assumption}
	& \vthead{{\sf Invertibility} assumption}
	
	& \vthead{{\sf RR}: Representation Robustness}
	& \vthead{{\sf CR}: Classification Robustness}
	& \vthead{{\sf DR}: Detection Robustness}
	& \vthead{{\sf TR}: Training Robustness}
	
	& \vthead{Windows Program}
	& \vthead{Android Package}
	& \vthead{PDF}
	
	\\\hline\hline
%
    \rowcolor{lightgray!30}
    Smutz and Stavrou \cite{smutz_2012} & 
    \checkmark &&&
    \specialcell{\pie{0}} & 
    \specialcell{\pie{0}} & 
    \specialcell{\pie{0}} &
    \specialcell{\pie{0}} & 
    \specialcell{\pie{0}} &
    $\mathbf{M}$ &
    {\sf OE2} &
    {\sf MI} &
    $\mathcal{X}_\mathbf{M}$ &
    & & \checkmark & & &
    & \checkmark & & &
    &  & \checkmark
    \\
    Biggio et al. \cite{Biggio:Evasion} &
    \checkmark &&& 
    \specialcell{\pie{0} \\ \pie{4}} & 
    \specialcell{\pie{0} \\ \pie{4}} & 
    \specialcell{\pie{0} \\ \pie{0}} & 
    \specialcell{\pie{0} \\ \pieg} & 
    \specialcell{\pie{0} \\ \pie{4}} &
    $\mathbf{M}$ &
    ${\sf OE2}$ &
    ${\sf GO}$ &
    $\mathcal{X}_{\mathbf{M}}$ &
    & \checkmark & \checkmark && \checkmark &
    & \checkmark & &&
    & & \checkmark 
    \\
%
    \rowcolor{lightgray!30}
    Maiorca et al. \cite{Maiorca:2013:LBE:2484313.2484327} &
 	\checkmark &&& 
 	\specialcell{\pie{4}} & 
    \specialcell{\pie{4}} & 
    \specialcell{\pie{4}} & 
    \specialcell{\pie{4}} &
    \specialcell{\pie{4}}  &
 	$\mathcal{M}$ &
 	${\sf BE}$ &
 	{\sf MI} &
 	$\mathcal{Z}_{\mathcal{M}}$ &
 	& \checkmark & & & &
 	& & \checkmark & &
 	&  & \checkmark 
 	\\
%
    {\v{S}}rndi\'{c} and Laskov \cite{rndic_laskov} &
 	\checkmark &&& 
 	\specialcell{\pie{4} \\ \pie{4} \\ \pie{0} \\ \pie{0}} & 
    \specialcell{\pie{4} \\ \pie{4} \\ \pie{4} \\ \pie{4}} & 
    \specialcell{\pieg \\ \pieg \\ \pieg \\ \pieg} & 
    \specialcell{\pie{4} \\ \pieg \\ \pie{4} \\ \pieg} &
    \specialcell{\pie{4} \\ \pie{4} \\ \pie{4} \\ \pie{4}}  &
 	\specialcell{$\mathbf{M}$ \\ $\mathcal{M}$ \\ $\mathbf{M}$ \\ $\mathcal{M}$} &
 	\specialcell{${\sf BE}$ \\ ${\sf BE}$ \\ ${\sf OE2}$ \\ ${\sf OE2}$ } &
 	\specialcell{${\sf TR}$ \\ ${\sf TR}$ \\ ${\sf TR}$ \\ ${\sf TR}$} &
 	\specialcell{$\mathcal{X}_{\mathbf{M}}$ \\ $\mathcal{Z}_{\mathcal{M}}$ \\ $\mathcal{X}_{\mathbf{M}}$ \\ $\mathcal{Z}_{\mathcal{M}}$} &
 	& \checkmark &&&\checkmark& 
 	&&\checkmark&&
 	& & \checkmark 
 	\\
%
    \rowcolor{lightgray!30}
    Xu et al.~\cite{316904628}  &
 	\checkmark &&& 
 	\specialcell{\pie{4}} & 
    \specialcell{\pie{4}} & 
    \specialcell{\pie{4}} & 
    \specialcell{\pie{4}} &
    \specialcell{\pie{0}} &
 	$\mathcal{M}$ &
 	${\sf BE}$ &
 	{\sf HS} &
 	$\mathcal{Z}_{\mathcal{M}}$ &
 	& & & & &
 	& & \checkmark &&
 	& & \checkmark 
 	\\
%
 	Carmony et al. \cite{DBLP:conf/ndss/CarmonyHYBZ16} &
 	\checkmark &&& 
 	\specialcell{\pie{4}} & 
 	\specialcell{\pie{4}} & 
 	\specialcell{\pie{4}} & 
 	\specialcell{\pie{4}} &
 	\specialcell{\pie{4}}  &
 	$\mathcal{M}$ &
 	${\sf BE}$ &
 	{\sf MI} &
 	$\mathcal{Z}_{\mathcal{M}}$ &
 	& \checkmark & & & &
 	& & \checkmark & &
 	&  & \checkmark 
 	\\
%
    \rowcolor{lightgray!30}
    Hu and Tan \cite{Hu2017} & 
    \checkmark &&& 
    \specialcell{\pie{4}} & 
    \specialcell{\pie{4}} & 
    \specialcell{\pie{0}} & 
    \specialcell{\pie{4}} &
    \specialcell{\pie{0}}  &
    $\mathbf{M}$ &
    ${\sf BE}$ &
    {\sf GM} &
    $\mathcal{X}_{\mathbf{M}}$ &
    \checkmark& \checkmark & & & \checkmark& 
    & & \checkmark &&
    \checkmark & & 
    \\
    %
    Hu and Tan \cite{hu2017black} & 
    \checkmark &&& 
    \specialcell{\pie{4}} & 
    \specialcell{\pie{4}} & 
    \specialcell{\pie{0}} & 
    \specialcell{\pie{4}} &
    \specialcell{\pie{0}}  &
    $\mathbf{M}$ &
    ${\sf BE}$ &
    {\sf GM} &
    $\mathcal{X}_{\mathbf{M}}$ &
    \checkmark& \checkmark & & & \checkmark &
    & & \checkmark &&
    \checkmark & & 
    \\
    %
    %
    \rowcolor{lightgray!30}
    Demontis et al. \cite{demontis2017yes} &
    \checkmark &&& 
    \specialcell{\pie{0} \\ \pie{4}} & 
    \specialcell{\pie{0} \\ \pie{4}} &
    \specialcell{\pie{0} \\ \pie{0}} & 
    \specialcell{\pie{0} \\ \pieg} &
    \specialcell{\pie{0} \\ \pie{4}} &
    $\mathbf{M}$ &
    ${\sf OE2}$ &
    {\sf SF} &
    $\mathcal{X}_{\mathbf{M}}$ &
    & \checkmark & \checkmark & & \checkmark &
    & \checkmark & &&
    & \checkmark &  
    \\
%
    Grosse et al. \cite{grosse2017adversarial} &
 	\checkmark &&& 
 	\specialcell{\pie{4}} & 
    \specialcell{\pie{0}} & 
    \specialcell{\pie{0}} & 
    \specialcell{\pie{0}} & 
    \specialcell{\pie{0}} &
 	$\mathbf{M}$ &
 	${\sf OE2}$ &
 	{\sf SF} &
 	$\mathcal{X}_{\mathbf{M}}$ &
 	& \checkmark &\checkmark & & \checkmark &
 	\checkmark & \checkmark & &&
 	& \checkmark &  
 	\\
%
    \rowcolor{lightgray!30}
    Chen et al. \cite{chen2017adversarial}  &
    \checkmark &&& 
    \specialcell{\pie{0}} & 
    \specialcell{\pie{0}} & 
    \specialcell{\pie{0}} & 
    \specialcell{\pie{0}} & 
    \specialcell{\pie{0}} &
    $\mathbf{M}$ &
    ${\sf OE2}$ &
    {\sf SF} &
    $\mathcal{X}_{\mathbf{M}}$ &
    & \checkmark & \checkmark & & \checkmark & 
    & \checkmark&&&
    \checkmark & & 
    \\
%
    Khasawneh et al. \cite{khasawneh2017rhmd}  &
    \checkmark &&& 
    \specialcell{\pie{4}} & 
    \specialcell{\pie{4}} & 
    \specialcell{\pieg} &
    \specialcell{\pie{4}} & 
    \specialcell{\pie{0}}  &
    \specialcell{ $\mathbf{M}$ \\ $\mathcal{M}$} &
    \specialcell{${\sf BE}$ \\ ${\sf BE}$} &
    \specialcell{{\sf TR} \\ {\sf TR}} &
    \specialcell{$\mathcal{X}_{\mathbf{M}}$ \\ $\mathcal{Z}_{\mathcal{M}}$} &
    & \checkmark & & & \checkmark & 
    & \checkmark & &&
    \checkmark & &  
    \\
%
    \rowcolor{lightgray!30}
    Dang et al. \cite{dang2017evading}  &
    \checkmark &&& 
    \specialcell{\pie{4}} & 
    \specialcell{\pie{4}} & 
    \specialcell{\pie{4}} & 
    \specialcell{\pie{4}} &
    \specialcell{\pie{4}} &
    $\mathcal{M}$ &
    ${\sf BE}$ &
    {\sf HS} &
    $\mathcal{Z}_{\mathcal{M}}$ &
    & & & & &
    & & \checkmark &&
    &  & \checkmark 
    \\
%
    Mu{\~n}oz-Gonz{\'a}lez et al. \cite{munoz2017towards} & 
    &&\checkmark& 
    \specialcell{\pie{0} \\ \pie{0}} & 
    \specialcell{\pie{0} \\ \pie{4}} & 
    \specialcell{\pie{0} \\ \pie{0}} & 
    \specialcell{\pie{0} \\ \pie{4}} & 
    \specialcell{\pie{0} \\ \pie{4}}  &
    $\mathbf{M}$ &
    ${\sf OP}$ &
    {\sf GO} &
    $\mathcal{X}_{\mathbf{M}}$ &
    & \checkmark & & &\checkmark &
    & & & \checkmark &
    \checkmark & & 
    \\
%
    \rowcolor{lightgray!30}
    Yang et al. \cite{yang2017malware}  &
    \checkmark &&& 
    \specialcell{\pie{4}} & 
    \specialcell{\pie{4}} & 
    \specialcell{\pie{4}} & 
    \specialcell{\pie{4}} &
    \specialcell{\pie{0}} &
    $\mathcal{M}$ &
    ${\sf BE}$ &
    ${\sf HS}$ &
    $\mathcal{Z}_{\mathcal{M}}$ &
    & \checkmark & \checkmark & & &
    & & \checkmark &&
    & \checkmark &  
    \\
%
    Rosenberg et al. \cite{rosenberg2017generic}  &
 	\checkmark &&& 
 	\specialcell{\pie{4}} & 
    \specialcell{\pie{4}} & 
    \specialcell{\pie{0}} &
    \specialcell{\pie{4}} & 
    \specialcell{\pie{0}}  &
 	\specialcell{$\mathbf{M}$ \\ $\mathcal{M}$} &
 	\specialcell{${\sf BE}$ \\ ${\sf BE}$} &
 	\specialcell{{\sf TR} \\ {\sf TR}} &
 	\specialcell{$\mathcal{X}_{\mathbf{M}}$ \\ $\mathcal{Z}_{\mathcal{M}}$} &
 	& \checkmark & & & \checkmark &
 	\checkmark & \checkmark & &&
 	\checkmark & &  
 	\\
%
    \rowcolor{lightgray!30}
    Anderson et al. \cite{anderson2018learning}  &
 	\checkmark &&& 
 	\specialcell{\pie{4}} & 
    \specialcell{\pie{4}} & 
    \specialcell{\pie{4}} & 
    \specialcell{\pie{4}} &
    \specialcell{\pie{0}} &
 	$\mathcal{M}$ &
 	${\sf OE2}$ &
 	{\sf GM} &
 	$\mathcal{Z}_{\mathcal{M}}$ 
 	&\checkmark& \checkmark& \checkmark & & 
 	& \checkmark & \checkmark  &&
 	& \checkmark & &  
 	\\
%
 	Kreuk et al. \cite{kreuk2018adversarial} &
 	\checkmark &&& 
 	\specialcell{\pie{4}} & 
 	\specialcell{\pie{0}} & 
 	\specialcell{\pie{0}} & 
 	\specialcell{\pie{0}} & 
 	\specialcell{\pie{0}} &
 	$\mathbf{M}$ &
 	${\sf OE2}$ &
 	{\sf GO} &
 	$\mathcal{X}_{\mathbf{M}}$ 
 	& &\checkmark& & & \checkmark
 	& \checkmark & \checkmark &&
 	& \checkmark & &  
 	\\
%
 	\rowcolor{lightgray!30}
 	Chen et al. \cite{chen2018automated} & 
 	\checkmark &&& 
 	\specialcell{\pie{0}\\\pie{4}\\ \pie{0}} & 
 	\specialcell{\pie{0} \\ \pie{4} \\ \pie{4}} & 
 	\specialcell{\pie{0} \\ \pie{4} \\ \pieg} & 
 	\specialcell{\pie{0} \\ \pie{0} \\ \pie{0}} & 
 	\specialcell{\pie{0}\\ \pie{0} \\ \pie{0}}  &
 	$\mathbf{M}$ &
 	{\sf BP} &
 	{\sf SF} &
 	$\mathcal{X}_\mathbf{M}$ &
 	& \checkmark & \checkmark & & \checkmark & 
 	& & & \checkmark &
 	& \checkmark & 
 	\\
%
    Al-Dujaili et al. \cite{al2018adversarial} & 
 	\checkmark &&& 
 	\specialcell{\pie{4}} & 
    \specialcell{\pie{0}} & 
    \specialcell{\pie{0}} & 
    \specialcell{\pie{0}} & 
    \specialcell{\pie{0}}  &
 	$\mathbf{M}$ &
 	${\sf OE2}$ &
 	{\sf GO} &
 	$\mathcal{X}_{\mathbf{M}}$ &
 	& \checkmark & & & \checkmark &
 	&& \checkmark &&
 	\checkmark & & 
 	\\
%
 	\rowcolor{lightgray!30}
 	Suciu et al. \cite{suciu2018does} & 
 	&\checkmark&& 
 	\specialcell{\pieg \\ \pie{0} \\ \pie{0} \\ \pie{0}} & 
 	\specialcell{\pie{4} \\ \pie{4} \\ \pie{4} \\ \pie{0}} & 
 	\specialcell{\pie{0} \\ \pieg \\ \pie{0} \\ \pie{0}} & 
 	\specialcell{\pieg \\ \pieg \\ \pie{4} \\ \pie{0}} & 
 	\specialcell{\pie{4} \\ \pie{4} \\ \pie{4} \\ \pie{0}}   &
 	\specialcell{$\mathbf{M}$ \\ $\mathcal{M}$} &
 	\specialcell{${\sf BP}$ \\ ${\sf BP}$} &
 	\specialcell{{\sf SF} \\ {\sf SF}} &
 	\specialcell{$\mathcal{X}_{\mathbf{M}}$ \\ $\mathcal{Z}_{\mathcal{M}}$} &
 	& \checkmark & \checkmark & & \checkmark & 
 	& & &\checkmark&
 	& \checkmark & 
 	\\
%
    Kolosnjaji et al. \cite{kolosnjaji2018adversarial}  &
 	\checkmark &&& 
 	\specialcell{\pie{4}} & 
    \specialcell{\pie{0}} & 
    \specialcell{\pie{0}} & 
    \specialcell{\pie{0}} & 
    \specialcell{\pie{0}} &
 	$\mathbf{M}$ &
 	${\sf OE2}$ &
 	{\sf GO} &
 	$\mathcal{X}_{\mathbf{M}}$ 
 	&& \checkmark&& & \checkmark
 	& \checkmark & \checkmark &&
 	& \checkmark & &  
 	\\
%
    \rowcolor{lightgray!30}
    Suciu et al. \cite{8844597}  &
 	\checkmark &&& 
 	\specialcell{\pie{4}} & 
    \specialcell{\pie{0}} & 
    \specialcell{\pie{0}} & 
    \specialcell{\pie{0}} & 
    \specialcell{\pie{0}} &
 	$\mathbf{M}$ &
 	${\sf OE2}$ &
 	{\sf GO} &
 	$\mathcal{X}_{\mathbf{M}}$ &
 	& \checkmark & & & \checkmark &
 	\checkmark & \checkmark &&&
 	\checkmark & &  
 	\\
%
    Chen et al. \cite{chen2018android} &
 	\checkmark &&& 
 	\specialcell{\pie{4} \\ \pie{4} \\ \pie{0} \\ \pie{0}} &
    \specialcell{\pie{4} \\ \pie{4} \\ \pie{4} \\ \pie{4}} &
    \specialcell{\pie{0} \\ \pie{0} \\ \pie{0} \\ \pie{0}} &
    \specialcell{\pie{4} \\ \pie{4} \\ \pie{4} \\ \pie{4}} &
    \specialcell{\pie{4} \\ \pie{0} \\ \pie{4} \\ \pie{0}} &
 	\specialcell{$\mathbf{M}$ \\ $\mathbf{M}$} &
 	\specialcell{${\sf OE1}$ \\ ${\sf OE2}$} &
 	\specialcell{{\sf GO} \\ {\sf SF}} &
 	\specialcell{$\mathcal{X}_{\mathbf{M}}$\\ $\mathcal{X}_{\mathbf{M}}$} &
 	& \specialcell{\checkmark } & \specialcell{\checkmark } & & \checkmark &
 	& \checkmark & &  &
 	& \checkmark &  
 	\\
%
	\rowcolor{lightgray!30}
	Pierazzi et al. \cite{pierazzi2020problemspace} &
	\checkmark &&& 
	\specialcell{\pie{0}} & 
	\specialcell{\pie{0}} &
	\specialcell{\pie{0}} & 
	\specialcell{\pie{0}} &
	\specialcell{\pie{0}} &
	$\mathcal{M}$ &
	${\sf OE2}$ &
	{\sf MI} &
	$\mathcal{Z}_{\mathbf{M}}$ &
	& \checkmark & & & &
	& & \checkmark & & 
	& \checkmark &  
	\\ 	 
%
	Li and Li \cite{DBLP:LiL20} & 
	\checkmark &&& 
	\specialcell{\pie{4}} & 
	\specialcell{\pie{0}} & 
	\specialcell{\pie{0}} & 
	\specialcell{\pie{0}} & 
	\specialcell{\pie{0}}  &
	\specialcell{$\mathbf{M}$ \\ $\mathcal{M}$} &
	\specialcell{${\sf OE2}$ \\ ${\sf OE2}$} &
	\specialcell{{\sf MS} \\ {\sf MS}} &
	\specialcell{$\mathcal{X}_{\mathbf{M}}$ \\ $\mathcal{Z}_{\mathcal{M}}$} &
	& \checkmark & & & \checkmark &
	&& \checkmark &&
	& \checkmark & 
	\\\hline
	\end{tabular}
}
	\label{tbl:attacks}
\end{table}

\subsubsection{Attacks using Gradient-based Optimization ({\sf GO})}

Biggio et al.~\cite{Biggio:Evasion} propose solving the problem of optimal evasion attacks by leveraging gradient-based optimization techniques. They focus on high-confidence evasion attacks with small perturbations (cf. Definition \ref{def:opt_evasion2}). Given a malware representation-label pair $(\mathbf{x},y=+)$, the optimization problem  specified in Eq.\eqref{eq:a8:evs2} with respect to {\sf GO} is instantiated as:
\begin{align*}
	\max_{\delta_{\mathbf{x}}} L_\mathcal{A}(\hat{F}_{\hat\theta}(\mathbf{x}+\delta_{\mathbf{x}}),y=+) & = \min_{\delta_{\mathbf{x}}} \left(L(F_\theta(\mathbf{x}+\delta_{\mathbf{x}}, y=-)) - \beta_a\mathcal{K}(\mathbf{x} + \delta_{\mathbf{x}})\right) \\ ~~\text{s.t.}~~(\delta_{\mathbf{x}}\in[\mathbf{0},\overline{\mathbf{u}}]) &\land (C(\mathbf{x},\mathbf{x}+\delta_{\mathbf{x}})\leq m),
\end{align*}
where $\beta_a\geq 0$ is a balance factor and $\mathcal{K}$ is a density estimation function for lifting $\mathbf{x}+\delta_{\mathbf{x}}$ to the populated region of benign examples. Since $\delta_{\mathbf{x}}\geq \mathbf{0}$, the manipulation only permits object injections to meet the requirement of preserving malicious functionalities. The attack is validated by using the PDF malware detector and the feature representation is the number of appearances of hand-selected keywords (e.g., JavaScript). Because the perturbation is continuous, the authors suggest searching a discrete point close to the continuous one and aligning the point with $\nabla L_\mathcal{A}(\mathbf{x}+\delta_{\mathbf{x}},y=+)$. 
This attack makes the invertibility Assumption \ref{assumption:inverse} because it operates in the feature space. Experimental results show that when $\mathcal{I}$ employs no countermeasures, knowing $\mathcal{I}$'s feature set $S$ and learning algorithm $F$ are sufficient for $\mathcal{A}$ to evade $\mathcal{I}$'s detector. This attack and its variants have been shown to evade PDF malware detectors \cite{rndic_laskov,biggio2014security,russu2016secure,zhang2016adversarial}, PE malware detectors \cite{kolosnjaji2018adversarial}, Android malware detectors \cite{DBLP:LiL20}, and Flash malware detectors \cite{maiorca2017adversarial}. The kernel density estimation item makes the perturbed representation $\mathbf{x}+\delta_\mathbf{x}$ similar to the representations of benign examples, explaining the successful evasion. In summary, the attack works under the ${\sf Oracle}$, ${\sf Measurability}$, and {\sf Invertibility} assumptions. $\mathcal{A}$'s input is, or $\mathcal{A}$ can be characterized as, $(a_1,\ldots,a_5|A_6,\cdots,A_9) = (1,1,1,1,1|\mathbf{M},{\sf OE2},{\sf GO},\mathcal{X}_{\mathbf{M}}) \lor (0,0,1,*,0|\mathbf{M},{\sf OE2},{\sf GO},\mathcal{X}_{\mathbf{M}})$ and $\mathcal{A}$ breaks the {\sf CR} property.

Al-Dujaili et al.~\cite{al2018adversarial} propose evasion attacks against DNN-based malware detectors in the feature space. In this attack,
$\mathcal{A}$ generates adversarial examples  with possibly large perturbations in the feature space. More precisely, given a representation-label pair $(\mathbf{x},y=+)$, the optimization problem of Eq.\eqref{eq:a8:evs2} with respect to {\sf GO} is instantiated as:
	$\max_{\delta_{\mathbf{x}}} L(F_{\theta}(\mathbf{x}+\delta_{\mathbf{x}}),y=+) ~~\text{s.t.}~~ \delta_{\mathbf{x}} \in [\mathbf{0}, \mathbf{1}-\mathbf{x}].$
The attack has four variants, with each perturbing the representation in a different direction (e.g., normalized gradient of the loss function using the $\ell_\infty$ norm). A ``random'' rounding operation is used to map continuous perturbations into a discrete domain. When compared with the basic rounding (which returns 0 if the input is smaller than 0.5, and returns 1 otherwise), the ``random'' rounding means that the threshold of rounding is sampled from the interval $[0,1]$ uniformly. For binary feature representation, the manipulation set $\mathbf{M}_\mathbf{x}=[\mathbf{0}, \mathbf{1}-\mathbf{x}]$ assures the flipping of 0 to 1. The effectiveness of the attack is validated using Windows malware detector in the feature space. In summary, the attack works under the {\sf Oracle} and {\sf Invertibility} assumptions with $\mathcal{A}$ input $(a_1,\ldots,a_5|A_6,\cdots,A_9) =(0,1,1,1,1|\mathbf{M}, {\sf OE2},{\sf GO},\mathcal{X}_\mathbf{M})$ and breaks the {\sf DR} property.

Kreuk et al. \cite{kreuk2018adversarial} propose an evasion attack in the feature space against MalConv, which is an end-to-end Windows PE malware detector (as reviewed in Section \ref{sec:ml}) \cite{raff2017malware}. Given a malware embedding code $\mathbf{x}$, the optimization problem of Eq.\eqref{eq:a8:evs2} with respect to {\sf GO} is instantiated as:
	$\min_{\delta_{\mathbf{x}}} L(F_{\theta}([\mathbf{x}|\delta_{\mathbf{x}}]),y=-) ~~\text{s.t.}~~ \Vert\delta_{\mathbf{x}}\Vert_p \leq \epsilon,$
where | means concatenation and $\Vert\cdot\Vert_p$ is the $p$ norm where $p\geq 1$. Because MalConv is learned from sequential data, perturbation means appending some content to the end of a PE file. Perturbations are generated in a single step by following the direction of the $\ell_\infty$ or $\ell_2$ normalized gradients of the loss function \cite{goodfellow_2015,DBLP:journals/corr/KurakinGB16a}. For instance, the attack based on the $\ell_{\infty}$ norm is
$\tilde{\mathbf{x}}'=\mathbf{x} - \epsilon\cdot\sign(\nabla_{\bf x}(L(F_\theta(\mathbf{x}), -))$, where $\sign(x)=+1~(-1)$ if $x\geq0~(x<0)$.
Since the embedding operation uses a look-up table to map discrete values (0, 1, $\ldots$, 255) to the learned real-value vectors, the attack uses a nearest neighbor search to look for the learned embedding code close to $\tilde{\mathbf{x}}'$.  
In summary, the attack works under the {\sf Oracle} and {\sf Invertibility} assumptions with input $(a_1,\cdots,a_5|A_6,\cdots,A_9)=(0,1,1,1,1|\mathbf{M},{\sf OE2},{\sf GO},\mathcal{X}_\mathbf{M})$ and breaks the {\sf RR} and {\sf CR} properties.

Kolosnjaji et al. \cite{kolosnjaji2018adversarial} and Suciu et al. \cite{8844597} independently propose gradient-based attacks in the feature space to evade MalConv \cite{raff2017malware}. Both studies also use the loss function exploited by Kreuk et al. \cite{kreuk2018adversarial}. Kolosnjaji et al. \cite{kolosnjaji2018adversarial} use the manipulation set $\mathcal{M}$ corresponding to appending instructions at the end of a file. This attack proceeds iteratively and starts with randomly initialized perturbations. In each iteration, continuous perturbations are updated in the direction of the $\ell_2$ normalized gradient of the loss function with respect to the input, and then a nearest neighbor search is applied to obtain discrete perturbations. Suciu et al. \cite{8844597} perturb embedding codes in the direction of the $\ell_\infty$ normalized gradient of the loss function, while adding instructions in the mid of a PE file (e.g., between PE {\em sections}) while noting that appended content could be truncated by MalConv. Both attacks work under the {\sf Oracle} and {\sf Invertibility} assumptions with input $(a_1,\cdots,a_5|A_6,\cdots,A_9)=(0,1,1,1,1|\mathbf{M},{\sf OE2},{\sf GO},\mathcal{X}_\mathbf{M})$ and break the {\sf RR} and {\sf CR} properties.

Mu{\~n}oz-Gonz{\'a}lez et al. \cite{munoz2017towards} propose the optimal poisoning {\sf OP} attack in the feature space (Definition \ref{def:optimal poisoning attack}), which is NP-hard. In this case, the optimization problem of Eq.\eqref{eq:a8:poi} is relaxed by supposing that the classifier is linear to render the optimization problem tractable \cite{biggio2012poisoning,xiao2015feature,munoz2017towards}. The attack is waged against Windows PE malware detectors. Feature set includes API calls, actions and modifications in the file system; each file is represented by a binary vector. The attack has two variants: one uses white-box input, where $\mathcal{A}$ derives $D_{poison}'$ from $\mathcal{I}$'s detector $f$; the other uses grey-box input, where $\mathcal{A}$ knows $\mathcal{I}$'s training set as well as feature set and trains a surrogate detector. The attack works under the {\sf Oracle} and {\sf Invertibility} assumptions with input $(a_1,\ldots,a_5|A_6,\cdots,A_9) =(1,1,1,1,1|\mathbf{M},{\sf OP}, {\sf GO}, \mathcal{X}_\mathbf{M})\lor(1,0,1,0,0|\mathbf{M},{\sf OP}, {\sf GO},\mathcal{X}_\mathbf{M})$ and breaks {\sf TR}.

\subsubsection{Attacks using Sensitive Features ({\sf SF})}

Demontis et al. \cite{demontis2017yes} propose the optimal evasion {\sf OE2} in the feature space to perturb important features in terms of their weights in the linear function $\varphi(\mathbf{x})={\mathbf{w}}^\top\mathbf{x} + b$, where $\mathbf{w}=(w_1,w_2,\cdots,w_d)$ is a weight vector, $b$ is the bias, and $d$ is the dimension of feature space. The attack is waged against Drebin malware detector (which is reviewed in Section \ref{sec:ml}). $\mathcal{A}$ manipulates the $x_i$'s with largest $|w_i|$'s as follows:
flip $x_i=1$ to $x_i=0$ if $w_i > 0$, flip $x_i=0$ to $x_i=1$ if $w_i < 0$, and do nothing otherwise, while obeying the manipulation set $\mathbf{M}$ corresponding to the injection or removal of features. The attack works under the {\sf Oracle}, {\sf Measurability}, and {\sf Invertiblity} assumptions with
input $(a_1,\cdots,a_5|A_6,\cdots,A_9)=(1,1,1,1,1|\mathbf{M},{\sf OE2},{\sf SF},\mathcal{X}_\mathbf{M}) \lor (0,0,1,*,0|\mathbf{M},{\sf OE2},{\sf SF},\mathcal{X}_\mathbf{M})$ and breaks the {\sf CR} property.

Grosse et al. \cite{grosse2017adversarial} propose a variant of the Jacobian-based Saliency Map Attack (JSMA) \cite{papernot_2016} in the feature space against the Drebin  malware detector (which is reviewed in Section \ref{sec:ml}). Instead of using SVM, Deep Neural Network (DNN) is used to build a detector. Important features are identified by leveraging the gradients of the softmax output of a malware example with respect to the input. A large gradient value indicates a high important feature. $\mathcal{A}$ only injects manifest features to manipulate Android Packages and generates adversarial files from $\mathcal{I}$'s detector $f$. The attack works under the {\sf Oracle}, {\sf Measurability}, and {\sf Invertiblity} assumptions
with input $(a_1,\cdots,a_5|A_6,\cdots,A_9)=(0,1,1,1,1|\mathbf{M},{\sf OE2},{\sf SF},\mathcal{X}_\mathbf{M})$ and breaks the {\sf RR} and {\sf CR} properties.

Chen et al. \cite{chen2017adversarial} propose an evasion attack in the feature space by perturbing the important features derived from a wrapper-based feature selection algorithm \cite{chen2017adversarial,Chen:2017:SES,zhang2016adversarial,peng2005feature}. The attacker's loss function $L_\mathcal{A}$ has two parts: (i) the classification error in the mean squared loss and (ii) the manipulation cost $C(\mathbf{x}, \mathbf{x}')=\sum_{i=1}^{d}c_i|x_i - x_i'|$, where $\mathbf{x}=(x_1,\ldots,x_d)$, $\mathbf{x}'=(x_1', \ldots,x_d')$, and $c_i$ is the hardness of perturbing the $i$th feature while preserving malware's functionality. The attack is waged against Windows PE malware detector that uses hand-crafted Windows API calls as features and the binary feature representation. However, there are no details about the composition of manipulation set. This attack works under the {\sf Oracle}, {\sf Measurability}, and {\sf Invertibility} assumptions with input $(a_1,\ldots,a_5|A_7,\cdots,A_9)=(1,1,1,1,1|\mathbf{M},{\sf OE2},{\sf SF}, \mathcal{X}_\mathbf{M}))$ and breaks {\sf CR}.

Chen et al. \cite{chen2018android} propose evasion attacks in the feature space against two Android malware detectors, MaMaDroid \cite{mariconti2016mamadroid} and Drebin \cite{Daniel:NDSS}. The manipulation set $\mathbf{M}$ corresponds to the injection of manifest features (e.g., {\em activities}) and 
API calls. $\mathcal{A}$ evades MaMaDroid by using the optimal evasion ${\sf OE1}$ (Definition \ref{def:opt_evasion1}) and ${\sf OE2}$ (Definition \ref{def:opt_evasion2}), and evades Drebin by using ${\sf OE2}$. The optimization problem of ${\sf OE1}$ Eq.\eqref{eq:a8:evs1} is solved using an advanced gradient-based method known as C\&W \cite{carliniW16a}. ${\sf OE2}$ is solved using JSMA \cite{papernot_2016}. Because JSMA perturbs sensitive features, we categorize this attack into the {\sf SF} group. The ${\sf OE2}$ attack works under the {\sf Oracle}, {\sf Measurability} and {\sf Invertibility} assumptions,
with four kinds of input 
$(a_1,\cdots,a_5|A_6,\cdots,A_9)=(0,0,1,0,0|\mathbf{M},{\sf OE2}, {\sf SF},\mathcal{X}_\mathbf{M}) \lor (0,0,1,0,1|\mathbf{M},{\sf OE2},{\sf SF},\mathcal{X}_\mathbf{M}) \lor (1,0,1,0,0|\mathbf{M},{\sf OE2},{\sf SF}, \mathcal{X}_\mathbf{M}) \lor (1,0,1,0,1|\mathbf{M},{\sf OE2},{\sf SF},\mathcal{X}_\mathbf{M})$,
and breaks the {\sf CR} property. The ${\sf OE1}$ attack works under the same assumptions with the same input except using attack technique {\sf GO}, and breaks the {\sf CR} property.

Chen et al. \cite{chen2018automated} propose a basic poisoning {\sf BP} attack in the feature space against Android malware detectors. The feature set contains syntax features (e.g., permission, hardware, {API}) and semantic features (e.g., sequence of pre-determined program behaviors such as {\tt getDevicedID}$\rightarrow$ URL$\rightarrow${\tt openConnection}). The ML algorithm used is SVM, random forest, or $K$-Nearest Neighbor (KNN) \cite{breiman2001random,shalev2014understanding}. The malware representations are perturbed using a JSMA variant \cite{papernot_2016} against the SVM-based classifier (while noting JSMA is applicable neither to random forests nor to KNN because they are gradient-free). Feature manipulation set ${\bf M}$ corresponds to the injection of syntax features. $\mathcal{A}$ poisons $\mathcal{I}$'s training set by injecting perturbed perturbations with label $-$. The attack works under the {\sf Oracle}, {\sf Measurability}, and {\sf Invertibility} assumptions with input
$(a_1,\ldots,a_5|A_6,\cdots,A_9)=(1,1,1,1,1|\mathbf{M},{\sf BP},{\sf SF},\mathcal{X}_\mathbf{M}) \lor (0,0,0,1,1|\mathbf{M},{\sf BP},{\sf SF},\mathcal{X}_\mathbf{M}) \lor (1,0,*,1,1|\mathbf{M},{\sf BP},{\sf SF},\mathcal{X}_\mathbf{M})$ and breaks {\sf TR}.

Suciu et al. \cite{suciu2018does} propose a basic poisoning attack in both feature and problem spaces. The authors obtain $D'_{poison}$ by applying small manipulations to non-adversarial benign files and then obtain their labels as given by VirusTotal service \cite{VirusTotal:Online}. 
$\mathcal{A}$'s objective is to make $\mathcal{I}$'s classifier $f$ mis-classify a targeted malware file $z_{mal}$ as benign. $\mathcal{A}$ proceeds as follow: (i) obtain an initial benign file $z_{ben}$, where $z_{ben} \approx z_{mal}$ in the feature space with respect to the $\ell_1$ norm; (ii) use the JSMA method \cite{papernot_2016} to manipulate $z_{ben}$ to $z'_{ben}$ by making a small perturbation so that they have similar feature representations; (iii) add $z'_{ben}$ and its label obtained from VirusTotal to $D'_{poison}$ and use $D_{train} \cup D'_{poison}$ to train classifier $f'$ (Definition \ref{definition:sf}); (iv) undo the addition if $z'_{ben}$ lowers the classification accuracy significantly, and accept it otherwise. The attack is waged against the Drebin malware detector and the manipulation set corresponds to the feature injection of permission, API, and strings. This attack works under the {\sf Oracle}, {\sf Measurability}, and {\sf Inversiability} assumptions with input 
$(a_1,\ldots,a_5|A_6,\ldots,A_9) =(*,0,1,*,0|\mathbf{M},{\sf BP},{\sf SF},\mathcal{X}_\mathbf{M})\lor(1,1,1,1,1|\mathbf{M},{\sf BP},{\sf SF},\mathcal{X}_\mathbf{M})\lor(1,0,*,*,0|\mathbf{M},{\sf BP},{\sf SF},\mathcal{X}_\mathbf{M}) \lor (1,0,1,0,0|\mathbf{M},{\sf BP},{\sf SF},\mathcal{X}_\mathbf{M})$
and breaks the {\sf TR} property. The study generates adversarial malware examples, but does not test their malicious functionalities.

\subsubsection{Attacks using MImicry ({\sf MI})} Smutz and Stavrou \cite{smutz_2012} propose a mimicry attack in the feature space to modify features of a malicious file to mimic benign ones, where $\mathcal{A}$ knows $\mathcal{I}$'s classifier $f$. Discriminative features are identified by observing their impact on classification accuracy. The attack perturbs features of malware examples by replacing their value with the mean of the benign examples. The attack is leveraged to estimate the robustness of PDF malware detectors without considering the preservation of malware functionality. The attack works under the {\sf Measurability} assumption with input $(a_1,\ldots,a_5|A_6,\cdots,A_9) = (1,1,1,1,1|\mathbf{M},{\sf OE2},{\sf MI},\mathcal{X})$ and breaks the {\sf CR} property.

Maiorca et al. \cite{Maiorca:2013:LBE:2484313.2484327} propose a reverse mimicry attack against PDF malware detectors in the problem space. Instead of modifying malicious files to mimic benign ones, $\mathcal{A}$ embeds malicious payload (e.g., JavaScript code) into a benign file. The attack can be enhanced by using parser confusion strategies, which make the injected objects being neglected by feature extractors when rendered by PDF readers \cite{DBLP:conf/ndss/CarmonyHYBZ16}. The attack works under the {\sf Oracle} assumption with input $(a_1,\ldots,a_5|A_6,\cdots,A_9) = (0,0,0,0,0|$ $\mathcal{M},{\sf BE},{\sf MI},\mathcal{Z}_\mathcal{M})$ and breaks the {\sf DR} property. 

Pierazzi et al. \cite{pierazzi2020problemspace} propose a white-box evasion attack against the Drebin malware detector and then an enhanced version of the detector in the problem space \cite{Daniel:NDSS,demontis2017yes}. They intend to bridge the gap between the attacks in the problem space and the attacks in the feature space.
In addition, four realistic constraints are imposed on the manipulation set $\mathcal{M}$, including available transformation, preserved semantics, robustness to preprocessing, and plausibility. In order to cope with the side-effect features when incorporating gradient information of $\mathcal{I}$'s classifier, the attacker first harvests a set of manipulations from benign files; 
Manipulations in the problem space are used to query $\mathcal{I}$'s feature extraction for obtaining perturbations in the feature space;
an adversarial malware example is obtained by using the manipulations corresponding to the perturbations that have a high impact on the classification accuracy. This attack works under the {\sf Oracle} assumption with input $(a_1,\cdots,a_5|A_6,\cdots,A_9)=(1,1,1,1,1|\mathcal{M},{\sf OE2},{\sf MI},\mathcal{Z}_\mathcal{M})$ and breaks the {\sf DR} property.

\subsubsection{Attacks using TRansferability ({\sf TR})} 

{\v{S}}rndi\'{c} and Laskov \cite{rndic_laskov} investigate the mimicry attack and the aforementioned gradient descent and kernel density estimation attack against the PDFrate service, where $\mathcal{A}$ knows some features used by $\mathcal{I}$. $\mathcal{A}$ makes the representation of an adversarial malware example similar to a benign one. Manipulation set $\mathcal{M}$ corresponds to adding objects into PDF files. Both attacks perturb feature vectors against a surrogate model, and then map the perturbed feature representations to the problem space by injecting manipulations between the body and the trailer of a PDF file. For the mimicry attack, $\mathcal{A}$ uses $N_{ben}>0$ benign examples to guide manipulations, resulting in $N_{ben}$ perturbed examples. The example incurring the highest classification error is used as an adversarial example. The attack works under the {\sf Oracle} and {\sf Invertibility} assumptions with input 
$( a_1,\ldots,a_5|A_6,\ldots,A_9) =(0,0,*,0,0|\mathbf{M},{\sf BE},{\sf TR},\mathcal{X}_\mathbf{M})\lor(0,0,*,*,0|\mathbf{M},{\sf BE},{\sf TR},\mathcal{X}_\mathbf{M})\lor(1,0,*,0,0|\mathbf{M},{\sf BE},{\sf TR},\mathcal{X}_\mathbf{M})\lor(1,0,*,*,0|\mathbf{M},{\sf BE},{\sf TR},\mathcal{X}_\mathbf{M})\lor(0,0,*,0,0|\mathcal{M},{\sf BE},{\sf TR},\mathcal{Z}_\mathcal{M})\lor(0,0,*,*,0|\mathcal{M},$ ${\sf BE},{\sf TR},\mathcal{Z}_\mathcal{M})\lor(1,0,*,0,0|\mathcal{M},{\sf BE},{\sf TR},\mathcal{Z}_\mathcal{M})\lor(1,0,*,*,0|\mathcal{M},{\sf BE},{\sf TR},\mathcal{Z}_\mathcal{M})$
and breaks {\sf DR}. The gradient-based attack neglects the constraint of small perturbations and works under the same assumptions with the same input except using the attack technique {\sf OE2}.

Khasawneh et al. \cite{khasawneh2017rhmd} propose an evasion attack in both the feature and problem spaces against malware detectors learned from dynamic hardware features (e.g., instruction frequency), where $\mathcal{A}$ knows some features used by $\mathcal{I}$. The attack proceeds as follows. $\mathcal{A}$ first queries $\mathcal{I}$'s classifier to obtain a surrogate model and then generates adversarial files against the surrogate model. Manipulation set $\mathcal{M}$ corresponds to the injection of some features because the others (e.g., memory access) are uncontrollable.  Perturbations are conducted to the important features that are identified by large weights in the model. The attack works under the {\sf Oracle} and {\sf Invertibility} assumptions with input $( a_1,\ldots,a_5|A_6,\cdots,A_9) =(0,0,*,0,1|\mathbf{M},{\sf BE},{\sf TR},\mathcal{X}_\mathbf{M})\lor(0,0,*,0,1|\mathcal{M},{\sf BE},{\sf TR},\mathcal{Z}_\mathcal{M})$ and breaks the {\sf CR} property.

Rosenberg et al. \cite{rosenberg2017generic} propose an evasion attack in both the feature and problem spaces against a Recurrent Neural Network (RNN) based surrogate model, which is learned from API call sequences. In this attack, $\mathcal{A}$'s training data is different from $\mathcal{I}$'s, but the labels are obtained by querying $\mathcal{I}$'s detector. In order to reduce the number of queries to $\mathcal{I}$'s detector, $\mathcal{A}$ augments its training data using the Jacobian-based augmentation technique \cite{papernotMGJCS16} and modifies the API sequence of an example in the direction of the $\ell_\infty$ normalized gradient of the loss function. Manipulation set ${\bf M}$ corresponds to inserting no-op API calls. Experimental results show that adversarial examples generated from a surrogate RNN model can evade SVM, DNN, and RNN detectors. The attack works under the {\sf Oracle} and {\sf Invertibility} assumptions with input $( a_1,\ldots,a_5|A_6,\cdots,A_9) =(0,0,1,0,1|\mathbf{M},{\sf BE},{\sf TR},\mathcal{X}_\mathbf{M}) \lor (0,0,1,0,1|\mathcal{M},{\sf BE},{\sf TR},\mathcal{Z}_\mathcal{M})$ and breaks the {\sf RR} and {\sf CR} properties.

\subsubsection{Attacks using Heuristic Search ({\sf HS})}

Xu et al.~\cite{316904628} propose black-box evasion attacks in the problem space against two PDF malware detectors known as PDFrate \cite{smutz_2012} and Hidost \cite{vsrndic2016hidost}, respectively. Given a malicious file $z$, $\mathcal{A}$ uses a genetic algorithm to iteratively generate $z'$ from $z$ as follows: (i) $\mathcal{A}$ manipulates a set of candidates (or $z$ in the initial iteration) via object deletion, insertion, or replacement. (ii) $\mathcal{A}$ queries these variants to $\mathcal{O}$ and $f$. (iii) $\mathcal{A}$ succeeds when obtaining a successful adversarial example $z'$, namely $({\tt True} \leftarrow \mathcal{O}(z,z')) \land (-\leftarrow f(z'))$; otherwise, $\mathcal{A}$ uses a score function to select candidates for the next iteration or aborts after reaching a threshold number of iterations. The score function $h$ varies with classifiers; for PDFrate, $h(\mathcal{O},f,z,z')=0.5-f(z')$ if $\mathcal{O}(z,z')={\sf true}$, and returns -0.5 if $\mathcal{O}(z,z')={\sf false}$. This attack models an {\sf Oracle} and works with the input $(a_1,\cdots, a_5|A_6,\cdots,A_9) = (0,0,0,0,1|\mathcal{M},{\sf BE},{\sf HS},\mathcal{Z}_\mathcal{M})$ and breaks the {\sf DR} property.

Yang et al. \cite{yang2017malware} propose evasion attacks against Android malware detectors in the problem space. In this attack, $\mathcal{A}$ also uses a genetic algorithm to perturb a malware example $z$ iteratively. In each iteration, $\mathcal{A}$ extracts some features and calculates similarity scores between the malicious APKs in the feature space; the features that have high impact on the similarity scores are selected; the manipulations are to perturb the selected features.
The attack works under the {\sf Oracle} and {\sf Measurability} assumptions with input $(a_1,\cdots, a_5|A_6,\cdots,A_9) = (0,0,0,0,1|\mathcal{M},\mathcal{Z},{\sf BE}, {\sf HS},\mathcal{Z}_\mathcal{M})$ and breaks the {\sf DR} property.

Dang et al. \cite{dang2017evading} propose a black-box evasion attack against malware detectors (e.g., PDFrate) in the problem space. Given a malicious file $z$, $\mathcal{A}$ uses the hill-climbing algorithm to iteratively generate adversarial file $z'$ from $z$. In each iteration, $\mathcal{A}$ generates a path of variants sequentially, each of which is perturbed from its predecessor using manipulations corresponding to object deletion, insertion, or replacement. A score function $h$ is leveraged to select candidates, such as $h(\mathcal{O},f,z,z')={\sf mal}_{z'}-{\sf clf}_{z'}$ or $h(\mathcal{O},f,z,z')={\sf mal}_{z'}/{\sf clf}_{z'}$, where ${\sf mal}_{z'}$ denotes the length of the first example turned from malicious to benign (obtaining by using an oracle) on the manipulation path (cf. Definition \ref{definition:hs}) and ${\sf clf}_{z'}$ denotes the length of the first malware example that has successfully misled the classifier $f$. Both examples of interest are obtained by a binary search, effectively reducing the number of queries to oracle $\mathcal{O}$ and $f$. The attack models an {\sf Oracle} and works with input $(a_1,\cdots, a_5|A_6,\cdots,A_9) = (0,0,0,0,0|\mathcal{M},{\sf BE},{\sf HS},\mathcal{Z}_{\mathcal{M}})$ and breaks the {\sf DR} property.

\subsubsection{Attacks using Generative Model ({\sf GM})}

Hu and Tan \cite {Hu2017} propose an evasion attack against Windows malware detectors in the feature space, by using Generative Adversarial Networks (GAN) \cite{goodfellow2014generative}. In this attack, $\mathcal{A}$ modifies the binary representation of Windows API calls made by malicious files, namely flipping some feature values from 0 to 1. $\mathcal{A}$ learns a generator $G_{\theta_{g}}$ and a discriminator from $\mathcal{A}$'s training dataset. The discriminator is a surrogate detector learned from feature vectors corresponding to $\mathcal{A}$'s benign files and those produced by $G_{\theta_{g}}$, along with labels obtained by querying $\mathcal{I}$'s detector $f$. An adversarial example feature vector is generated by using
$\mathbf{x}' = \max(\mathbf{x}, \round(G_{\theta_{g}}(\mathbf{x}, \mathbf{a})))$,
where $\mathbf{a}$ is a vector of noises, $\round$ is the round function, and $\max$ means element-wise maximum.
Hu and Tan~\cite{hu2017black} also propose another evasion attack using the Seq2Seq model \cite{cho2014learning}. Both attacks work under the {\sf IID}, {\sf Oracle} and {\sf Invertibility} assumptions with input $(a_1,\cdots,a_5|A_6,\cdots,A_9)=(0,0,1,0,1|\mathbf{M},{\sf BE},{\sf GM},\mathcal{X}_\mathbf{M})$ and break the {\sf DR} property.

Anderson et al. \cite{anderson2018learning} propose a Reinforcement Learning (RL)-based evasion attack against Windows PE malware detectors in the problem space. Manipulation set $\mathcal{M}$ is the RL action space, which includes some bytecode injections (e.g., API insertion) and some bytecode deletion. Attacker $\mathcal{A}$ learns an RL agent on $\mathcal{A}$'s data, with labels obtained by querying defender $\mathcal{I}$'s detector $f$. The learned agent predicts manipulations sequentially for a given malware example. Moreover, $\mathcal{A}$ is restricted by only applying a small number of manipulations to a malicious PE file. Experimental results show that the attack is not as effective as others (e.g., gradient-based methods). The attack works under the {\sf Oracle} and {\sf Measurability} assumptions with input $(a_1,\cdots, a_5|A_6,\cdots,A_9) = (0,0,0,0,1|\mathcal{M},{\sf OE2},{\sf GM},\mathcal{Z}_\mathcal{M})$ and breaks the {\sf RR} and {\sf CR} properties.

\subsubsection{Attacks using Mixture Strategy ({\sf MS})} 

Li and Li \cite{DBLP:LiL20} propose evasion attacks against DNN-based Android malware detectors in both feature and problem spaces. Given four gradient-based attack methods, the attack looks for the best one to perturb malware representations. $\mathcal{A}$ can iteratively perform this strategy to modify the example obtained in the previous iteration. Experimental results show that the mixture of attacks can evade malware detectors effectively. The attack works under the {\sf IID}, {\sf Oracle} and {\sf Invertibility} assumptions with input $(a_1,\cdots, a_5|A_6,\cdots,A_9) = (1,1,1,1,1|\mathbf{M},{\sf BE},{\sf MS},\mathcal{X}_\mathbf{M})\lor(1,1,1,1,1|\mathcal{M},{\sf BE},{\sf MS},\mathcal{Z}_\mathcal{M})$ and breaks the {\sf DR} property. 


\subsubsection{Drawing Observations and Insights} 

We summarize the preceding reviews with the following observations.
(i) Indiscriminate attacks have been much more extensively investigated than targeted attacks and availability attacks. (ii) Evasion attacks have been much more extensively studied than poisoning attacks. (iii) The {\sf Oracle} assumption has been widely made. 
In addition, we draw the following insights.
\begin{insight} 
(i) Knowing the defender's feature set is critical to the success of transfer attacks, highlighting the importance of keeping the defender's feature set secret (e.g., randomizing the defender's feature set).
(ii) The effectiveness of evasion attacks largely depends on the attacker's degree of freedom in conducting manipulations in the problem space (i.e., a smaller degree of freedom means it is harder for the attack to succeed).
\end{insight}

\subsection{Systematizing Defense Literature} \label{sec:defense}

\begin{table}[htbp]
	\caption{Summary of AMD defenses (\checkmark means applicable, \protect\fullcirc means 0, \protect\emptycirc means 1, \protect\dotcirc means a value in $[0,1]$)}
	\resizebox{1.\columnwidth}{!}{
	\begin{threeparttable}
	\centering
	\setlength{\tabcolsep}{0.4em}
	\begin{tabular}{l|c|ccccc|cccc|ccccc|cccc|ccc}
	\hline
	\multicolumn{1}{c|}{\specialcell{Defense \\ (in chronological order)}}& 
	\multicolumn{1}{c|}{\specialcell{Defense \\ Objective}} &
	\multicolumn{9}{c|}{\specialcell{Defense Input}} & 
	\multicolumn{5}{c|}{\specialcell{Assumptions}} & 
	\multicolumn{4}{c|}{\specialcell{Achieved \\ Properties}} & 
	\multicolumn{3}{c}{\specialcell{Malware\\Detector}}
	\\
	
	& \vthead{malware detection} 
	
	& \vthead{$A_1$: Training set $D_{train}$}
	& \vthead{$A_2$: Defense technique}
	& \vthead{$A_3$: Feature set}
	& \vthead{$A_4$: Learning algorithm}
	& \vthead{$A_5$: Response}
	& \vthead{$A_6$: Manipulation set}
	& \vthead{$A_7$: Attack tactic}
	& \vthead{$A_8$: Attack technique}
	& \vthead{$A_9$: Adversarial example set}
	
	& \vthead{${\sf IID}$ assumption}
	& \vthead{${\sf Oracle}$ assumption}
	& \vthead{${\sf Measurability}$ assumption}
	& \vthead{${\sf Smoothness}$ assumption}
	& \vthead{${\sf Invertibility}$ assumption}
	
	& \vthead{{\sf RR}: Representation Robustness}
	& \vthead{{\sf CR}: Classification Robustness}
	& \vthead{{\sf DR}: Detection Robustness}
	& \vthead{{\sf TR}: Training Robustness}
	
	& \vthead{Windows Program}
	& \vthead{Android Package}
	& \vthead{PDF}
	\\
	\hline\hline
%
    \rowcolor{lightgray!30}
    Biggio et al. \cite{biggio2015one} &
    \checkmark&
 	$D_{train}$ &
 	${\sf EL}$ &
 	$S$ &
 	$F_\theta$ &
 	{\sf FQ} &
 	\pie{4} & 
 	\pie{0} & 
 	\pie{4} &
 	\pie{4} &
 	\checkmark& & & & &
 	&\checkmark&&&
 	& &\checkmark
 	\\
%
    Smutz and Stavrou \cite{smutz2016tree} &
    \checkmark&
 	$D_{train}$ &
 	${\sf SE}$ &
 	$S$ &
 	$F_\theta$ &
 	{\sf FQ} &
 	\pie{4} & 
 	\pie{0} & 
 	\pie{4} &
 	\pie{4} &
 	\checkmark&&&&&
 	&&\checkmark&&
 	& & \checkmark 
 	\\
%
    \rowcolor{lightgray!30}
    Zhang et al. \cite{zhang2016adversarial} &
    \checkmark&
 	$D_{train}$ &
 	${\sf RF}$ &
 	$S$ &
 	$F_\theta$ &
 	{\sf FQ} &
 	\pie{0} & 
 	\pie{0} & 
 	\pie{4} &
 	\pie{4} &
 	\checkmark&&\checkmark&\checkmark&&
 	\checkmark&\checkmark&&&
 	& & \checkmark
 	\\
%
    Demontis et al. \cite{demontis2017yes} &
    \checkmark&
 	$D_{train}$ &
 	${\sf WR}$ &
 	$S$ &
 	$F_\theta$ &
 	{\sf FQ} &
 	\pie{0} & 
 	\pie{0} & 
 	\pie{4} &
 	\pie{4} &
 	\checkmark& &\checkmark& & &
 	&\checkmark&&&
 	& \checkmark &
 	\\
%
 	\rowcolor{lightgray!30}
 	Wang et al. \cite{wang_2017} &
 	\checkmark&
 	$D_{train}$ &
 	${\sf IT}$ &
 	$S$ &
 	$F_\theta$ &
 	{\sf FQ} &
 	\pie{4} & 
 	\pie{0} & 
 	\pie{4} &
 	\pie{4} &
 	\checkmark&&&&&
 	&\checkmark&&&
 	\checkmark & &
 	\\
%
    Grosse et al. \cite{grosse2017adversarial} &
    \checkmark&
 	$D_{train}$ &
 	${\sf WR}$ &
 	$S$ &
 	$F_\theta$ &
 	{\sf FQ} &
 	\pie{4} & 
 	\pie{0} & 
 	\pie{4} &
 	\pie{4} &
 	\checkmark&&&&&
 	\checkmark&\checkmark&&&
 	& \checkmark &
 	\\
%
    \rowcolor{lightgray!30}
    Grosse et al. \cite{grosse2017adversarial} &
    \checkmark&
 	$D_{train}$ &
 	${\sf AT}$ &
 	$S$ &
 	$F_\theta$ &
 	{\sf FQ} &
 	\pie{0} & 
 	\pie{0} & 
 	\pie{0} &
 	\pie{4} &
 	\checkmark& &\checkmark&&&
 	\checkmark&\checkmark&&&
 	& \checkmark &
 	\\
%
    Chen et al. \cite{chen2017adversarial} &
    \checkmark&
 	$D_{train}$ &
 	${\sf AT}$ &
 	$S$ &
 	$F_\theta$ &
 	{\sf FQ} &
 	\pie{0} & 
 	\pie{0} & 
 	\pie{0} &
 	\pie{4} &
 	\checkmark& &\checkmark&&&
 	&\checkmark&&&
 	\checkmark & &
 	\\
%
    \rowcolor{lightgray!30}
    Khasawneh et al. \cite{khasawneh2017rhmd} &
    \checkmark&
 	$D_{train}$ &
 	${\sf CD}$ &
 	$S$ &
 	$F_\theta$ &
 	{\sf FQ} &
 	\pie{4} & 
 	\pie{0} & 
 	\pie{0} &
 	\pie{4} &
 	\checkmark&&&&&
 	&\checkmark&&&
 	\checkmark & &
 	\\  
%
 	Dang et al. \cite{dang2017evading} &
 	\checkmark&
 	$D_{train}$ &
 	${\sf SE}$ &
 	$S$ &
 	$F_\theta$ &
 	{\sf LQ} &
 	\pie{4} & 
 	\pie{0} & 
 	\pie{4} &
 	\pie{4} &
 	\checkmark&&&&&
 	&&\checkmark&&
 	& & \checkmark 
 	\\  
%
    \rowcolor{lightgray!30}
    Yang et al. \cite{yang2017malware} &
    \checkmark&
 	\specialcell{$D^\ast_{train}$} &
 	${\sf AT}$ &
 	$S$ &
 	$F_\theta$ &
 	{\sf FQ} &
 	\pie{4} & 
 	\pie{0} & 
 	\pie{4} &
 	\pieg &
 	\checkmark&&&&&
 	&&\checkmark&&
 	& \checkmark &
 	\\
%
    Yang et al. \cite{yang2017malware} &
    \checkmark&
 	\specialcell{$D_{train}$} &
 	${\sf SE}$ &
 	$S$ &
 	$F_\theta$ &
 	{\sf FQ} &
 	\pie{0} & 
 	\pie{0} & 
 	\pie{0} &
 	\pie{4} &
 	\checkmark&&&&&
 	&&\checkmark&&
 	& \checkmark &
 	\\
%
    \rowcolor{lightgray!30}
    Chen et al. \cite{Chen:2017:SES} &
    \checkmark&
 	$D_{train}$ &
 	${\sf RF}$ &
 	$S$ &
 	$F_\theta$ &
 	{\sf FQ} &
 	\pie{0} & 
 	\pie{0} & 
 	\pie{4} &
 	\pie{4} &
 	\checkmark& &\checkmark&\checkmark& &
 	\checkmark&\checkmark&&&
 	& \checkmark & 
 	\\
%
    Incer et al. \cite{incer2018adversarially} &
    \checkmark&
 	$D_{train}$ &
 	${\sf VL}$ &
 	$S$ &
 	$F_\theta$ &
 	{\sf FQ} &
 	\pie{0} & 
 	\pie{0} & 
 	\pie{4} &
 	\pie{4} &
 	\checkmark& & & & &
 	\checkmark&\checkmark&\checkmark& & 
 	\checkmark& & 
 	\\ 
%
 	\rowcolor{lightgray!30}
 	Chen et al. ~\cite{chen2018automated} &
 	\checkmark&
 	$D_{train}$ &
 	${\sf SE}$ &
 	$S$ &
 	$F_\theta$ &
 	{\sf FQ} &
 	\pie{4} & 
 	\pie{0} & 
 	\pie{4} &
 	\pie{4} &
 	&&\checkmark&&&
 	&&&\checkmark&
 	&\checkmark &
 	\\
%
    Al-Dujaili et al.  \cite{al2018adversarial} &
    \checkmark&
 	$D_{train}$ &
 	${\sf AT}$ &
 	$S$ &
 	$F_\theta$ &
 	{\sf FQ} &
 	\pie{0} & 
 	\pie{0} & 
 	\pie{4} &
 	\pie{4} &
 	\checkmark&&&& &
 	&&\checkmark&&
 	\checkmark & &
 	\\
%
    \rowcolor{lightgray!30}
    Chen et al. \cite{YeFOSINT-SI-2018} &
    \checkmark&
 	$D_{train}$ &
 	${\sf IT}$ &
 	$S$ &
 	$F_\theta$ &
 	{\sf FQ} &
 	\pie{4} & 
 	\pie{0} & 
 	\pie{4} &
 	\pie{4} &
 	\checkmark&&&&&
 	&\checkmark&&&
 	& \checkmark &
 	\\
%
    Jordan et al. \cite{jordan2018safe} &
    \checkmark&
 	$D_{train}$ &
 	${\sf RF}$ &
 	$S$ &
 	$F_\theta$ &
 	{\sf FQ} &
 	\pie{0} & 
 	\pie{0} & 
 	\pie{4} &
 	\pie{4} &
 	&&&&&
 	\checkmark&\checkmark&\checkmark&&
 	& & \checkmark
 	\\ 
%
    \rowcolor{lightgray!30}
    Li et al. \cite{li2019enhancing} &
    \checkmark&
 	$D_{train}$ &
 	${\sf AT}$ &
 	$S$ &
 	$F_\theta$ &
 	{\sf FQ} &
 	\pie{4} & 
 	\pie{0} & 
 	\pie{4} &
 	\pie{4} &
 	\checkmark&&\checkmark&\checkmark& &
 	\checkmark&\checkmark&&&
 	\checkmark & &
 	\\
%
 	Tong et al. \cite{DBLP:conf/uss/Tong0HXZV19} &
 	\checkmark&
 	$D_{train}$ &
 	${\sf RF}$ &
 	$S$ &
 	$F_\theta$ &
 	{\sf FQ} &
 	\pie{0} & 
 	\pie{0} & 
 	\pie{4} &
 	\pie{4} &
 	\checkmark&&&&&
 	\checkmark&\checkmark&\checkmark&&
 	& & \checkmark
 	\\ 
%
 	\rowcolor{lightgray!30}
 	Li and Li  \cite{DBLP:LiL20} &
 	\checkmark&
 	$D_{train}$ &
 	${\sf AT}$ &
 	$S$ &
 	$F_\theta$ &
 	{\sf FQ} &
 	\pie{0} & 
 	\pie{0} & 
 	\pie{4} &
 	\pie{4} &
 	\checkmark&&&& &
 	&&\checkmark&&
 	& \checkmark &
 	\\
%
    Chen et al. \cite{DBLP:conf/uss/Chen0SJ20} &
    \checkmark&
    $D_{train}$ &
    ${\sf VL}$ &
    $S$ &
    $F_\theta$ &
    {\sf FQ} &
    \pie{0} & 
    \pie{0} & 
    \pie{4} &
    \pie{4} &
    \checkmark& &\checkmark & \checkmark& &
    \checkmark&\checkmark& & & 
    & & \checkmark
    \\
    
    \rowcolor{lightgray!30}
    Li et al. \cite{DBLP:journals/tnse/LiLYX21} & 
 	\checkmark &
 	$D_{train}$ &
 	\small{{\sf EL}+{\sf AT}+{\sf RF}} &
 	$S$ &
 	$F_\theta$ &
 	{\sf FQ} &
 	\pie{0} & 
 	\pie{0} & 
 	\pie{4} &
 	\pie{4} &
 	\checkmark&&\checkmark&& &
 	&&\checkmark&&
 	& \checkmark &
 	\\
    \hline
	\end{tabular}
	\begin{tablenotes}
		\small
		\item $D^\ast_{train}$ contains $D_{train}$ and a portion of $\mathcal{A}$'s adversarial examples.
	\end{tablenotes}
    \end{threeparttable}
	}
	\label{tbl:defenses}
\end{table}


\subsubsection{Defenses using Ensemble Learning ({\sf EL})} \label{sec:defense-el}

Biggio et al. \cite{biggio2015one} propose a one-and-a-half-class SVM classifier against evasion attacks, by leveraging an interesting observation (i.e., decision boundaries of one-class SVM classifiers are tighter than that of two-class SVM classifiers) to facilitate outlier detection. Specifically, the authors propose an ensemble of a two-class classifier and two one-class classifiers, and then combine them using another one-class classifier. 
The defense can enhance PDF malware detectors against gradient-based attacks \cite{Biggio:Evasion}, which can be characterized as $(a_1,\cdots,a_5|A_6,\cdots,A_9)=(1,1,1,1,1|\mathbf{M},{\sf OP2}, {\sf GO},\mathcal{X}_\mathbf{M})$. However, the defense cannot thwart attacks incurring large perturbations. Independent of this study, other researchers propose using the random subspace and bagging techniques to enhance SVM-based malware detectors, dubbed Multiple Classifier System SVM (MCS-SVM), which leads to evenly distributed weights \cite{DBLP:journals/mlc/BiggioFR10,demontis2017yes}.
These defenses work under the {\sf IID} assumption with input $(A_1,\cdots,A_5|a_6,\cdots,a_9)=(D_{train},{\sf EL},S,F_\theta,{\sf FQ}|0,1,0,0)$ and achieves the {\sf CR} property.

\subsubsection{Defenses using Weight Regularization ({\sf WR})} \label{sec:attack-wr}

Demontis et al. \cite{demontis2017yes} propose enhancing the Drebin malware detector $\varphi(\mathbf{x})=\mathbf{w}^\top\mathbf{x} + b$ by using box-constraint weights. The inspiration is that the classifier's sensitivity to perturbations based on the $\ell_1$ norm is bounded by the $\ell_\infty$ norm of the weights. This defense
hardens the Drebin detector against a mimicry attack with input $(a_1,\cdots,a_5|A_6,\cdots,A_9)=(0,0,1,0,0|\mathbf{M},{\sf BE},{\sf MI},\mathcal{X}_\mathbf{M})$, obfuscation attack \cite{dexguard:Online} with input $(a_1,\cdots,a_5|A_6,\cdots,A_9)=(0,0,0,0,$ $0|\mathcal{M},{\sf BE},-, \mathcal{Z}_\mathcal{M})$, and the attack that modifies important features  \cite{demontis2017yes} with input $(a_1,\cdots,a_5|A_6,\cdots,$ $A_9)=(1,1,1,1,1|\mathbf{M},{\sf OE2},{\sf SF},\mathcal{X}_\mathbf{M})$, here `$-$' means inapplicable. Experimental results show this defense outperforms MSC-SVM \cite{DBLP:journals/mlc/BiggioFR10}. The defense works under the {\sf IID}, {\sf Oracle} and {\sf Measurability} assumptions with input $(A_1,\cdots,A_5|a_6,\cdots,a_9)=(D_{train},{\sf WR},S,F_\theta,{\sf FQ}|1,1,0,0)$ and achieves {\sf CR}.

Grosse et al. \cite{grosse2017adversarial} investigate how to apply two defense techniques known as distillation \cite{DBLP:conf/sp/PapernotM0JS16} and retraining \cite{szegedyZSBEGF13} to enhance the DNN-based Drebin malware detector. The distillation technique can decrease a model's generalization error by leveraging a teacher to relabel the training data represented by real-value vectors (rather than one-hot encoding). It uses retraining to tune a learned model with respect to an augmented training set with adversarial examples. Both defenses are estimated against a variant of JSMA and can be characterized by their input as $(A_1,\cdots,A_5|a_6,\cdots,a_9)=(0,1,1,1,1|\mathbf{M},{\sf OE2},{\sf SF}, \mathcal{X}_\mathbf{M})$. Experimental results show the two defenses achieve limited success. The defense based on the distillation technique works under the {\sf IID} assumption with input $(A_1,\cdots,A_5|a_6,\cdots,a_9)=(D_{train},{\sf WR},S,F_\theta,{\sf FQ}|0,1,0,0)$ and  achieves the {\sf RR} and {\sf CR} properties. The defense based on the retraining technique works under the {\sf IID} and {\sf Measurability} assumptions with input $(A_1,\cdots,A_5|a_6,\cdots,a_9)=(D_{train},{\sf AT},S,F_\theta,{\sf FQ}|1,$ $1,1,0)$ and achieves the {\sf RR} and {\sf CR} properties.

\subsubsection{Defenses using Adversarial Training ({\sf AT})}

Chen et al. \cite{chen2017adversarial} adapt a generic retraining framework proposed in the AML context \cite{li2016general} to enhance linear malware detectors. The defense uses a label smoothness regularization technique to mitigate the side-effect of adversarial training \cite{yang2015min}. The defense is evaluated using Windows malware detectors against ``feature selection''-based evasion attacks, which can be characterized as $(A_1,\cdots,A_5|a_6,\cdots,a_9)=(1,1,1,1,1|\mathbf{M},{\sf OE2},{\sf SF},\mathcal{X}_\mathbf{M})$. The defense works under the {\sf IID} and {\sf Measurability} assumptions and can be characterized as $(A_1,\cdots,A_5|a_6,\cdots,a_9)=(D_{train},{\sf AT},S,F_\theta,{\sf FQ}|1,1,1,0)$, while assuring the {\sf CR} property.

Yang et al. \cite{yang2017malware} propose a defense against genetic algorithm-based evasion attacks that can be characterized as $(A_1,\cdots,A_5|a_6,$ $\cdots,a_9)=(0,0,0,0,1|\mathcal{M},{\sf BE},{\sf HS},\mathcal{Z}_\mathcal{M})$. The defense uses three techniques: adversarial training, sanitizing examples, and weight regularization \cite{demontis2017yes}. The adversarial training uses one half of $\mathcal{A}$'s adversarial examples. The defense works under the {\sf IID} assumption and can be characterized as
$(A_1,\cdots,A_5|a_6,\cdots,a_9)=(D'_{train},{\sf AT},S,F_\theta,{\sf FQ}|0,1,0,*)$, where $D'_{train}$ is the union of $D_{train}$ and a portion (e.g., one half) of $\mathcal{A}$'s adversarial examples. The defense of sanitizing examples is learned from manipulations used by the attacker and works under the IID assumption with input $(A_1,\cdots,A_5|a_6,\cdots,a_9)=(D_{train},{\sf SE},S,F_\theta,{\sf FQ}|1,1,1,0)$. Both defenses achieve the {\sf DR} property. The defense of wight regularization is reviewed in Section \ref{sec:attack-wr}.

Al-Dujaili et al.~\cite{al2018adversarial} adapt the idea of minmax adversarial training (proposed in the AML context) to enhance DNN-based malware detectors. In this defense, the inner-layer optimization generates adversarial files by maximizing the classifier's loss function; the outer-layer optimization searches for the parameters $\theta$ (of DNN $F_\theta$) that minimize the classifier's loss with respect to the adversarial files.
The defense enhances Windows malware detectors against attacks with input $(A_1,\cdots,A_5|a_6,\cdots,a_9)=(0,1,1,1,1|\mathbf{M},{\sf OE2},{\sf GO},\mathcal{X}_\mathbf{M})$. 
Experimental results show that malware detectors that are hardened to resist one attack may not be able to defend against other attacks. By observing this phenomenon, researchers propose using a mixture of attacks to harden DNN-based malware detectors \cite{DBLP:LiL20}.
The defense works under the {\sf IID} assumption and can be characterized as  $(A_1,\cdots,A_5|a_6,\cdots,a_9)=(D_{train},{\sf AT},S,F_\theta,{\sf FQ}|1,1,0,0)$. The defense assures the {\sf DR} property. 

Li et al. \cite{li2019enhancing} propose a DNN-based attack-agnostic framework to enhance adversarial malware detectors. The key idea is dubbed adversarial regularization, which enhances malware detectors via the (approximately) optimally small perturbation. The framework wins the AICS'2019 adversarial malware classification challenge organized by MIT Lincoln Lab researcher \cite{aics:Online}, without knowing anything about the attack. 
The defense works under the {\sf IID}, {\sf Measurability}, and {\sf Smoothness} assumptions with input $(A_1,\cdots,A_5|a_6,\cdots,a_9)=(D_{train},{\sf AT},S,F_\theta,{\sf FQ}|0,1,0,0)$ and assures the {\sf RR} and {\sf CR} properties. In the extended study \cite{DBLP:journals/tnse/LiLYX21}, the authors further enhance the framework with 6 defense principles, including ensemble learning, adversarial training, and robust representation learning. The enhanced defense is validated with 20 attacks (including 11 grey-box attacks and 9 white-box attacks) against Android malware detectors. The enhanced defense works under the {\sf IID} and {\sf Measurability} assumptions with input $(A_1,\cdots,A_5|a_6,\cdots,a_9)=(D_{train},{\sf AT}+{\sf EL}+{\sf RF},S,F_\theta,{\sf FQ}|1,1,0,0)$ and assures the {\sf DR} property.

\subsubsection{Defenses using Verifiable Learning ({\sf VL})}

Incer et al. \cite{incer2018adversarially} propose using monotonic malware classifiers to defend  against evasion attacks, where {\em monotonic} means
$\varphi(\mathbf{x}) \leq \varphi(\mathbf{x}')$ when ${\mathbf{x}} \leq {\mathbf{x}}'$ \cite{gupta2016monotonic}. Technically, this can be achieved by using (i) robust features that can only be removed or added but not both and (ii) monotonic classification function (e.g., linear models with non-negative weights). The resulting classifier can thwart any attack that perturbs feature values monotonically. The defense works under the {\sf IID} assumption with input $(A_1,\cdots,A_5|a_6,\cdots,a_9)=(D_{train},{\sf VL},S,F_\theta,{\sf FQ}|1,1,0,0)$ and assures the {\sf RR}, {\sf CR} and {\sf DR} properties.

Chen et al. \cite{DBLP:conf/uss/Chen0SJ20} propose a defense to enhance PDF malware detectors against evasion attacks, by leveraging the observation that manipulations on PDF files are subtree additions and/or removals. They also propose new metrics for quantifying such structural perturbations. This allows to adapt the 
{\em symbolic interval analysis} technique proposed in the AML context  \cite{DBLP:conf/uss/WangPWYJ18} to enhance the PDF malware detectors. The defense can cope with attacks leveraging small perturbations in the training phase. This defense works under the {\sf IID}, {\sf Measurability}, and {\sf Smoothness} assumptions with input $(A_1,\cdots,A_5|a_6,\cdots,a_9)=(D_{train},{\sf VL},S,F_\theta,{\sf FQ}|1,1,0,0)$ and achieves the {\sf RR} and {\sf CR} properties.

\subsubsection{Defenses using Robust Features ({\sf RF})}

Zhang et al. \cite{zhang2016adversarial} propose leveraging optimal adversarial attacks for feature selection. The defense enhances PDF malware detectors against gradient-based attacks, which can be characterized as $(A_1,\cdots,A_5|a_6,\cdots,a_9)=(1,1,1,1,1|\mathbf{M},{\sf OE2},{\sf GO},\mathcal{X}_{\mathbf{M}})$. The defense works under the {\sf IID}, {\sf Measurability}, and {\sf Smoothness} assumptions and can be characterized as $(A_1,\cdots,A_5|a_6,\cdots,a_9)=(D_{train},{\sf RF},S,F_\theta,{\sf FQ}|1,1,0,0)$. The defense assures the {\sf RR} and {\sf CR} properties.

Tong et al. \cite{DBLP:conf/uss/Tong0HXZV19} propose refining features into invariant ones to defend against genetic algorithm-based evasion attacks with input $(A_1,\cdots,A_5|a_6,\cdots,a_9)=(0,0,0,0,1|\mathcal{M},{\sf BE},{\sf HS},\mathcal{Z}_\mathcal{M})$. Experimental results show that adversarial training can be further leveraged to enhance the robustness of the defense. The defense works under {\sf IID} assumption with input $(A_1,\cdots,A_5|a_6,\cdots,a_9)$ $=(D_{train},{\sf RF},S,F_\theta,{\sf FQ}|1,1,0,0)$, and achieves {\sf RR}, {\sf CR} and {\sf DR}.

Chen et al. \cite{Chen:2017:SES} propose mitigating evasive attacks by filtering features according to their importance $|w_i|/c_i$  with respect to the linear function $\varphi(\mathbf{x})=\mathbf{w}^\top\mathbf{x} + b$, where $x_i$, $w_i$ and $c_i$ denote respectively the $i$th component of $\mathbf{x}$, $\mathbf{w}$ and the constraint on manipulation cost $\mathbf{c}$. The defense enhances Android malware detectors against three attacks: a random attack with input $(A_1,\cdots,A_5|a_6,\cdots,a_9)=(0,0,1,0,0|\mathbf{M},{\sf BE},-,\mathcal{X}_\mathbf{M})$, a variant of the mimicry attack with input $(0,0,1,0,0|\mathbf{M},{\sf BE},{\sf MI},$ $\mathcal{X}_\mathbf{M})$, and the attack that modifies important features with input $(A_1,\cdots,A_5|a_6,\cdots,a_9)=(1,1,1,1,1|\mathbf{M},{\sf BE},{\sf SF},$ $\mathcal{X}_\mathbf{M})$, where `$-$' means inapplicable. The defense works under the {\sf IID}, {\sf Measurability}, and {\sf Smoothness} assumptions with input $(A_1,\cdots,A_5|a_6,$ $\cdots,a_9)=(D_{train},{\sf RF},S,F_\theta,{\sf FQ}|1,1,0,0)$ and achieves {\sf RR} and {\sf CR}.

Jordan et al. \cite{jordan2018safe} propose a robust PDF malware detector against evasion attacks by interpreting JavaScript behaviors using static analysis. A PDF file is classified as malicious when it calls a vulnerable API method or when it exhibits potentially malicious or unknown behaviors. The defense is validated against the {\em reverse mimicry} attack \cite{Maiorca:2013:LBE:2484313.2484327} with input $(A_1,\cdots,A_5|a_6,\cdots,a_9)=(0,0,0,0,0|\mathcal{M},{\sf BE},{\sf MI},\mathcal{Z}_\mathcal{M})$. The defense has input $(A_1,\cdots,A_5|a_6,\cdots,a_9)=(D_{train},{\sf RF},S,F_\theta,{\sf FQ}|1,$ $1,0,0)$ and achieves {\sf RR}, {\sf CR} and {\sf DR}.

\subsubsection{Defenses using Input Transformation ({\sf IT})}

Wang et al. \cite{wang_2017} propose the {\em random feature nullification} to enhance DNN-based malware detectors against the attack of Fast Gradient Sign Method (FGSM) \cite{goodfellow_2015} by nullifying (or dropping) features randomly in both training and testing phases. This offers a probabilistic assurance in preventing a white-box attacker from deriving adversarial files by using gradients of the loss function with respect to the input. 
The defense enhances Windows malware detectors against the FGSM attack with input $(A_1,\cdots,A_5|a_6,\cdots,a_9)=(0,1,1,1,1|\mathbf{M},{\sf OE2},{\sf GO},\mathcal{X}_\mathbf{M})$. The defense works under {\sf IID} assumption with input $(A_1,\cdots,A_5|a_6,$ $\cdots,a_9)=(D_{train},{\sf IT},S,F_\theta,{\sf FQ}|0,1,0,0)$ and achieves {\sf CR}.

DroidEye \cite{YeFOSINT-SI-2018} defends Android malware detectors against evasion attacks by quantizing binary representations, namely transforming binary representations into real values and then using compression to reduce the effect of adversarial manipulations. The defense enhances linear malware detectors against a ``feature selection''-based attack with input
$(A_1,\cdots,A_5|a_6,\cdots,a_9)=(1,1,1,1,1|\mathbf{M},{\sf OE2},{\sf SF},\mathcal{X}_\mathbf{M})$ \cite{chen2017adversarial} and the FGSM attack with input $(A_1,\cdots,A_5|a_6,\cdots,a_9)=(0,1,1,1,$ $1|\mathbf{M},{\sf OE2},{\sf GO},\mathcal{X}_\mathbf{M})$ \cite{goodfellow_2015}. The defense works under {\sf IID} assumption with input $(A_1,\cdots,A_5|a_6,\cdots,a_9)$ $=(D_{train},{\sf IT},S,F_\theta,{\sf FQ}|0,1,0,0)$ and  achieves {\sf CR}.

\subsubsection{Defenses using Classifier Randomization ({\sf CD})}

Khasawneh et al. \cite{khasawneh2017rhmd} propose randomizing classifiers (i.e., using one randomly chosen from a pool of classifiers that use heterogeneous features) to defend against transfer attacks. The defense is validated against an attack which perturbs important features with input
$(a_1,\dots,a_5|a_6,\cdots,a_9)=(0,0,*,0,1|\mathcal{M},{\sf BE},{\sf SF},\mathcal{Z}_\mathcal{M})$.
The defense works under the {\sf IID} assumption with input $(a_1,\dots,a_5|a_6,\cdots,a_9)=(D_{train},{\sf CD},S,F_\theta,$ ${\sf FQ}|0,1,1,0)$ and  achieves the {\sf CR} property.

\subsubsection{Defenses using Sanitizing Examples ({\sf SE})}

Smutz and Stavrou \cite{smutz2016tree} propose an ensemble classifier to defend against grey-box evasion attacks by returning classification results as benign, uncertain and malicious according to the voting result (e.g., $[0\%,25\%]$ classifiers saying malicious can be treated as benign, $[25\%,75\%]$ saying malicious can be treated as uncertain, and $[75\%,100\%]$ saying malicious can be treated as malicious). The defense enhances a PDF malware detector against three types of evasion attacks: gradient-based attack \cite{rndic_laskov} with input $(a_1,\ldots,a_5|A_6,\cdots,A_9) =(1,0,*,*,0|\mathcal{M},{\sf OE2},{\sf TR},\mathcal{Z}_\mathbf{M})$, mimicry attack with input $(1,0,*,*,0|\mathcal{M},{\sf BE},{\sf TR},\mathcal{Z}_\mathbf{M})$, and reverse mimicry attack with input $(0,0,0,0,0|\mathcal{M},{\sf BE},{\sf MI},\mathcal{Z}_\mathbf{M})$ \cite{Maiorca:2013:LBE:2484313.2484327}.
The defense works under the {\sf IID} assumption with input $(A_1,\cdots,A_5|a_6,\cdots,a_9)$ $=(D_{train},{\sf SE},S,F_\theta,{\sf FQ}|0,1,0,0)$ and achieves {\sf DR}.

Dang et al. \cite{dang2017evading} propose enhancing PDF malware detectors by lowering the classification threshold $\tau$ and restricting the maximum query times, rendering genetic algorithm-based evasion attacks harder to succeed. This defense works under the {\sf IID} assumption with input $(A_1,\cdots,A_5|a_6,\cdots,a_9)=(D_{train},{\sf SE},S,F_\theta,{\sf LQ}|0,1,0,0)$ and achieves {\sf DR}.

Chen et al. \cite{chen2018automated} investigate defending Android malware detectors against poisoning attacks with input $(a_1,\ldots,a_5|A_6,\cdots,A_9) =(1,1,1,1,1|\mathbf{M},{\sf BP},{\sf SF},\mathcal{X}_\mathbf{M})$
The idea is to filter adversarial files that are distant from non-adversarial ones, where distance is measured by the Jaccard index, Jaccard-weight similarity and cosine similarity.
The defense works under the {\sf Measurability} assumption with input 
$(A_1,\cdots,A_5|a_6,\cdots,a_9)=(D_{train},{\sf SE},S,F_\theta,$ ${\sf FQ}|0,1,0,0)$
and achieves {\sf TR}.

\subsubsection{Drawing Observations and Insights} 

Summarizing the preceding discussions, we draw the following observations.
(i) Most studies focus on black-box defenses (i.e., the defender knows little about the attacker), which is against the principle of ``knowing yourself and knowing your enemy". (ii) Most studies focus on defenses against evasion attacks rather than poisoning attacks. (iii) There is no silver bullet defense against evasion attacks or poisoning attacks, at least for now. (iv) Sanitizing adversarial files as outliers is effective against black-box and grey-box attacks, but not white-box attacks. (v) The security properties achieved by defenses have been evaluated empirically rather than rigorously proven (despite that provable security is emerging on the small degree of perturbations; see for example \cite{gilmer2018adversarial,DBLP:conf/uss/Chen0SJ20}). (vi) There is no theoretical evidence to support that the effectiveness of defense tactics on the training set (e.g., adversarial training and verifiable learning) can generalize to other adversarial examples.
In addition, we draw the following insights:
\begin{insight}
(i) Effective defenses often require the defender to know the attacker's manipulation set. In the real world, it is hard to achieve this, explaining from one perspective why it is hard to design effective defenses.
(ii) The effectiveness of adversarial training depends on the defender's capability in identifying the most powerful attack.
\end{insight}

\subsection{Systematizing AMD Arms Race}

\afterpage{
\begin{rotatepage}
\begin{figure*}[ht!]
\centering
\rotatebox{90}{
\scalebox{0.8}{
\begin{tikzpicture}

\node(rbot1)[label=below:{}, label=left:{},align=center] at (20,3.6) {$ (*,0,1,*,0)$\\$ (1,0,1,0,0)$\\$ (1,0,*,*,0)$};
\node(rbot1_attr)[label=below:{},align=center] at (20,2.6) {$(\mathbf{M},{\sf BP},{\sf SF},\mathcal{X}_\mathbf{M})$ \\ $(\mathcal{M},{\sf BP},{\sf SF},\mathcal{Z}_\mathcal{M})$};
\node(rbot1_tag)[label=below:{},align=center] at (20,2.1) {\cite{suciu2018does}};


\node(rbot)[label=below:{}, label=left:{}] at (20,1) {$ (1,1,1,1,1)$};
\node(rbot_attr)[label=below:{},align=center] at (20,0.6) {$(\mathbf{M},{\sf OE2},{\sf MI},\mathcal{X})$};
\node(rbot_tag)[label=below:{},align=center] at (20,0.2) {\cite{smutz_2012}};

\node(rbot0)[label=below:{}, label=left:{}] at (20,-2) {$ (1,1,1,1,1)$};
\node(rbot0_attr)[label=below:{},align=center] at (20,-2.4) {$(\mathbf{M},{\sf OE2},{\sf SF},\mathcal{X}_\mathbf{M})$};
\node(rbot0_refer)[label=below:{},align=center] at (20,-2.8) {\cite{chen2017adversarial}};



\node(lbot3)[label=below:{}, label=left:{}] at (0,3) {$ (0,0,0,0,1)$};
\node(lbot3_attr)[label=below:{},align=center] at (0,2.6) {$(\mathcal{M},{\sf BE},{\sf HS},\mathcal{Z}_\mathcal{M})$};
\node(lbot3_refer)[label=below:{},align=center] at (0,2.2) {\cite{316904628}};

\node(lbot2)[label=below:{}, label=left:{}] at (0,1) {$ (0,0,0,0,0)$};
\node(lbot2_attr)[label=below:{},align=center] at (0,0.6) {$(\mathcal{M},{\sf BE},{\sf HS},\mathcal{Z}_\mathcal{M})$};
\node(lbot2_refer)[label=below:{},align=center] at (0,0.2) {\cite{dang2017evading}};

\node(lbot1)[label=below:{}, label=left:{}] at (0,-1) {$ (0,0,0,0,0)$};
\node(lbot1_attr)[label=below:{},align=center] at (0,-1.4) {$(\mathcal{M},{\sf BE},{\sf MI},\mathcal{Z}_\mathcal{M})$};
\node(lbot1_refer)[label=below:{},align=center] at (0,-1.8) {\cite{DBLP:conf/ndss/CarmonyHYBZ16}};

\node(lbot0)[label=below:{}, label=left:{}] at (0,-3) {$ (0,0,0,0,0)$};
\node(lbot0_attr)[label=below:{},align=center] at (0,-3.4) {$(\mathcal{M},{\sf BE},{\sf MI},\mathcal{Z}_\mathcal{M})$};
\node(lbot0_refer)[label=below:{},align=center] at (0,-3.8) {\cite{Maiorca:2013:LBE:2484313.2484327}};


\node(above7_attr)[label=below:{},align=center] at (20,6) {$(\mathbf{M},{\sf OE2},{\sf GO},\mathcal{X}_\mathbf{M})$};
\node(above7)[label=below:{}, label=left:{}] at (20,6.4) {$ (0,1,1,1,1)$};
\node(above7_refer)[label=below:{},align=center] at (20,6.8) {\cite{8844597,kreuk2018adversarial,kolosnjaji2018adversarial}};

\node(above6_attr)[label=below:{},align=center] at (16.8,6) {$(\mathbf{M},{\sf OE2},{\sf MS},\mathcal{X}_\mathbf{M})$};
\node(above6)[label=below:{}, label=left:{}] at (16.8,6.4) {$ (0,1,1,1,1)$};
\node(above6_refer)[label=below:{},align=center] at (16.8,6.8) {\cite{DBLP:LiL20}};

\node(above5_attr)[label=below:{},align=center] at (14,6) {$(\mathbf{M},{\sf OE2},{\sf GO},\mathcal{X}_\mathbf{M})$};
\node(above5)[label=below:{}, label=left:{}] at (14,6.4) {$ (0,1,1,1,1)$};
\node(above5_refer)[label=below:{},align=center] at (14,6.8) {\cite{al2018adversarial}};

\node(above4_attr)[label=below:{},align=center] at (11.2,6) {$(\mathbf{M},{\sf OE2},{\sf SF},\mathcal{X}_\mathbf{M})$}; 
\node(above4)[label=below:{}, label=left:{}] at (11.2,6.4) {$ (0,1,1,1,1)$};
\node(above4_refer)[label=below:{},align=center] at (11.2,6.8) {\cite{grosse2017adversarial}};

\node(above3_attr)[label=below:{},align=center] at (8.4,6) {$(\mathcal{M},{\sf OE2},{\sf MI},\mathcal{Z}_\mathcal{M})$};
\node(above3)[label=below:{}, label=left:{}] at (8.4,6.4) {$ (1,1,1,1,1)$};
\node(above3_refer)[label=below:{},align=center] at (8.4,6.8) {\cite{pierazzi2020problemspace}};

\node(above2_attr)[label=below:{},align=center] at (5.6,6) {$(\mathcal{M},{\sf BE},{\sf HS},\mathcal{Z}_\mathcal{M})$};
\node(above2)[label=below:{}, label=left:{}] at (5.6,6.4) {$ (0,0,0,0,1)$};
\node(above2_refer)[label=below:{},align=center] at (5.6,6.8) {\cite{yang2017malware}};

\node(above1_attr)[label=below:{},align=center] at (2.8,6) {$(\mathbf{M},{\sf OE2},{\sf SF},\mathcal{X}_\mathbf{M})$};
\node(above1)[label=below:{}, label=left:{}] at (2.8,6.4) {$ (0,0,1,*,0)$};
\node(above1_refer)[label=below:{},align=center] at (2.8,6.8) {\cite{demontis2017yes}};

\node(above0_attr)[label=below:{},align=center] at (0,5.8) {$(\mathbf{M},{\sf OE1},{\sf GO},\mathcal{X}_\mathbf{M})$ \\ $(\mathbf{M},{\sf OE2},{\sf SF},\mathcal{X}_\mathbf{M})$ };
\node(above0)[label=below:{}, label=left:{}] at (0,6.4) {$ (0,0,1,0,0)$};
\node(above0_refer)[label=below:{},align=center] at (0,6.8) {\cite{chen2018android}};

\node(below7)[label=below:{}, label=left:{}] at (20,-6.4) {$ (1,0,1,0,0)$};
\node(below7_attr)[label=below:{},align=center] at (20,-6) {$(\mathbf{M},{\sf OP},{\sf GO},\mathcal{X}_\mathbf{M})$};
\node(below7_refer)[label=below:{},align=center] at (20,-6.8) {\cite{munoz2017towards}};

\node(below6)[label=below:{}, label=left:{}] at (16.8,-6) {$ (0,0,0,1,1)$};
\node(below6_attr)[label=below:{},align=center] at (16.8,-6.4) {$(\mathbf{M},{\sf BP},{\sf SF},\mathcal{X}_\mathbf{M})$};
\node(below6_tag)[label=below:{},align=center] at (16.8,-6.8) {\cite{chen2018automated}};

\node(below5)[label=below:{}, label=left:{}] at (14,-6) {$ (0,0,1,0,1)$};
\node(below5_attr)[label=below:{},align=center] at (14,-6.4) {$(\mathcal{M},{\sf BE},{\sf TR},\mathcal{Z}_\mathcal{M})$};
\node(below5_refer)[label=below:{},align=center] at (14,-6.8) {\cite{rosenberg2017generic}};

\node(below4)[label=below:{}, label=left:{}] at (11.2,-6) {$ (0,0,1,0,1)$};
\node(below4_attr)[label=below:{},align=center] at (11.2,-6.4) {$(\mathbf{M},{\sf BE},{\sf GM},\mathcal{X}_\mathbf{M})$};
\node(below4_refer)[label=below:{},align=center] at (11.2,-6.8) {\cite{hu2017black,Hu2017}};

\node(below3)[label=below:{}, label=left:{}] at (8.4,-6) {$ (0,0,1,*,0)$};
\node(below3_attr)[label=below:{},align=center] at (8.4,-6.4) {$(\mathbf{M},{\sf OE2},{\sf GO},\mathcal{X}_\mathbf{M})$};
\node(below3_refer)[label=below:{},align=center] at (8.4,-6.8) {\cite{Biggio:Evasion}};

\node(below2)[label=below:{}, label=left:{}] at (5.6,-6) {$ (0,0,*,0,1)$};
\node(below2_attr)[label=below:{},align=center] at (5.6,-6.4) {$(\mathcal{M},{\sf BE},{\sf TR},\mathcal{Z}_\mathbf{M})$};
\node(below2_refer)[label=below:{},align=center] at (5.6,-6.8) {\cite{khasawneh2017rhmd}};


\node(below1)[label=below:{}, label=left:{}] at (2.8,-6) {$ (0,0,0,0,1)$};
\node(below1_attr)[label=below:{},align=center] at (2.8,-6.4) {$(\mathcal{M},{\sf OE2},{\sf GO},\mathcal{Z}_\mathcal{M})$};
\node(below1_refer)[label=below:{},align=center] at (2.8,-6.8) {\cite{anderson2018learning}};

\node(below0)[label=below:{}, label=left:{}] at (0,-6) {$ (0,0,*,0,0)$};
\node(below0_attr)[label=below:{},align=center] at (0,-6.4) {$(\mathcal{M},{\sf BE},{\sf TR},\mathcal{Z}_\mathbf{M})$};
\node(below0_refer)[label=below:{},align=center] at (0,-6.8) {\cite{rndic_laskov}};


\fill [lightgray!30] (1.5,5) rectangle (18.6,-5);


\node(mid4)[label=below:{},align=left] at (15.5,0) {\cite{yang2017malware}:$(D_{train},{\sf SE},S,F_\theta,{\sf FQ})$ \\
\cite{grosse2017adversarial}:$(D_{train},{\sf AT},S,F_\theta,{\sf FQ})$\\
\cite{chen2017adversarial}:$(D_{train},{\sf AT},S,F_\theta,{\sf FQ})$
};
\node(mid4-attr)[align=center] at (15.5,-0.85) {$(1,1,1,0)$};

\node(mid3)[label=below:{$(1,1,0,0)$},align=left] at (10,1.88) {
\cite{demontis2017yes}:$(D_{train},{\sf WR},S,F_\theta,{\sf FQ})$ \\
\cite{DBLP:conf/uss/Tong0HXZV19}:$(D_{train},{\sf RF},S,F_\theta,{\sf FQ})$ \\
\cite{incer2018adversarially}:$(D_{train},{\sf RF},S,F_\theta,{\sf FQ})$ \\
\cite{jordan2018safe}:$(D_{train},{\sf RF},S,F_\theta,{\sf FQ})$ \\
\cite{al2018adversarial}:$(D_{train},{\sf AT},S,F_\theta,{\sf FQ})$  \\
\cite{DBLP:LiL20}:$(D_{train},{\sf AT},S,F_\theta,{\sf FQ})$ \\
\cite{DBLP:conf/uss/Chen0SJ20}:$(D_{train},{\sf AT},S,F_\theta,{\sf FQ})$ \\
\cite{Chen:2017:SES}:$(D_{train},{\sf RF},S,F_\theta,{\sf FQ})$ \\
\cite{zhang2016adversarial}:$(D_{train},{\sf RF},S,F_\theta,{\sf FQ})$
}; 

\node(mid2)[label=below:{$(0,1,1,0)$},align=left] at (10,-1.5) {\cite{khasawneh2017rhmd}:$(D_{train},{\sf CD},S,F_\theta,{\sf FQ})$};

\node(mid1)[label=below:{$(0,1,0,*)$},align=center] at (10,-3) {\cite{yang2017malware}:$(D^\ast_{train},{\sf AT},S,F_\theta,{\sf FQ})$};

\node(mid0)[label=below:{$(0,1,0,0)$},align=left] at (4.5,0) {
\cite{dang2017evading}:$(D_{train},{\sf SE},S,F_\theta,{\sf LQ})$\\
\cite{li2019enhancing}:$(D_{train},{\sf AT},S,F_\theta,{\sf FQ})$ \\
\cite{wang_2017}:$(D_{train},{\sf IT},S,F_\theta,{\sf FQ})$ \\
\cite{grosse2017adversarial}:$(D_{train},{\sf WR},S,F_\theta,{\sf FQ})$ \\
\cite{chen2018automated}:$(D_{train},{\sf SE},S,F_\theta,{\sf FQ})$ \\
\cite{smutz2016tree}:$(D_{train},{\sf SE},S,F_\theta,{\sf FQ})$ \\
\cite{YeFOSINT-SI-2018}:$(D_{train},{\sf IT},S,F_\theta,{\sf FQ})$ \\ 
\cite{biggio2015one}:$(D_{train},{\sf EL},S,F_\theta,{\sf FQ})$
};

\node(det3)[label=below:{},align=center, fill=blue!20,fill opacity=0.8] at (18.2,3) {\cite{Daniel:NDSS}};

\node(det2)[label=below:{},align=center, fill=blue!20,fill opacity=0.8] at (18.1,4.65) {\cite{raff2017malware}};

\node(dnn1)[label=below:{},align=center, fill=blue!20,fill opacity=0.8] at (15.2, 3.7) {DNN-based malware \\ detector \cite{grosse2017adversarial}};
\node(dnn1_label)[align=center,color=blue!50] at (15.2,2.8) {$(D_{train},\emptyset,S,F_\theta,{\sf FQ})$\\$(0,0,0,0)$};

\node(apk1)[label=right:{},align=center, fill=blue!20,fill opacity=0.8] at (6, 3.9) {Drebin \cite{Daniel:NDSS}};
\node(apk1_label) [align=center,color=blue!50] at (6.2,3.2) {$(D_{train},\emptyset,S,F_\theta,{\sf FQ})$\\$(0,0,0,0)$};

\node(det1)[label=below:{},align=center, fill=blue!20,fill opacity=0.8] at (3.05,3.9) {\cite{Daniel:NDSS,mariconti2016mamadroid}};

\node(pdf2)[label=below:{},align=center, fill=blue!20,fill opacity=0.8] at (3.6, 3.2) {PDFrate \cite{smutz_2012}};
\node(pdf2_label)[align=center,color=blue!50] at (3.6,2.4) {$(D_{train},\emptyset,S,F_\theta,{\sf FQ})$\\$(0,0,0,0)$};
\node(pdf1)[label=below:{},label=below:{},align=left, fill=blue!20,fill opacity=0.8] at (3.6, -3.8) {PDFrate \cite{smutz_2012}};
\node(pdf1_label)[align=center,color=blue!50] at (3.6,-4.6) {$(D_{train},\emptyset,S,F_\theta,{\sf FQ})$\\$(0,0,0,0)$};

{
\draw [-latex,line width=1] (3.3,-5.7) -- (3.3, -5) -- (2.3, -5) -- (2.3, -5.7);
\draw [-latex,line width=1] (11.7,-5.7) -- (11.7, -5) -- (10.7, -5) -- (10.7, -5.7);
\draw [-latex,line width=1] (14.5,-5.7) -- (14.5, -5) -- (13.5, -5) -- (13.5, -5.7);
\draw [-latex,line width=1] (20.5,-5.7) -- (20.5,-5) -- (19.5,-5) -- (19.5,-5.7);
\draw [-latex,line width=1] (19.5,1.2) -- (19.5,1.4) -- (18.6,1.4) -- (18.6,0) -- (19.5,0) -- (19.5,0.4); 

\draw [-latex,line width=1] (above0_attr) -- (0, 3.9) -- (det1);
\draw [-latex,line width=1] (20,5.7) -- (20,4.65) -- (det2);
\draw [-latex,line width=1] (19,3.) -- (det3);

\draw [-latex,line width=1] (6.1,-5.7) -- (6.1, -5) -- (5.1, -5) -- (5.1, -5.7);
\draw [-latex,line width=1] (17.3,-5.7) -- (17.3, -5) -- (16.3, -5) -- (16.3, -5.7);

\ignore{
\draw [-latex,line width=1] (6.4,-0.2)node[red,right]{\cite{chen2018automated}}--(5.8,-0.2);
\draw [-latex,line width=1] (6.4,0.2)node[red,right]{\cite{grosse2017adversarial}}--(5.9,0.2);
\draw [-latex,line width=1] (9.2,0.7)node[red,above]{\cite{khasawneh2017rhmd}}--(9.2,0.2);
\draw [-latex,line width=1] (6.4,-3.2)node[red,left]{\cite{yang2017malware}}--(6.9,-3.2);
\draw [-latex,line width=1] (13.6,-3.2)node[red,left]{\cite{chen2017adversarial}}--(14.1,-3.2);
}

\draw [-latex,line width=1] (below3) -- (8.4,-5.0) -- (7.25,-5.0) -- (7.25,-1.5) -- (6.25,-1.5);
\draw [-latex,line width=1] (7.25,-1.5) -- (7.25,0.2) -- (8.1,0.2);
\draw [line width=1] (18.7,-2.4) -- (18.6,-2.4) -- (18.6,-1.05) -- (7.35,-1.05); 
\draw [-latex,line width=1] (7.15,-1.05) --(6.25,-1.05);
\draw [-latex,line width=1] (12.55, -1.05) -- (12.55, 0.65) -- (11.75, 0.65);
\tkzDefPoint(7.25,-1.05){c1}
\tkzDefPoint(7.15,-1.05){c2}
\tkzDrawArc[rotate,color=black,line width=1](c1,c2)(180);

\draw [-latex,line width=1] (11.1,5.7) -- (11.1,4.3) -- (13.15,4.3) -- (13.15,3.7) -- (13.6, 3.7); 
\draw [blue,densely dashdotted,-latex,line width=1] (16.8,3.7) -- (17.5,3.7) -- (17.5,0.1) -- (17.2,0.1); 
\draw [red,dashed,-latex,line width=1] (11.3,5.7) -- (11.3,5.0) -- (13.9,5.0) -- (13.9,5.7);
\draw [-latex,line width=1] (14.1,5.7) -- (14.1,4.3) -- (17.7,4.3) -- (17.7,-0.1) -- (17.2,-0.1); 
\draw [blue,densely dashdotted,-latex,line width=1] (13.55,0) -- (12.75,0) -- (12.75,1.8) -- (11.6,1.8);
\draw [red,dashed,-latex,line width=1] (14.3,5.7) -- (14.3,5.0) -- (16.7,5.0) -- (16.7,5.7);
\draw [line width=1] (17.3,5.7) -- (17.3,4.4);
\draw [-latex,line width=1] (17.3,4.2) -- (17.3,2) -- (11.6,2);
\draw [blue,densely dashdotted,-latex,line width=1] (8.05, 1.9) -- (7.25,1.9) -- (7.25,1.45) -- (8.1,1.45) ;
\tkzDefPoint(17.3,4.3){c3}
\tkzDefPoint(17.3,4.2){c4}
\tkzDrawArc[rotate,color=black,line width=1](c3,c4)(180);

\draw [-latex,line width=1] (2.7,5.7) -- (2.7,4.3) -- (4.7,4.3) -- (4.7,3.9) -- (5.2,3.9); 
\draw [blue,densely dashdotted,-latex,line width=1] (6.9,3.9) -- (7.05,3.9) -- (7.05,3.58) -- (8.1,3.58); 
\draw [red,dashed,-latex,line width=1] (2.9,5.7) -- (2.9,5.0) -- (8.5,5.0) --(8.5,5.7); 
\draw [red,dashed,-latex,line width=1] (5.7,5.0) -- (5.7,5.7);
\draw [-latex,line width=1] (5.5,5.7) -- (5.5,4.3) -- (8.3,4.3) -- (8.3,3.7); 
\draw [line width=1] (8.3,5.7) -- (8.3,4.3); 
\draw [blue,densely dashdotted,-latex,line width=1] (11.9,3.7) -- (12.95,3.7) -- (12.95,0.43) -- (13.55,0.43);
\draw [blue,densely dashdotted,-latex,line width=1] (11.9,3.5) -- (12.75,3.5) -- (12.75,2.7) -- (11.8,2.7);

\draw [-latex,line width=1] (0.1, 3.1) -- (0.1,3.3) -- (2.5,3.3); 
\draw [blue,densely dashdotted,-latex,line width=1] (2.4,3.1) -- (1.7,3.1) -- (1.7, 1.55) -- (2.5,1.55); 
\draw [red,dashed,-latex,line width=1] (-0.1,2.1) -- (-0.1,1.1);
\draw [-latex,line width=1] (0.1,1.1) -- (0.1, 1.35) -- (2.5,1.35); 
\draw [blue,densely dashdotted,-latex,line width=1] (4.7,3.2) -- (8.1,3.2);

\draw [-latex,line width=1] (0.1,-5.7) -- (0.1,-5) -- (1.5,-5) -- (1.5,-5) -- (1.5,-3.9) -- (2.5,-3.9); 
\draw [blue,densely dashdotted,-latex,line width=1] (2.4,-3.7) -- (1.7,-3.7) -- (1.7, -0.75) -- (2.5, -0.75); 
\draw [red,dashed,-latex,line width=1] (-0.1,-5.7) -- (-0.1,-3.9); 
\draw [-latex,line width=1](0.1,-2.9) -- (0.1,-2) -- (1.5,-2) -- (1.5,-0.55) -- (2.5,-0.55); 
\draw [blue,densely dashdotted,-latex,line width=1] (6.4,-0.6) -- (7.05,-0.6) -- (7.05,2.3) -- (8.1,2.3); 
\draw [red,dashed,-latex,line width=1] (-0.1,-2.9) -- (-0.1,-2); 
\draw [-latex,line width=1](0.1,-0.9) -- (0.1,-0.55) -- (2.5,-0.55);
}

\draw [-latex,line width=1] (12.3,7.45) -- (12.9,7.45) node[right,align=center]{Waging \\ Attack};
\draw [blue,densely dashdotted,-latex,line width=1] (14.5,7.45) -- (15.1,7.45);
\node [right,align=center] at (15.1,7.45) {Defense \\ Escalation};
\draw [red,dashed,-latex,line width=1] (17,7.45) -- (17.6,7.45);
\node [right,align=center] at (17.6, 7.45) {Attack \\ Escalation};
\draw (11,7) -- (20,7) -- (20,7.9) -- (11,7.9) -- node [right]{Legend:}  (11,7);
\end{tikzpicture}
}
}
\caption{Arms race in AMD attack and defense escalations.}
\label{fig:attacks-defenses}
\end{figure*}
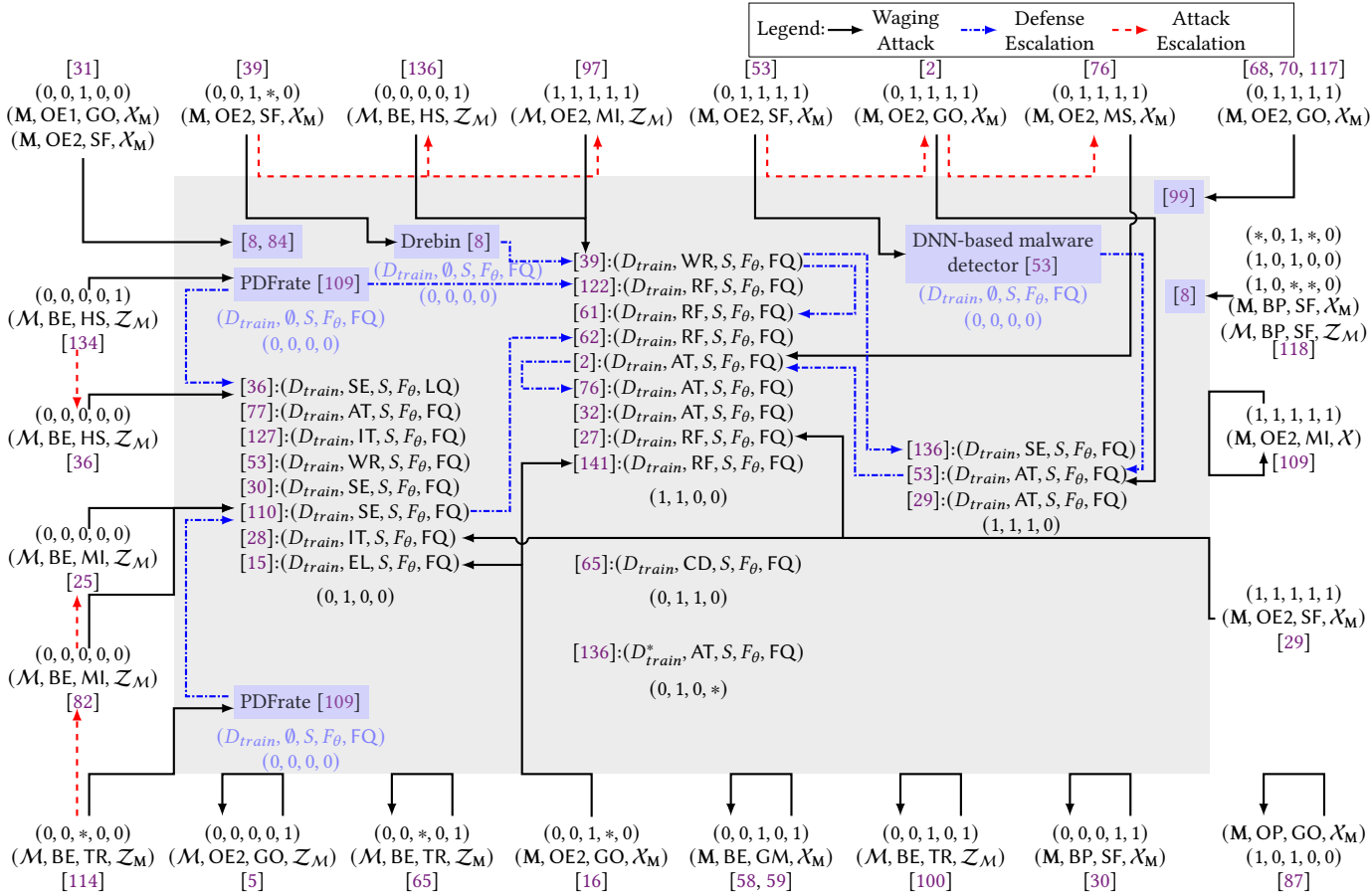
\end{rotatepage}
}

Figure \ref{fig:attacks-defenses} displays AMD attack-defense arms race surrounding three malware detectors: PDFrate, Drebin, and DNN-based detector. For a better visual effect, we group papers that proposed defense methods in terms of a common input $(a_6,\ldots,a_9)$. For example, we group
\cite{yang2017malware},\cite{grosse2017adversarial} and \cite{chen2017adversarial} together because the defenders in both papers have input $(a_6,\ldots,a_9)=(1,1,1,0)$, while noting that their input on $(A_1,\ldots,A_5)$ may or may not be different.
We also simplify attack and defense inputs while preserving the {\em critical information} when an attack (defense) works for multiple inputs. 
For example, $(a_1,\ldots,a_5|A_6,\cdots,A_9)=(0,0,1,0,0|\mathbf{M},{\sf OE2}, {\sf SF}, \mathcal{X}_\mathbf{M})$ is the {\em critical information} for attack input  
$(a_1,\cdots,a_5|A_6,\cdots,A_9)=(0,0,1,0,0|\mathbf{M},{\sf OE2}, {\sf SF}, \mathcal{X}_\mathbf{M}) \lor (0,0,1,0,1|\mathbf{M},{\sf OE2}, {\sf SF}, \mathcal{X}_\mathbf{M}) \lor (1,0,1,0,0|\mathbf{M},{\sf OE2}, {\sf SF}, \mathcal{X}_\mathbf{M}) \lor (1,0,1,0,1|\mathbf{M},{\sf OE2}, {\sf SF},\mathcal{X}_\mathbf{M})$
because it is the weakest attack input in the partial order formulated by these $(a_1,\ldots,a_5)$'s. This suggests us to focus on attack input $(0,0,1,0,0|\mathbf{M},{\sf OE2}, {\sf SF},\mathcal{X}_\mathbf{M})$ because it is already able to break some defense and automatically implies that a stronger input can achieve the same (while noting some special cases, see discussion in Section \ref{sec:defense-el}). Multiple defense inputs are simplified in the same manner.

{\em Arms race in PDF malware detection}: We summarize two sequences of escalations caused by PDFrate \cite{smutz_2012}. In one sequence, PDFrate is defeated by transfer attacks, which are realized by gradient-based and mimicry methods against surrogate models \cite{rndic_laskov}. These attacks trigger the defense escalation to an ensemble detector built on top of some diversified classifiers \cite{smutz2016tree}. This defense \cite{smutz2016tree} triggers attack escalation to {\em reverse mimicry attacks} \cite{Maiorca:2013:LBE:2484313.2484327}, 
which trigger the defense escalation of using robust hand-crafted features \cite{jordan2018safe}. This defense represents the state-of-the-art PDF malware detector, but still incurs a high false-positive rate. In the other sequence of arms race, PDFrate is defeated by genetic algorithm-based attacks \cite{316904628}. These attacks trigger the defense escalation to \cite{dang2017evading} and \cite{DBLP:conf/uss/Tong0HXZV19}. The former defense \cite{dang2017evading} restricts the responses to attacker queries, but can be defeated by the escalated attack that leverages the hill-climbing algorithm (also shown in \cite{dang2017evading}). The latter defense \cite{DBLP:conf/uss/Tong0HXZV19} uses invariant features to thwart the attacks and represents another state-of-the-art PDF malware detectors.

{\em Arms race in android malware detection}: Drebin is defeated by the attack that modifies a limited number of important features \cite{demontis2017yes}, which also proposes the new defense to defeat the escalated attack. This defense \cite{demontis2017yes} triggers the attack escalation to, and is defeated by, the genetic algorithm-based attack \cite{yang2017malware} and the mimicry-alike attack \cite{pierazzi2020problemspace}. The former attack \cite{yang2017malware} triggers the escalated defense (also presented in \cite{yang2017malware}) that leverages attack mutations to detect adversarial examples \cite{yang2017malware}. The latter attack \cite{pierazzi2020problemspace} injects objects in APKs and (in principle) can be defeated by the monotonic classifier \cite{pierazzi2020problemspace,incer2018adversarially}. These escalated defenses represent the state-of-the-art Android malware detectors, but still incur a high false-positive rate.

{\em Arms race in DNN-based malware detection}: The DNN-based detector \cite{grosse2017adversarial} triggers four gradient-based evasion attacks presented in \cite{al2018adversarial}, which also hardens the DNN malware detector by using an {\em minmax adversarial training} instantiation to incorporates the $\ell_\infty$ normalized gradient-based attack. This escalated defense \cite{al2018adversarial} triggers the mixture of attacks presented in \cite{DBLP:LiL20}. The defense of {\em minmax adversarial training} incorporating a mixture of attacks can defeat a broad range of attacks, but still suffers from the mimicry attack and other mixtures of attacks \cite{DBLP:LiL20}. 
As a consequence, there are no effective defenses can thwart all kinds of attacks.

{\em Independent arms race}:
There are studies that have yet to trigger cascading arms races, including: (i) Studies \cite{chen2018automated,wang_2017,chen2017adversarial,suciu2018does} propose independent attacks and then show how to defeat these attacks.
(ii) Studies \cite{chen2018android,Hu2017,hu2017black,rosenberg2017generic,kreuk2018adversarial,kolosnjaji2018adversarial,8844597,munoz2017towards} propose attacks to defeat naive malware detectors. (iii) Studies propose defenses to counter some attacks \cite{biggio2015one,Chen:2017:SES,YeFOSINT-SI-2018,li2019enhancing,DBLP:conf/uss/Chen0SJ20}.

\section{Future Research Directions (FDRs)} \label{sec:cd}

\noindent{\bf FRD 1}: {\em Pinning down the root cause(s) of adversarial malware examples}.
Speculations on root cause(s) include: (i) invalidity of the {\sf IID} assumption because of distribution drifting, namely that testing files and training files are drawn from different distributions   \cite{Biggio:Evasion,Biggio6494573,DBLP:journals/corr/GrosseMP0M17});
(ii) incompetent feature extraction  \cite{316904628,demontis2017yes};
(iii) high dimensionality of malware representations \cite{gilmer2018adversarial};
(iv) insufficient scale of training data \cite{schmidt2018adversarially};
(v) low-probability ``pockets'' in data manifolds \cite{szegedyZSBEGF13};
(vi) linearity of DNNs \cite{goodfellow_2015}; and
(vii) large curvature of decision boundaries \cite{fawzi2016robustness,moosavi2018robustness}.Although these speculations may be true, more studies are needed in order to (in)validate them. 

\smallskip

\noindent{\bf FRD 2}: {\em Characterizing the relationship between transferability and vulnerability}.
In the AMD context, an attacker may use a surrogate model to generate adversarial examples and a defender may use a surrogate model for adversarial training. Transferability is related to the extent at which knowledge gained by a surrogate model may be the same as, or similar to, what is accommodated by a target model. The wide use of surrogate models in the AMD context suggests that there may be a fundamental connection between knowledge transferability and model vulnerability.

\smallskip

\noindent{\bf FRD 3}: {\em Investigating adversarial malware examples in the wild}. In the AMD context, it is challenging to generate practical adversarial malware examples to correspond to perturbations conducted in the feature space, owing to realistic constraints. On the other hand, an attacker can directly search for manipulations in the problem space. This may cause large perturbations, putting the value of studies on small perturbations in question. This represents a fundamental open problem that distinguishes the field of AMD from its counterparts in other application settings. 
This issue is largely unaddressed by assuming that there is an oracle for telling whether manipulated or perturbed features indeed correspond to a malware sample or not.

\smallskip

\noindent{\bf FRD 4}: {\em Quantifying the robustness and resilience of malware detectors}. Robustness and resilience of malware detectors against adversarial examples need to be quantified, ideally with a provable guarantee. For this purpose, one may adapt the reduction-based paradigm underlying the provable security of cryptographic primitives and protocols.

\smallskip

\noindent{\bf FRD 5}: {\em Designing malware detectors with provable robustness and resilience guarantees}.
Having understood the root cause(s) of adversarial examples, characterized the effect of transferability, investigated the effectiveness of practical attacks, and designed metrics for quantifying the robustness and resilience of malware detectors, it is imperative to investigate robust malware detectors with provable robustness, ideally as rigorous as what has been achieved in the field of cryptography. In this regard, robust feature extraction, adversarial learning, and verifiable learning are promising candidates for making breakthroughs.

\smallskip

\noindent{\bf FRD 6}: {\em Forecasting the arms race in malware detection}. Arms race is a fundamental phenomenon inherent to the cybersecurity domain. In order to effectively defend against adversarial malware, one approach is to deploy proactive defense, which requires the capability to forecast the arms race between malware writers and defenders. For instance, it is important to predict how attacks will evolve and what kinds of information would be necessary in order to defeat such attacks.

\section{Conclusion} \label{sec:conc}

We have presented a framework for systematizing the field of AMD through the lens of assumptions, attacks, defenses and security properties. This paves the way for precisely relating attacks and defenses. We have also shown how to apply the framework to systematize the AMD literature, including the arms race between AMD attacks and defenses. We have reported a number of insights.

The study leads to a set of future research directions. In addition to the ones described in Section \ref{sec:cd}, we mention the following two, which are discussed here because there are rarely studies on these aspects. (i) To what extent explainability (or interpretability) of ML models can be leveraged to cope with adversarial malware examples? It is intuitive that explainability could be leveraged to recognize adversarial examples because they may not be explainable \cite{DBLP:conf/itasec/DemetrioBLRA19}. 
(ii) To what extent uncertainty quantification can be leveraged to cope with adversarial malware examples? If the uncertainty associated with detectors' predictions on adversarial malware examples are inherently and substantially higher than the uncertainty associated with non-adversarial malware examples, this fact can be leveraged to recognize adversarial malware examples.
Finally, we reiterate that the research community should seek to establish a solid foundation for AMD. Although this foundation can leverage ideas and techniques from AML, the unique characteristics of AMD warrant the need of a unique foundation.

\bibliographystyle{ACM-Reference-Format}
\bibliography{mal_attacks,mal_defenses,aml,cyber_security,data_mining,mal_detection}

\end{document}